\documentclass[journal]{IEEEtran}
\usepackage{amsmath}
\usepackage{amssymb}
\usepackage{amsfonts}
\usepackage{color}
\usepackage{graphicx} 
\usepackage{dsfont}
\usepackage{mathtools}
\usepackage[noadjust]{cite}

\allowdisplaybreaks

\DeclareMathOperator*{\argmax}{arg\,max}

\newcommand{\ba}{{\boldsymbol{a}}}
\newcommand{\bc}{{\boldsymbol{c}}}
\newcommand{\bd}{{\boldsymbol{d}}}
\newcommand{\be}{{\boldsymbol{e}}}
\newcommand{\bg}{{\boldsymbol{g}}}
\newcommand{\bh}{{\boldsymbol{h}}}

\newcommand{\bk}{{\boldsymbol{k}}}
\newcommand{\bm}{{\boldsymbol{m}}}

\newcommand{\bp}{{\boldsymbol{p}}}

\newcommand{\bt}{{\boldsymbol{t}}}
\newcommand{\bu}{{\boldsymbol{u}}}
\newcommand{\bv}{{\boldsymbol{v}}}
\newcommand{\bw}{{\boldsymbol{w}}}
\newcommand{\bx}{{\boldsymbol{x}}}
\newcommand{\by}{{\boldsymbol{y}}}
\newcommand{\bz}{{\boldsymbol{z}}}

\newcommand{\bC}{{\boldsymbol{C}}}
\newcommand{\bD}{{\boldsymbol{D}}}

\newcommand{\bK}{{\boldsymbol{K}}}
\newcommand{\bM}{{\boldsymbol{M}}}
\newcommand{\bQ}{{\boldsymbol{Q}}}
\newcommand{\bR}{{\boldsymbol{R}}}
\newcommand{\bS}{{\boldsymbol{S}}}

\newcommand{\bU}{{\boldsymbol{U}}}
\newcommand{\bV}{{\boldsymbol{V}}}
\newcommand{\bW}{{\boldsymbol{W}}}
\newcommand{\bX}{{\boldsymbol{X}}}
\newcommand{\bY}{{\boldsymbol{Y}}}
\newcommand{\bZ}{{\boldsymbol{Z}}}
\newcommand{\Tau}{\mathrm{T}}

\newcommand{\bbE}{{\mathbb E}}
\newcommand{\bbZ}{{\mathbb Z}}

\newcommand{\bdel}{{\boldsymbol{\delta}}}

\newcommand{\bSgm}{{\boldsymbol{\Sigma}}}
\newcommand{\bthe}{{\boldsymbol{\theta}}}
\newcommand{\bThe}{{\boldsymbol{\Theta}}}

\newcommand{\bzero}{{\boldsymbol{0}}}
\newcommand{\bone}{{\boldsymbol{1}}}
\newcommand{\btwo}{{\boldsymbol{2}}}

\newcommand{\bri}{{\Bar{i}}}
\newcommand{\brx}{{\Bar{x}}}
\newcommand{\brI}{{\Bar{I}}}
\newcommand{\brR}{{\Bar{R}}}
\newcommand{\brX}{{\Bar{X}}}

\newcommand{\bbri}{{\boldsymbol{\bri}}}
\newcommand{\bbrI}{{\boldsymbol{\brI}}}
\newcommand{\bbrR}{{\boldsymbol{\brR}}}

\newcommand{\cA}{{\cal A}}
\newcommand{\cB}{{\cal B}}

\newcommand{\cD}{{\cal D}}
\newcommand{\cE}{{\cal E}}
\newcommand{\cG}{{\cal G}}
\newcommand{\cM}{{\cal M}}
\newcommand{\cN}{{\cal N}}
\newcommand{\cL}{{\cal L}}
\newcommand{\cQ}{{\cal Q}}
\newcommand{\cR}{{\cal R}}
\newcommand{\cT}{{\cal T}}
\newcommand{\cU}{{\cal U}}
\newcommand{\cV}{{\cal V}}
\newcommand{\cW}{{\cal W}}
\newcommand{\cX}{{\cal X}}
\newcommand{\cY}{{\cal Y}}
\newcommand{\cZ}{{\cal Z}}

\newcommand{\rmV}{{\rm V}}
\newcommand{\rmT}{{\rm T}}

\newcommand{\bcD}{{\boldsymbol{\cD}}}
\newcommand{\bcM}{{\boldsymbol{\cM}}}

\newcommand{\bcY}{{\boldsymbol{\cY}}}

\newcommand{\oP}{\overline{P}}
\newcommand{\oS}{\overline{S}}

\newcommand{\obS}{\overline{\bS}}

\newcommand{\defeq}{\stackrel{\triangle}{=}}
\newcommand{\sgn}{{\small\rm sgn}}
\newcommand{\GF}{{\rm GF}}
\newcommand{\hg}{\hat{g}}
\newcommand{\hp}{\hat{p}}

\newcommand{\hbw}{\hat{\bw}}

\newcommand{\ind}{\mathds{1}}

\newtheorem{lem}{Lemma}
\newtheorem{thm}{Theorem}
\newtheorem{remark}{Remark}
\newtheorem{cor}{Corollary}
\newtheorem{defin}{Definition}
\begin{document}
\title{Finite-Blocklength and Error-Exponent Analyses for LDPC Codes in Point-to-Point and Multiple Access Communication} 
\author{
  \IEEEauthorblockN{Yuxin Liu and Michelle Effros}\\
  \IEEEauthorblockA{Department of Electrical Engineering, California Institute of Technology, Pasadena 91125, USA.\\Email:\{yuxinl, effros\}@caltech.edu}
  \thanks{This material is based upon
work supported by the National Science Foundation under Grant No. 1817241. The work of Y. Liu is supported in part by the Oringer
Fellowship Fund in Information Science and Technology.}
}

% \markboth{IEEE Transactions on Information Theory,~Vol.~??, No.~?, Month~Year}%
% {Author \MakeLowercase{\textit{et al.}}: Paper Title}

% Use for publisher's ID mark


\maketitle
\begin{abstract}
This paper applies error-exponent and dispersion-style analyses 
to derive finite-blocklength achievability bounds 
for low-density parity-check (LDPC) codes 
over the point-to-point channel (PPC) and multiple access channel (MAC).  
The error-exponent analysis applies Gallager's error exponent to bound achievable 
symmetrical and asymmetrical rates in the MAC. 
The dispersion-style analysis begins 
with a generalization of the random coding union (RCU) bound 
from random code ensembles with i.i.d. codewords 
to random code ensembles in which codewords may be statistically dependent; 
this generalization is useful since the codewords of random linear codes 
such as random LDPC codes are dependent. 
Application of the RCU bound yields improved
finite-blocklength error bounds and asymptotic achievability results 
for i.i.d.~random codes and new finite-blocklength error bounds and achievability results for LDPC codes. 
For discrete, memoryless channels, 
these results show that LDPC codes achieve first- and second-order performance 
that is optimal for the PPC 
and identical to the best-prior results for the MAC. 
\end{abstract}
% \begin{abstract}
% This paper studies the finite-blocklength achievability performance of low-density parity-check (LDPC) codes in the point-to-point channel and multiple access channel (MAC) scenarios, using both error exponent analysis and dispersion-style analyses. For the error exponent analysis, we present achievability results for symmetrical rates in the $K$-transmitter MAC and asymmetrical rates in the $2$-transmitter MAC using Gallager's error exponent. For the dispersion-style approach, we first generalize the random coding union (RCU) bound from i.i.d. codeword design to a more general family of randomly designed codes that includes LDPC code design. We then derive a finite-blocklength error bound and asymptotic achievability result for both i.i.d. random codes and LDPC codes. We demonstrate that LDPC codes achieve optimal performance up to the second order for both point-to-point and two-transmitter discrete memoryless MACs
% \end{abstract}

\section{Introduction}
\IEEEPARstart{L}{ow}-density parity-check (LDPC) codes are linear codes designed with sparse parity-check
matrices for the purpose of enabling low complexity decoding strategies. Introduced along with corresponding iterative decoding algorithms by Gallager in 1962 \cite{Gallager:62} and largely overlooked until their rediscovery with the introduction of turbo codes \cite{Berrou:93} in the 1990s, LDPC codes are now in widespread use, playing a role in commercial standards like 10 Gb/s Ethernet (IEEE 803.3an),
WiFi (IEEE 802.11n), WiMAX (IEEE 802.16e), and the 5G standard~\cite{Richardson:18}.

\begin{table}
\caption{Summary of notations}
\centering
\begin{tabular}{|c|c|}
\hline
$n$ & blocklength/number of LDPC variable nodes \\ \hline
$r$ & number of LDPC check nodes     \\ \hline
$\lambda$ & variable node degree of regular LDPC code  \\ \hline
$\rho$ & check node degree of regular LDPC code     \\ \hline
$\bc$ & single-transmitter codebook  \\ \hline
$\bd$ & MAC codebook     \\ \hline
$\cQ$  &$\GF(q)^K$      \\ \hline
$\cG$ & bipartite LDPC graph \\ \hline
$\cV$ & vertex set of a graph $\cG$ \\ \hline
$\cE$ & edge set of a graph $\cG$ \\ \hline
$i(x;y)$ & information density \\ \hline
$C$ & channel capacity \\ \hline
$V$ & channel dispersion \\ \hline
$T$ & third-order centered moment of information density \\ \hline
$Q$ & complementary Gaussian CDF \\ \hline
$\cT_q^n$ & set of all possible types for $n$ elements from $\GF(q)$ \\ \hline
$\cT_\cQ^n$ & set of all possible types for $n$ elements from $\cQ$ \\ \hline
$\bv$ & LDPC coset vector \\ \hline
$\delta$ & LDPC quantizer \\ \hline
$\oS^n(\bt)$ & ensemble-average number of type-$\bt$ codewords/codematrices \\ \hline
$\obS^n$ & ensemble-average spectrum \\ \hline
$\cD(g)$ & Bhattacharyya parameter for input $g$ \\ \hline
$B(n,\bt)$ & multinomial coefficient \\ \hline
$E_p(R)$ & Gallager's error exponent for distribution $p$ \\ \hline
\end{tabular}
\end{table}

This paper presents achievability bounds
for the finite-blocklength performance 
of LDPC codes 
over the point-to-point channel (PPC) 
and the multiple access channel (MAC). 
Proofs employ two types of analyses.  
\begin{enumerate}
    \item Error-exponent analyses 
    generalize the techniques in~\cite{BennatanB:04} 
    to demonstrate that average error probability $\epsilon$ 
    decays exponentially in blocklength $n$ 
    with an error exponent bounded below by Gallager's error exponent. 
    This technique yields tighter bounds when $\epsilon$ is very small. 
    \item Dispersion-style analyses generalize~\cite{Polyanskiy:10}, 
    bounding the log size of the codebook achievable for a given 
    average error probability $\epsilon$ and blocklength $n$.  
    This method yields tighter bounds when $n$ is very small. 
\end{enumerate}

We begin with a brief overview of prior LDPC and linear coding analyses.

% LDPC codes, a type of linear codes with sparse parity-check
% matrices, is introduced by Gallager back in 1962 \cite{Gallager:62}, along with the corresponding iterative decoding algorithms. However, limited computing powers prevent Gallager from demonstrating the effectiveness of LDPC codes for blocklength over 500, and thus his work was mostly ignored. LDPC codes are rediscovered in 1990s after the introduction of turbo codes \cite{Berrou:93}. More and more LDPC codes have now been adopted for different commercial standards, including 10 Gb/s Ethernet (IEEE 803.3an),
% WiFi (IEEE 802.11n), and WiMAX (IEEE 802.16e).

In his 1968 text \cite[Section 6.2]{Gallager:68}, 
Gallager describes a random coset parity-check matrix code ensemble. 
Each element of the parity-check matrix 
is chosen uniformly and independently from $\{0,1\}$. 
The coset ensemble is formed by adding the same random vector 
to all codewords defined by the parity-check matrix. 
For PPCs with non-binary input alphabets, 
a ``quantization'' mapping 
maps one or more binary vectors to each channel input symbol. 
Gallager shows that the proposed code
can achieve the capacity of an arbitrary discrete, memoryless PPC (DM-PPC) 
under maximum likelihood (ML) decoding.

In~\cite{Davey:98}, 
Davey and MacKay generalize binary LDPC codes 
to finite field $\GF(q), q\geq2$, 
showing empirically that $q$-ary codes 
can significantly improve binary code performance 
for binary-input PPCs under belief propagation decoding.

The first analysis of the standard $\GF(q)$ LDPC code ensemble  appears in~\cite{BennatanB:04}.
The standard $\GF(q)$ LDPC code ensemble 
employs a random Tanner graph 
that maps the vector of variable-node edge sockets 
to a random permutation of the vector of check-node edge sockets; 
edge weights are independent and identically distributed (i.i.d.) 
uniformly on $\GF(q) \setminus \{0\}$. 
For the DM-PPC under ML decoding,~\cite{BennatanB:04}
derives an upper bound on the average error probability 
using Gallager's error exponent, 
showing that the random code has a high probability 
under sufficiently large connectivity and blocklength 
of achieving vanishing error probability 
at rates arbitrarily close to the channel capacity. 
Independently of~\cite{BennatanB:04}, 
the authors in~\cite{Erez:05} 
analyze the performance over modulo-additive PPCs 
of two different $\GF(q)$-LDPC code ensembles under ML decoding. 
The error exponents for most codes in their design 
are bounded below asymptotically 
by the random coding error exponent \cite{Erez:05}.

While the above studies focus on asymptotic behavior of LDPC  code ensembles, the increasing prevalence of delay sensitive applications motivate finite-blocklength (non-asymptotic) code analyses. For example, blocklengths of current 5G LDPC and polar codes typically range from 100 to 20000.

In~\cite{Di:02}, Di et al. analyze the finite-blocklength performance 
of LDPC codes over the binary erasure channel (BEC), 
where finite-blocklength analysis boils down to a combinatorial problem. 
The paper derives the exact average bit- and block-erasure probability 
for a given regular ensemble of LDPC codes under an iterative decoding algorithm 
and presents upper bounds on the average bit- and block-erasure probability 
for standard binary LDPC  code ensembles and the random parity-check ensemble 
under ML decoding. Other studies that focus on the BEC include \cite{Richrdson:02,Amraoui:05, Amraoui:09}. The work in \cite{Yazdani:09,Mei:19} extends the finite-blocklength analysis to general (not necessarily symmetric) binary-input channels.

Unfortunately, the above-described non-asymptotic analyses yield expressions that are either difficult to evaluate or depend on empirical performance. As a result, they provide less insight than the dispersion-style bounds (with corresponding converse results) found in \cite{Polyanskiy:10}, which accurately characterize the backoff from channel capacity using the channel dispersion $V$ and target error $\epsilon$ for blocklengths as short as 100. This observation motivates our generalization of the dispersion-style analyses to the standard LDPC code ensemble.

Yang and Meng~\cite{Yang:15} study 
Gallager's independent, uniform parity-check ensemble 
and the standard binary LDPC code ensemble 
under modified Feinstein's threshold decoding.  
Noting that codewords under these ensembles are not pairwise independent 
and therefore that Shannon-style random coding arguments do not apply, 
they derive new achievability bounds 
for memoryless binary-input output-symmetric PPCs, 
demonstrating that Gallager's parity-check ensemble bound 
is asymptotically tight up to the second order and 
that the standard LDPC code ensemble is capacity achieving. 

% The authors in \cite{Yang:15} revisit Gallager's independent, uniform parity-check ensemble, and standard binary LDPC code ensemble over memoryless binary-input output-symmetric channels. They derive new non-asymptotic achievability bounds under modified Feinstein’s threshold decoding, demonstrating that Gallager's parity-check ensemble bound is asymptotically tight up to the second order and that the standard LDPC code ensemble is capacity achieving. The authors in \cite{Yang:15} note that Shannon's random coding type argument cannot be applied to these ensembles since the codewords are not pairwise independent. 

Fewer analyses are available for LDPC codes over MACs. In \cite{Roumy:07} and \cite{Sharifi:15}, the authors study the two-user Gaussian MAC with BPSK modulation using LDPC codes. The main results in \cite{Roumy:07} are two different approximations for the density evolution, which lead to a simple linear programming optimization for MAC LDPC code design. The authors of \cite{Sharifi:15} adopt a belief propagation (BP) algorithm, and derive the probability density function (PDF) of the log-likelihood-ratios (LLRs) fed to the component LDPC decoders. The authors of \cite{Yagi:11} consider LDPC coset codes in a compound MAC with common information and analyze the performance of the proposed coset codes by deriving a lower bound on error exponents. In \cite{Lahouti:19}, Ebrahimi et al. introduce a two-layer coded channel access framework and analyze its performance over erasure adder MACs and a random access network where the number of active users is known at the receiver. The paper presents density evolution analysis in cases where the outer layer is a
long-blocklength LDPC code.

The finite-blocklength performance of the standard LDPC code ensemble under either an arbitrary DM-PPC or discrete, memoryless MAC (DM-MAC) remains an open problem. 

This paper analyzes the finite-blocklength performance of the standard $\GF(q)$ LDPC code ensemble under ML decoding using both the error-exponent approach from \cite{BennatanB:04} and dispersion-style approach from \cite{Polyanskiy:10}. 

For the error-exponent analysis, we extend the result of \cite{BennatanB:04} from the DM-PPC to symmetrical rates in the $K$-transmitter DM-MAC (DM-$K$-MAC) and arbitrary rates in the DM-$2$-MAC using Gallager's error exponent; the latter generalizes to $K$-transmitter MACs for $K>2$. We then refine the result by providing a non-asymptotic expansion of Gallager's error exponent using \cite[Exercise 5.23]{Gallager:68}.

 For the dispersion-style approach, we derive finite-blocklength error bounds and asymptotic third-order achievability results for the DM-PPC and the DM-$2$-MAC for i.i.d. codes; the achievability result is optimal up to the third order in the DM-PPC case, improving the corresponding bound on the number of codewords achievable under a desired error probability bound from a third-order term $O(\log n)$ in \cite[Th. 49]{Polyanskiy:10} to $\frac{1}{2}\log n - O(1)$ and matching the corresponding converse bound \cite[Th. 48]{Polyanskiy:10} up to the third order. For the DM-$2$-MAC, our bound improves the third-order MAC achievability bound from $-\nu\log n\bone$ with $\nu \geq 2|\cX_1||\cX_2||\cY|$ in \cite{Tan:14} to $\frac{1}{2}\log n \bone - O(1)\bone$. As noted in \cite{Yang:15}, random LDPC code are random linear codes, and the use of an underlying parity-check matrix results in statistically dependent codewords. We therefore need to generalize the random coding union (RCU) bound \cite[Th. 16]{Polyanskiy:10} from codes employing i.i.d. codeword design to a more general family of randomly designed codes that includes codes with statistically dependent codewords. We use our generalized RCU bound to derive an upper bound for the standard LDPC code ensemble with coset vector and quantization, showing that LDPC codes achieve first- and second-order performance 
that is optimal for the DM-PPC 
and identical to the best-prior results for the DM-MAC. 
 
% For the dispersion-style approach, we first prove a general random coding union (RCU) bound from \cite[Th. 16]{Polyanskiy:10}, which can be used on the standard LDPC code ensemble. We then derive finite-blocklength error bound, and asymptotic achievability results for point-to-point and two-user multiple access DMCs if the codewords are independently and identically distributed (i.i.d.), and the achievability result is optimal up to the third-order in the point-to-point case. However, the lack of pairwise codewords independence prevents us to apply the same approach to standard LDPC code ensemble. Our contribution is we derive an RCU-type upper bound for standard LDPC code ensemble with coset vector and quantization, and the result shows there is no loss in performance up to the second order, for both point-to-point and two-user multiple access DMCs.

% {\color{red} Add random access related papers if our result is completed.}

\begin{remark}
Although practical implementations of LDPC codes typically employ fast but sub-optimal decoders, it is instructive to study the performance of LDPC codes under ML decoding in order to distinguish how much performance penalty, if any, results from the application of a low density encoder and separate this impact from the impact of sub-optimal decoding.
\end{remark}

The organization of this paper is as follows. Section~\ref{sec:not} defines notation. Section~\ref{sec:PPC_MAC} introduces our channel models. Section~\ref{sec:ensemble} defines the quantized coset LDPC codes used in our study. Sections~\ref{sec:ml} and \ref{sec:2mac_ldpc_ee} apply the error-exponent approach to bound the performance of quantized coset LDPC codes with ML decoding on the DM-MAC; the analysis treats both communication at a symmetrical rate point in an arbitrary symmetrical DM-$K$-MAC, and communication at an asymmetrical rate point for an arbitrary DM-$2$-MAC. Section~\ref{sec:exp} relates the error exponent results to the dispersion-style results, revealing that the error-exponent analysis achieves a sub-optimal second-order coefficient in blocklength $n$ but a superior bound when target error probability $\epsilon$ is small. Sections~\ref{sec:rcu_p2p_px} and \ref{sec:rcu_2mac_px} present the performance of standard i.i.d. codes for the DM-PPC and DM-MAC using the RCU bound; the resulting bounds are optimal to the third-order for the DM-PPC and the tightest result to date for the DM-MAC. In Section~\ref{sec:rcu_p2p_ldpc}, we apply the generalized RCU bound to quantized coset LDPC codes, which lack the property of codeword independence used in bounding code performance in the DM-PPC. We present both a finite-blocklength error bound and an asymptotic achievability result that is optimal up to the second order. Section~\ref{sec:rcu_2mac_ldpc} extends the result to the DM-$2$-MAC, showing that LDPC codes achieve first- and second-order performance that is identical to the best-prior results for the DM-$2$-MAC.

The main results of this paper are Theorems \ref{thm:Pe},~\ref{thm:ePe}, and \ref{thm:Pe_2mac}, which bound the error exponent performance of the quantized coset LDPC code; Theorems \ref{thm:rcu_p2p_px} and \ref{thm:rcu_mac_px}, which give the finite-blocklength error bound and asymptotic achievability result for standard i.i.d. codes; and Theorems \ref{thm:rcu_p2p_ldpc} and \ref{thm:rcu_mac_ldpc}, which present a finite-blocklength error bound and asymptotic achievability result for the quantized coset LDPC code.

\section{Definitions and Notation}
\subsection{Notation} \label{sec:not}
Throughout this paper, we denote the set of integers $\{1,2,\ldots, k_1\}$ as $[k_1]$, and $\{k_1,\ldots,k_2\}$ as $[k_1:k_2]$ for any positive integers $k_1$ and $k_2$, where $[k_1:k_2] = \emptyset$ when $k_1 >k_2$. We use uppercase letters (e.g., $X$ and $Y$) for random variables, lowercase letters (e.g., $x$ and $y$) for realizations of the corresponding random variables, and calligraphic uppercase letters (e.g., $\cX$ and $\cY$) for sample spaces. To represent vectors, we use both superscripts (e.g., $x^n$ and $X^n$) and bold face (e.g., $\boldsymbol{\mathrm{x}}$ and $\bone = (1, \ldots, 1)$) when the length of the vector is clear from the context. We use both $X_i$ and $\bX[i]$ to represent the $i$th element of the vector $\bX = X^n$. For any scalar function $f(\cdot)$ and any vector $\boldsymbol{\mathrm{x}} \in \mathbb{R}^n$, $f(\boldsymbol{\mathrm{x}})$ is the vector of function values, defined as $f(\boldsymbol{\mathrm{x}}) \defeq (f(x_i),i \in [n])$. Given a set $\cZ \subseteq \mathbb{R}^n$, a vector $\bv \in \mathbb{R}^n$, and a scalar $a \in \mathbb{R}$, $a\cZ + \bv \defeq \{a\bz+\bv,\bz \in \cZ\}$.

For any joint distribution $P_{XY}$ on discrete alphabet $\cX \times \cY$, we denote the information density by
\begin{align}
%     i(x;y) \defeq \log \frac{d P_{XY}}{d(P_X \times P_Y)} (x,y) = \log \frac{dP_{Y|X=x}}{dP_Y}(y). \label{eqn:id_gen} %information density - general
% \end{align}
% Note that if $\cX$ and $\cY$ are , \eqref{eqn:id_gen} is equivalent to % \begin{align}
    i(x;y) \defeq \log \frac{P_{XY}(x,y)}{P_X(x)P_Y(y)} = \log \frac{P_{Y|X}(y|x)}{P_Y(y)}.  \label{eqn:id_d} %information density - discrete alphabet
\end{align}

Given a set $\cX$, we denote the $n$-fold Cartesian product of $\cX$ as $\cX^n$ and indicate a probability distribution on $\cX^n$ by $P_{X^n}$. For any alphabets $\cX_i, i\in [n]$ and any countable ordered set $\cA\subseteq [n]$, we define $\cX_\cA \defeq \prod_{i \in \cA} \cX_i$ and let $P_{X_\cA}$ denote a distribution on the alphabet $\cX_\cA$. We say $x_\cA \geq y_\cA$ if $x_a \geq y_a$ for all $a \in \cA$. For any joint distribution $P_{X^nY}$ on $\cX^n$, $\cY$ and any ordered sets $\cA$ and $\cB$ with $\cA \cap \cB = \emptyset$, and any $x_\cA \in \cX_\cA, x_\cB \in \cX_\cB$, and $y \in \cY$ %We say $x_\cA \geq y_\cA$ if $x_a \geq y_a, \forall a \in \cA$.
\begin{align}
    i(x_\cA ; y) &\defeq \log \frac{P_{Y| X_\cA}(y|x_\cA)}{P_{Y}(y)} \\
    i(x_\cA ; y|x_\cB) &\defeq \log \frac{P_{Y| X_\cA,X_\cB}(y|x_\cA,x_\cB)}{P_{Y|X_\cB}(y|x_\cB)}.
\end{align}

The mutual informations, dispersions, conditioned dispersions, and third centered moments of information are 
\begin{align}
    I(P_{X_\cA}) &\defeq \bbE[i(X_\cA ; Y)]\\
    I(P_{X_\cA} |P_{X_\cB}) &\defeq \bbE[i(X_\cA ; Y|X_\cB))]\\
    V(P_{X_\cA}) &\defeq \text{Var}[i(X_\cA;Y)] \\
    V(P_{X_\cA}|P_{X_\cB}) &\defeq \text{Var}[i(X_\cA;Y|X_\cB)]  \\
    V^Y(P_{X_\cA}) &\defeq \text{Var}[i(X_\cA;Y)|Y] \\
    V^Y(P_{X_\cA}|P_{X_\cB}) &\defeq \text{Var}[i(X_\cA;Y|X_\cB)|Y]  \\
    T(P_{X_\cA}) &\defeq \bbE[|i(X_\cA;Y) - I(P_{X_\cA})|^3]\\
    T(P_{X_\cA}|P_{X_\cB}) &\defeq \bbE[|i(X_\cA;Y|X_\cB) -  I(P_{X_\cA} |P_{X_\cB})]|^3].
\end{align}

The cumulative distribution function (CDF) and PDF for standard Gaussian distribution $\cN(0,1)$ are denoted by
\begin{align}
    \Phi(x) &\defeq \frac{1}{\sqrt{2\pi}}\int_{-\infty}^x e^{-\frac{u^2}{2}}du, \\
    \phi(x) &\defeq \frac{1}{\sqrt{2\pi}}e^{-\frac{x^2}{2}},
\end{align}
respectively. The function $Q(\cdot)$ denotes the standard Gaussian complementary CDF
\begin{align}
    Q(x) \defeq 1 - \Phi(x) =\frac{1}{\sqrt{2\pi}}\int_{x}^\infty e^{-\frac{u^2}{2}}du,
\end{align}
and $Q^{-1}(\cdot)$ is the inverse function of $Q(\cdot)$.

We use $P_{X(1)\ldots X(M)}$ to denote the distribution of a codebook with $M$ codewords. For any ordered set $\cA \subseteq [M]$, the notation $X(\cA) = (X(i),i \in \cA)$ captures a subset of the codewords.

Throughout this paper, the base of all logarithms and exponentials, unless otherwise indicated, is $q$, where prime power $q$ specifies the alphabet for the $\GF(q)$-LDPC code defined in the next section. We employ standard $o(\cdot)$ and $O(\cdot)$ notations writing $f(n) = o(g(n))$ if $\lim_{n\rightarrow \infty} |\frac{f(n)}{g(n)}| = 0$ and $f(n) = O(g(n))$ if there exist constants $a$ and $n_0$ such that $|f(n)|\leq a|g(n)| $ for all $n>n_0$.
% $\lim \sup_{n\rightarrow \infty} |\frac{f(n)}{g(n)}| < \infty$.
% Throughout this paper, the base of all logarithms, including these used in entropy function $H(\cdot)$, is always $q$ (unless otherwise indicated). We employ standard $o(\cdot)$ and $O(\cdot)$ notations: $f(n) = o(g(n))$ if $\lim_{n\rightarrow \infty} |\frac{f(n)}{g(n)}| = 0$, and $f(n) = O(g(n))$ if $\lim \sup_{n\rightarrow \infty} |\frac{f(n)}{g(n)}| < \infty$.
\subsection{Channel Models: DM-PPC and DM-MAC} \label{sec:PPC_MAC}
% In this subsection, we define the DM-PPC and the DM-MAC 
\begin{defin} (DM-PPC)
A DM-PPC is described by
\[(\cX,P_{Y|X},\cY),\]
where $\cX$ and $\cY$ are the discrete channel input and output alphabets, respectively, and $P_{Y|X}(y|x)$ specifies the channel transition probability for all $x\in\cX$ and $y\in \cY$. The $n$-th order extension $(\cX^n,P_{Y^n|X^n},\cY^n)$ of $(\cX,P_{Y|X},\cY)$ satisfies $\Pr[y_k|x^k,y^{k-1}] = \Pr[y_k|x_k]$ for all $k \in [n]$.
\end{defin} 
\begin{defin} (DM-$K$-MAC)
A DM-$K$-MAC is defined by
% \[(\cX_1\times \cX_2\times \cdots \times \cX_K,P_{Y|\bX},\cY)\]
\[\left(\prod_{i=1}^K \cX_i,P_{Y|\bX},\cY\right)\]
where $\cX_i,i\in[K]$, and $\cY$ are the discrete channel input and output alphabets, respectively, and $P_{Y|\bX} = P_{Y|X_1,X_2,\ldots,X_K}$ is the channel transition probability. A DM-$K$-MAC is called {\bf symmetric} if all transmitters have the same input alphabet $\cX_i = \cX$ for all $i\in[K]$ and
\[P_{Y|\bX}(y|\bx) = P_{Y|\bX}(y|\pi(\bx))\]
for all $y\in\cY$, $\bx \in \cX^K$, and permutations $\pi$ on $[K]$.
\end{defin}
% \section{Codes with Identical Encoders}\label{sec:ensemble}
\subsection{Quantized Coset Codes}\label{sec:ensemble}
%Consider an arbitrary, symmetrical $K$-transmitter discrete memoryless MAC.
We begin with a formal definition of the quantized coset $\GF(q)$-LDPC code used in our study.

\begin{figure}  [h]
        \includegraphics[width=0.48\textwidth]{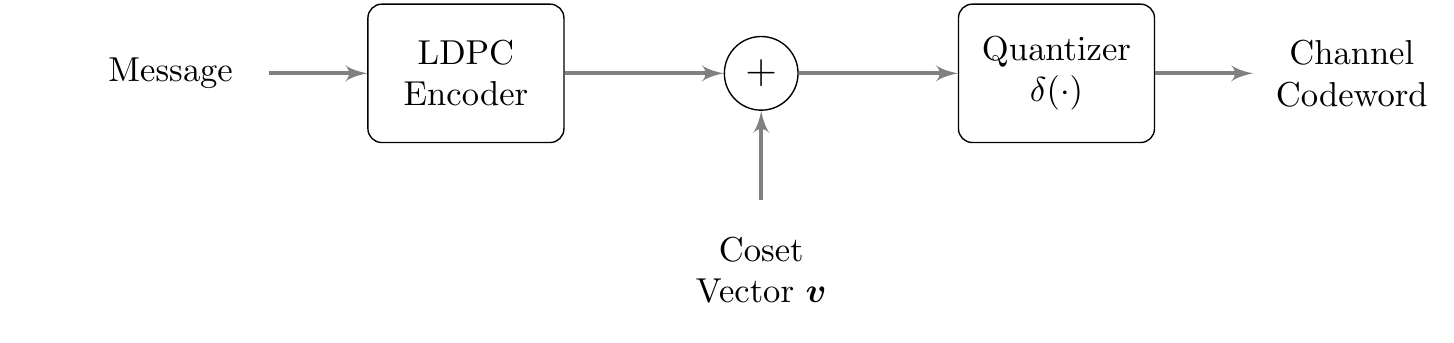} 
      \caption{Encoding of Quantized Coset LDPC Code}
      \label{fig:qc_ldpc_design}
\end{figure}

For any prime power $q$ and finite field $\GF(q)$, a quantized coset $\GF(q)$-LDPC code is defined by three components: a standard LDPC encoder, a coset vector $\bv$, and a quantizer $\delta$, defined below and illustrated in Figure \ref{fig:qc_ldpc_design}.

\begin{defin} (Standard $\GF(q)$-LDPC code)
A {\bf standard $\GF(q)$-LDPC code} is defined using a bipartite Tanner graph $\cG=(\cV,\cE)$ with $n$ variable nodes, $r$ check nodes, and edge set $\cE\subseteq[n]\times[r]$. 
For each $(i,j)\in \cE$, $(i,j)$ represents an undirected edge 
connecting the $i$th variable node and the $j$th check node; 
each edge $(i,j)\in \cE$ carries a constant $g_{i,j}\in\GF(q)\setminus\{0\}$.  
The notation 
\begin{eqnarray*}
% \cN(i) & = & \{j:(i,j)\in \cE\}, \\
\cN(j) & \defeq & \{i:(i,j)\in \cE\},
\end{eqnarray*}
captures the neighborhood of check node $j\in[r]$ 
resulting from edge set $\cE$. 

The $n$ variable nodes hold a column vector $\bu$ from $\GF(q)^n$. Vector $\bu$ is a {\bf codeword} if it satisfies all check nodes, 
giving   
\[
\sum_{i\in\cN(j)}g_{i,j}u_i = 0\ \ \forall j\in[r];
\]
the linear equation operates in $\GF(q)$. The set of all $M = |\bc|$ codewords constitute the {\bf codebook}
\begin{align*}
    \bc = \{\bc_1,\ldots, \bc_M\} \subseteq \GF(q)^n
\end{align*}
for the given Tanner graph $\cG=(\cV,\cE)$.
\end{defin}

Following~\cite{Elias:55,BennatanB:06},
we do not transmit codewords from the LDPC encoder 
but instead apply quantized coset coding.  
\begin{defin} (Coset $\GF(q)$-LDPC Code)
    Given a Tanner graph $\cG=(\cV,\cE)$ and the corresponding LDPC codebook $\bc$, we obtain the coset LDPC code by adding a constant vector $\bv$, called the {\bf coset vector}, to each codeword $\bc_i \in \bc$. The addition $$\bc_i + \bv, i \in [M]$$ is performed component-wise in $\GF(q)$. The set $\{\bc_i+\bv,i \in [M]\}$ is the codebook for the {\bf coset $\GF(q)$-LDPC code}.
\end{defin}

\begin{defin} (Quantized Coset $\GF(q)$-LDPC Code)
    Given an LDPC codebook $\bc$ and a coset vector $\bv$, we map each symbol from $\bc_i +\bv$ to a symbol from the channel input alphabet $\cU$ using {\bf quantizer} $\delta$:
    \begin{align}
        \delta: \GF(q) \rightarrow \cU.
    \end{align}
    Mapping $\delta$ is applied component-wise; we therefore employ notation
    \begin{align*}
        \delta(\bc_i+\bv) \defeq [\delta(\bc_i[j]+\bv[j])]_{j \in [n]} = [\delta((\bc_i+\bv)[j])]_{j \in [n]}
    \end{align*}
    for coset codeword $\bc_i + \bv$.
    The set $\{\delta(\bc_i+\bv),i \in [M]\}$ is the codebook for the {\bf quantized coset $\GF(q)$-LDPC code}.
\end{defin}

The quantizer $\delta$ enables us to approximate, using a code on $\GF(q)$, any rational probability mass function $P_U$, for which $P_U(u)$ is an integer multiple $N_u$ of $1/q$ for every $u\in \cU$ (giving $P_U(u) = N_u/q$). This is achieved by mapping $N_u$ elements to each channel input symbol $u\in \cU$.

\begin{remark}
The quantization $\delta(\cdot)$ is an essential component in code designs for arbitrary (not necessarily symmetric) DM-PPCs since unequal channel transition probabilities between input and
output symbols can lead to non-uniform capacity-achieving
input distributions. The performance penalty for using a uniform input distribution in place of the optimal input distribution is called the {\bf shaping gap}.
\end{remark}

Our analysis focuses on a random ensemble of quantized coset $\GF(q)$-LDPC codes. 
We restrict attention to regular Tanner graphs, in which 
all left nodes have degree $\lambda$ and 
all right nodes have degree $\rho$. A random graph is chosen by 
first labeling the $|\cE|$ edge sockets from left nodes from $1$ to $|\cE|$, 
then labeling the $|\cE|$ edge sockets from right nodes from $1$ to $|\cE|$, 
and finally choosing a permutation $\pi$ 
uniformly at random from the set of permutations on $[|\cE|]$. 
The graph connects each left node edge socket $i$  to the right-node edge socket $\pi_i$.
The edge constant $g_{i,j}$ for each edge $(i,j)\in \cE$ 
is chosen uniformly and independently at random from $\GF(q)\setminus\{0\}$.
%Denote this ensemble by ${\rm LDPC}(\lambda,\rho;n)$.
\begin{remark}
An attracting property of regular LDPC codes is that the minimum distance grows linearly with blocklength~\cite{Gallager:62}, therefore regular LDPC codes achieve superior performance than irregular LDPC codes under ML decoding. In contrast, lower iterative decoding threshold makes irregular LDPC codes outperform regular LDPC codes under iterative decoding~\cite{Costello:14}.
\end{remark}
The design rate of the described ensemble
is $R$ $q$-ary symbols per channel use, where 
	\[
	R \defeq1-\frac{r}{n} = 1 - \frac{\lambda}{\rho}.
	\]
The actual number of legitimate codewords is $q^{nR}$ if the parity-check matrix corresponding to the randomly drawn Tanner graph has full rank and larger if that parity-check matrix does not have full rank. We restrict the operational rate to equal the design rate by choosing exactly $q^{n R}$ active codewords for use in coding.
Before communication begins, the codebook, coset vector, and quantizer are revealed to all parties, so that the receiver knows which $M = q^{nR}$ codewords are employed and how they are processed. We refer to the process of selecting precisely $q^{nR}$ codewords for active use and effectively removing others from the codebook as {\bf codeword removal}.

\begin{defin} (Codeword Removal)
Given an ensemble of $\GF(q)$-LDPC codes with design rate $R$, the codeword removal process generates an ensemble by dividing the probability of each code in the original ensemble equally among all code(s) corresponding to a distinct combination of $q^{nR}$ codewords from the original code.
\end{defin}
We denote the random ensemble of $\GF(q)$-LDPC codes resulting from random Tanner graph design by ${\rm LDPC}({\rm Full},\lambda,\rho;n)$; the random ensemble of $\GF(q)$-LDPC codes -- after codeword removal but before coset addition or application of quantizer $\delta$ -- by ${\rm LDPC}(\lambda,\rho;n)$; and the random ensemble of quantized coset $\GF(q)$-LDPC codes by ${\rm LDPC}(\lambda,\rho,\delta;n)$.
% To unify our notation,
% \begin{itemize}
%     \item ${\rm LDPC}(\lambda,\rho;n)$ is the ensemble from the random $(\lambda,\rho)$ LDPC graph and uniform codeword removal;
%     \item ${\rm LDPC}(\lambda,\rho,\delta;n)$ is the ensemble from the random $(\lambda,\rho)$ LDPC graph, uniform codeword removal, uniform coset selection, and fixed quantizer $\delta$.
% \end{itemize}
\section{Error-exponent bounds for LDPC Code ensemble on MAC}
\subsection{Error-Exponent Bound for LDPC Code Ensemble on the DM-$K$-MAC with Identical Encoders}\label{sec:ml}
In this section, we consider an arbitrary, symmetric DM-$K$-MAC and derive the expected ensemble error probability under ML decoding. In this analysis, we assume that all transmitters employ the same random codebook from the ${\rm LDPC}(\lambda,\rho;n)$ ensemble, but each is offset by an independent random coset vector $\bv_j, j\in [K]$. All transmitters employ the same quantizer $\delta(\cdot)$. 

For a fixed LDPC graph with $M=q^{nR}$ codewords $\bc_1,\ldots,\bc_M\in\GF(q)^n$, 
the {\bf single-transmitter codebook} for transmitter $k$ is 
\[
\bc_{(k)}=\{\bc_1,\ldots,\bc_M\}\subseteq\GF(q)^{n}
\]
for each $k\in[K]$. 
The {\bf MAC codebook} is the set of codematrices
\[
\bd=\{\bd_{\bm}:\bm\in[M]^k\} \subseteq \GF(q)^{n \times K}
\]
that result from those codewords, 
where for any $\bm=(m(1),\ldots,m(K))\in[M]^K$, 
$\bd_\bm=(\bc_{m(1)},\ldots,\bc_{m(K)})$.  

We denote the MAC ensemble before restriction of codematrices by ${\rm LDPC}_K({\rm Full},\lambda,\rho;n)$. After random selection of $q^{nR}$ codewords from which we build MAC codematrices, we denote the MAC ensemble before and after applying the random coset matrix and fixed quantization by ${\rm LDPC}_K(\lambda,\rho;n)$ and ${\rm LDPC}_K(\lambda,\rho,\delta;n)$, respectively.

Let $\bv$ denote the coset matrix formed by combining the $K$ coset vectors column-wise, giving $\bv = [\bv_1 \bv_2 \cdots \bv_K]$. We map each symbol from the matrix $\bd_\bm+\bv \in \GF(q)^{n\times K}$ to a symbol from the channel input alphabet $\cU$ using the (component-wise) quantizer $\delta$. The resulting channel input is 
\[
\delta(\bd_\bm+\bv).
\]

\begin{remark}
As noted in \cite{Recep:20_ARXIV}, using the same codebook from the ${\rm LDPC}(\lambda,\rho;n)$ ensemble for all transmitters has practical advantages. In our case, each device is the same except for its unique random coset vector $\bv_j$. When considering an arbitrary (not necessarily symmetric) DM-$K$-MAC or an arbitrary rate vector, a different quantized coset LDPC code ${\rm LDPC}(\lambda_j,\rho_j,\delta_j;n), j\in[K]$ can be applied to each transmitter. For simplicity of notation, we assume in this section that both the MAC and the desired rate are symmetric. General MACs and rate vectors are studied in Section~\ref{sec:2mac_ldpc_ee} for the case of $K=2$. %, but the proof goes through unchanged in the general case.
\end{remark}

In order to analyze the expected ensemble error probability for some fixed value $(\lambda,\rho)$, 
% In order to analyze the expected error probability 
% of a code randomly chosen from the ensemble described in Section~\ref{sec:ensemble}
% for some fixed value $(\lambda,\rho)$, 
we require a means of describing the distribution over the types of codematrices. 
The following definitions are useful for that discussion.  

For any matrix $\ba\in\GF(q)^{n\times K}$, 
recall that $\cQ\defeq\GF(q)^K$ specifies the alphabet of each row of $\ba$. Let $\cT^n_\cQ(\ba)$ denote 
the {\bf type} that results when we view $\ba$ as a list of $n$ 
elements from alphabet $\cQ$, 
giving 
\begin{eqnarray*}
\cT_\cQ^n(\ba) & = & (t(g):g\in\cQ), \\
t(g) & = & \sum_{i=1}^n1(a[i,*]=g).
\end{eqnarray*}
If $\ba=\bd_\bm$ for some codematrix $\bd_\bm$, 
then $\cT_Q^n(\ba)$ captures, for each $g\in\cQ$, the number of time steps 
when the ($K$-dimensional) row of codematrix $\bd_\bm$ takes value $g$.
The {\bf set of possible types} is 
\[
\cT_\cQ^n \defeq\{\cT_\cQ^n(\ba):\ba\in\GF(q)^{n\times K}\}\subset \bbZ_+^{|\cQ|}.
\]

For any MAC codebook $\bd$, let 
\[
\bS_\bd^n = (S^n_\bd(\bt):\bt\in\cT_\cQ^n)
\]
represent the {\bf spectrum of codebook $\bd$}, 
where for any type $\bt\in\cT_\cQ^n$, 
\begin{equation}\label{eqn:defS}
S_\bd^n(\bt)  = \sum_{\bm}1(\cT_\cQ^n(\bd_\bm)=\bt)
\end{equation}
is the number of codematrices of type $\bt$ in MAC codebook $\bd$.
When the code is chosen at random 
(e.g., through random LDPC graph design and random codeword removal), 
we use 
\begin{align}
\obS^n \defeq E_\bD[\bS_\bD^n]=(\oS^n(\bt):\bt \in \cT_\cQ^n)  \label{eqn:obs_n} %=(\oS^n(\bt):\bt \in \cT_\cQ^n)  
\end{align}
to represent the {\bf ensemble-average spectrum} 
of the random codebook $\bD$, where $E_\bD[\cdot]$ here captures the expectation with respect to the random choice of codebook $\bD$.

The following notation 
is used in the statement of Theorem~\ref{thm:Pe}.
Given a discrete, memoryless $K$-transmitter MAC  
with input alphabet $\cX=\cU^K$, channel transition $P_{Y|X}$, and quantizer $\delta(\cdot)$, 
let $\bcD=(\cD(g):g\in\cQ)$, where 
\begin{equation}\label{eqn:defD}
\cD(g)\defeq \frac1{q^K}\sum_{g'\in\cQ}\sum_y\sqrt{P_{Y|X}(y|\delta(g'))P_{Y|X}(y|\delta(g'+g))}
\end{equation}
is the extension of Bhattacharyya parameter to non-binary channels.

For any type $\bt\in\cT^n_\cQ$, let $\bcD^\bt$ be the product of terms $\cD(g)$ resulting from type $\bt$, giving
\begin{equation}\label{eqn:defDt}
\bcD^\bt \defeq \prod_{g\in\cQ}\cD(g)^{t(g)},
\end{equation}
and let $B(n,\bt)$ denote the number of distinct matrices $\ba\in\GF(q)^{n\times K}$ of type $\bt$, which is the multinomial coefficient
\[
B(n,\bt) \defeq  \frac{n!}{\prod_{g\in \cQ} t_g!}.
\]

Theorem~\ref{thm:Pe} derives an upper bound on the ensemble-average error probability for the LDPC  code ensemble as a function of the product of Bhattacharyya parameter $\bcD^\bt$, ensemble-average number of codematrices $\oS^n(\bt)$, and Gallager's error exponent $E_p(\cdot)$, defined below.
\begin{thm}\label{thm:Pe}
Let $P_{Y|X}$ be the transition probability 
for a symmetric DM-$K$-MAC 
with input alphabet $\cX=\cU^K$ and output alphabet $\cY$.  
Let the MAC's maximal symmetrical rate vector 
be the $K$-vector $(C,\ldots,C)$, 
and fix any $\bR=(R,\ldots,R)$ with $R<C$. 
Let $P_U$ be a pmf on $\cU$ for which $P_U(u)= N_u/q$ for some integer $N_u$ for each $u\in\cU$,
and let $\delta:\GF(q)\rightarrow\cU$ be a quantization matched to $P_U$. Consider any ensemble of random $K$-MAC LDPC codes, denoted by $\cL$, with codeword removal and blocklength $n$, symmetrical rate $\bR$, and ensemble-average spectrum $\obS^n$. 
% Consider the ${\rm LDPC}_K(\lambda,\rho,\delta;n)$ ensemble of blocklength $n$ and symmetrical rate $\bR$. Denote the ensemble-average spectrum of ${\rm LDPC}_K(\lambda,\rho;n)$ by $\obS_L^n$. %(before applying the coset vector and quantization).
% Consider a random ensemble $K$-transmitter MAC codes, in which each transmitter employs the same code from the ${\rm LDPC}(\lambda,\rho;n)$ ensemble, offset by an independent coset vector and same quantizer $\delta$, of symmetrical rate $\bR$, and ensemble-average spectrum $\obS^n$.  

Let $\Tau\subseteq\cT_\cQ^n$ be any fixed set of types. Then for any blocklength $n$, the ensemble-average error probability of the quantized coset-shifted ensemble of $\cL$
under ML decoding is bounded as 
\[
E[P_e^{(n)}]\leq \sum_{\bt\in \Tau}\oS^n(\bt) \bcD^\bt
+q^{-nE_p(KR+(\log\alpha_{\rm \scalebox{0.4}{MAC}})/n)}, 
\]
% \[
% {\color{red} +q^{-nE_p(KR+(\log\alpha_{\rm {MAC}})/n)}}
% \]
where $E_p(\cdot)$ is Gallager's error exponent for the input distribution $P_X = P_{U^K}=P_{U}^K$, defined using 
\begin{align}
E_p(R) & \defeq  \max_{0\leq \rho\leq 1} [E_0(\rho,P_X)-\rho R], \\
E_0(\rho,P_X) & \defeq  -\log\sum_y\left[\sum_{x\in\cU^K}P_X(x)P_{Y|X}(y|x)^{1/(1+\rho)}\right]^{1+\rho},
\end{align}
and 
\begin{equation}\label{eqn:alpha}
\alpha_{\rm \scalebox{0.4}{MAC}} =  \max_{\bt\in \Tau^c}\frac{\oS^n(\bt)}{(M^K-1)B(n,\bt)q^{-nK}}.
\end{equation}
Here $\Tau^c=\cT_\cQ^n\setminus \Tau\setminus \{\cT_{\cQ}^n(\bzero)\}$, where
$\cT_\cQ^n(\bzero)$ is the type of the all zero codematrix, 
and $M=q^{nR}$.  
\end{thm}

{\em Proof:}  See Appendix~\ref{app:Pe}.

\begin{remark}
The tightest bound for each blocklength $n$ in Theorem~\ref{thm:Pe} can be obtained by optimizing over the set of types $\Tau$. 
\end{remark}

\begin{remark}
The error probability expression in Theorem~\ref{thm:Pe} takes the same form for different ensembles of $K$-MAC LDPC codes, but the ensemble-average spectrum, $\obS^n$, and consequently $\frac{\log \alpha_{\rm \scalebox{0.4}{MAC}}}{n}$ vary for different ensembles.
\end{remark}

Theorem~\ref{thm:Pe} captures the error bound in two terms. In Theorem~\ref{thm:ePe} below, we demonstrate that the first term in Theorem~\ref{thm:Pe} can be made equal to zero for some non-trivial choice of $\Tau$ provided that we first expurgate (remove) codes with small minimum distance. The definition of our expurgated code ensemble follows.

% Recall that each codematrix $\bc_\bm$ in a codebook $\bc$ 
% has a type $\bt=(t_g:g\in\cQ)=\cT_\cQ^n(\bc_\bm)$, 
% where $t_g$ describes 
% the number of time steps $i$ for which $\bc[i,*]=g$. 
% Dividing this number by $n$ gives a rational pmf on $\cQ$.  
% The spectrum $\bS_\bc^n=(S_\bc(\bt):\bt\in\cT_\cQ^n)$ 
% of codebook $\bc$ 
% then describes the number of codematrices of each type 
% in codebook $\bc$.

% Before proceeding to these analyses, we pause to compare the derived random LDPC code bound to the corresponding bound in the traditional random coding argument.

\begin{defin} (${\rm LDPC}_K$-${\rm Ex}_\sigma(\lambda,\rho,\delta;n)$ ensemble) \label{def:exp_code}
Let $P_L(\bD)$ denote the probability of observing a randomly chosen code $\bD$ from the ${\rm LDPC}_K({\rm Full},\lambda,\rho;n)$ ensemble. The expurgated MAC LDPC  code ensemble {\bf ${\rm LDPC}_K$-${\rm Ex}_\sigma({\rm Full},\lambda,\rho;n)$} is the ensemble obtained by placing probability zero on all codes of minimum distance less than or equal to $\sigma n$, and probability $\Pr_L(\bD|d_{\min}(\bD)>\sigma n)$ on the remaining codes, giving
\begin{align}
    \Pr\nolimits_{{\rm ex},\sigma}(\bD) =\begin{cases}
    0,& \mbox{if } d_{\min}(\bD)\leq \sigma \\
    \Pr_L(\bD|d_{\min}(\bD)>\sigma n),& \mbox{otherwise}.
    \end{cases}
\end{align}
Here the distance between two $n\times K$ codematrices $\bd_1$ and $\bd_2$, denoted by $d(\bd_1,\bd_2)$, is the number of rows that differ,
\[d(\bd_1,\bd_2)=\sum_{i=1}^n \ind(\bd_1[i,*] \neq \bd_2[i,*])\]
and the minimum distance of codebook $\bd$ is 
\[d_{\min}(\bd) = \min_{\bm \neq \bm'} d(\bd_\bm,\bd_{\bm'}).\]
Applying the codeword removal process to the ${\rm LDPC}_K$-${\rm Ex}_\sigma({\rm Full},\lambda,\rho;n)$ generates the ${\rm LDPC}_K$-${\rm Ex}_\sigma(\lambda,\rho;n)$ ensemble, and applying the coset addition and quantization to the ${\rm LDPC}_K$-${\rm Ex}_\sigma(\lambda,\rho;n)$ ensemble gives the the quantized coset MAC LDPC  code ensemble ${\rm LDPC}_K$-${\rm Ex}_\sigma(\lambda,\rho,\delta;n)$.
\end{defin}

For any $\lambda\geq3$, the probability that an LDPC code drawn from the ${\rm LDPC}({\rm Full},\lambda,\rho;n)$ ensemble has a small minimum distance decays exponentially to zero as the blocklength $n$ grows \cite[Th. 6]{BennatanB:04}. In Appendix~\ref{app:dMin}, we show that the same bound applies after restriction to our fixed-rate code.

Since expurgation eliminates the first term in Theorem~\ref{thm:Pe}, the remainder of Theorem~\ref{thm:ePe} works to demonstrate that the second term in Theorem~\ref{thm:Pe} has the desired property.

In the DM-PPC, Gallager's error exponent has the property that $E_p(R) >0$ for all $R<C$. Here, similarly, $E_p(KR) > 0$ for all $KR<KC$ in the DM-$K$-MAC (where $KR$ is the symmetrical sum-rate and $KC$ is the maximum symmetrical sum-rate). Notice, however, that the second term in Theorem~\ref{thm:Pe} employs $E_p(KR+\frac{\log\alpha_{\rm \scalebox{0.4}{MAC}}}{n})$ rather than $E_p(KR)$. Theorem~\ref{thm:ePe} therefore also seeks to evaluate the rate offset $\frac{\log \alpha_{\rm \scalebox{0.4}{MAC}}}{n}$ in Gallager's error exponent. Using a series of supporting theorems provided in Appendix~\ref{app:log_a}, Theorem~\ref{thm:ePe} shows that this rate offset can be made arbitrarily small. More precisely, Theorem~\ref{thm:ePe} shows that if $\rho = \kappa n$ and $\kappa \rightarrow 0$ no more quickly than $\bThe(\frac{\log n}{n})$, then $\frac{\log\alpha_{\rm \scalebox{0.4}{MAC}}}{n}$ decays as $O(\frac{\log n}{n})$. Therefore, our proposed code design is asymptotically capacity achieving. 

% Theorem~\ref{thm:ePe} derives the ensemble-average error probability for the expurgated ensemble ${\rm LDPC}_K$-${\rm Ex}_\sigma(\lambda,\rho,\delta;n)$.
\begin{thm} \label{thm:ePe}
Let $P_{Y|X}$ be the transition probability 
for a discrete memoryless $K$-transmitter MAC 
with input alphabet $\cX=\cU^K$ 
and output alphabet $\cY$.  
Let the MAC's maximal symmetrical rate vector 
be the $K$-vector $(C,\ldots,C)$, 
and fix any $\bR=(R,\ldots,R)$ with $R<C$. 
Let $P_U$ be a pmf on $\cU$ for which $P_U(u)$ is an integer multiple of $1/q$ for each $u\in\cU$,
and let $\delta:\GF(q)\rightarrow\cU$ be a quantization matched to $P_U$. Let $\Delta R>0$ be some arbitrary number. Then for large enough $\rho$ and $n$, there exists LDPC parameters $(\lambda,\rho)$ for which the ensemble-average error probability for ${\rm LDPC}_K$-${\rm Ex}_\sigma(\lambda,\rho,\delta;n)$ ensemble under ML decoding is bounded as: 
% Let $C, R, p(u)$ and $\delta(\cdot)$ be defined as in Theorem \ref{thm:Pe}.
\[
E_{{\rm ex}}[P_e^{(n)}]\leq q^{-nE_p(KR+\Delta R)},
\]
where $E_p(\cdot)$ is Gallager's error exponent defined in Theorem \ref{thm:Pe}. Further, if $\rho = \kappa n$for some $\kappa$ that approaches zero no more quickly than $\bThe(\frac{\log n}{n})$, then the minimum rate offset $\Delta R$ decays as $O(\frac{\log n}{n})$.
% for large enough $n$.
\end{thm}

{\em Proof:}  See Appendix~\ref{app:ePe}.
\begin{remark}
Theorem~\ref{thm:ePe} provides an upper bound on the ensemble-average error probability in terms of Gallager's error exponent $E_p(R)$ for input distribution $P_X = P_U^K$. Here $P_U$ is restricted to be a rational pmf for which, for all $u\in \cU$, $P_U(u)= N_u/q$ for some integer $N_u$. By choosing $P_U$ to approximate the capacity-achieving input distribution, we obtain $E_p(R)>0$ for all $R<C$. Therefore, the ensemble-average error probability of ${\rm LDPC}_K$-${\rm Ex}_\sigma(\lambda,\rho,\delta;n)$ asymptotically approaches 0, and the existence argument of a deterministic capacity-achieving quantized coset-shifted LDPC MAC code follows. However, note that the nature of the quantizer $\delta(\cdot)$ restrict achievable $P_U$ to be integer multiples of $\frac{1}{q}$. When the optimal input distribution $P_X^*$ is irrational or not an integer multiple of $\frac{1}{q}$, then a large alphabet size $q$ may be required to closely approximate $P_X^*$.
\end{remark}

Our study chooses $M=q^{nR}$ codewords uniformly at random from the set of $q^{nR_C}\geq q^{nR}$ valid parity-check solutions. This approach differs from most other studies of LDPC codes, which assume that the parity-check matrix of a code randomly chosen from the ${\rm LDPC}(\lambda,\rho;n)$ ensemble has full rank, giving $R_\bC=R$. This assumption is not precise, but it does become increasingly probable in the limit of large parity-check matrices. The following theorem formalizes this observation and demonstrates that the probability that the actual rate $R_C$ deviates from the design rate $R$ decays exponentially in the blocklength $n$.

\begin{thm}\label{thm:full_R}
Consider the ensemble ${\rm LDPC}(\lambda,\rho;n)$ without random codeword removal. Let $R \defeq  1 - \frac{\lambda}{\rho}$ denote the design rate of the ensemble and let $R_\bC$ denote the actual rate of a code $\bC$ from the ensemble using the full collection of legitimate codewords. For any $\epsilon >0$, there exists some integer $n(\epsilon)$ such that for $n > n(\epsilon)$
\begin{equation} \label{eq:full_R}
    \Pr[R_\bC - R >\epsilon] \leq q^{-n\epsilon/2}.
\end{equation}
In addition, for any $\epsilon>0$, there exists a $T_0>0$ such that for all $n>n(\epsilon)$
\begin{equation}  \label{eq:full_R_2}
    \bbE[R_\bC - R] \leq T_0\frac{\log n}{n}.
\end{equation}
\end{thm}
{\em Proof:}  See Appendix~\ref{app:full_R}. %The proof of Theorem \ref{thm:full_R} uses result from Theorem \ref{thm:BML}. Therefore its proof is shown after the proof of Theorem \ref{thm:BML}.

\subsection{Error-Exponent Bound for LDPC Code Ensemble on the DM-$2$-MAC}\label{sec:2mac_ldpc_ee}
While the previous section treats the ensemble-average error probability for a symmetrical $K$-transmitter MAC with a symmetrical rate vector, this section gives the corresponding bound for a general $2$-transmitter MAC with an arbitrary rate vector.

We first define the achievable rate region of a $2$-transmitter MAC under a fixed input distribution. We then present the main error-exponent bound when LDPC code ensembles are employed. 

\begin{defin}
Let $P_{Y|X_1,X_2}$ be the transition probability 
for an arbitrary DM-$2$-MAC. Let $\cR(P_{X_1},P_{X_2})$ be the set of $(R_1,R_2)$ such that 
\begin{align}
    R_1 &< I(X_1;Y|X_2) \\
    R_2 &< I(X_2;Y|X_1) \\
    R_1+R_2 &< I(X_1,X_2;Y),
\end{align}
where the mutual informations are evaluated according to distribution $P_{Y|X_1,X_2}P_{X_1}P_{X_2}$.
\end{defin}

\begin{thm}\label{thm:Pe_2mac}
Let $P_{Y|X_1,X_2}$ be the transition probability 
for an arbitrary DM-$2$-MAC
with input alphabet $\cX=\cX_1\times \cX_2$ and output alphabet $\cY$.  
Let $P_{X_i}$ be a pmf on $\cX_i$ for which $P_{X_i}(x_i)$ is an integer multiple of $1/q$ for each $x_i\in\cX_i$ and $i\in \{1,2\}$. Let $\delta_i:\GF(q)\rightarrow\cX_i$ be the corresponding quantization matched to $P_{X_i}$, $i \in \{1,2\}$.  
Assume transmitter $i$ employs a random code from the ${\rm LDPC}(\lambda_i,\rho_i,\delta_i;n)$ ensemble $i\in\{1,2\}$ with independent coset vector $\bv_i$, such that the rate vector $(R_1,R_2) \in \cR(P_{X_1},P_{X_2})$. Then for any blocklength $n$, the ensemble-average error probability under ML decoding is bounded as 
\begin{align}
E[P_e^{(n)}]&\leq q^{-nE_{p_1}(R_1+\frac{\log\alpha_1}{n})}+q^{-nE_{p_2}(R_2+\frac{\log\alpha_2}{n})} \nonumber \\
&\hphantom{=}+q^{-nE_{p_{12}}(R_1+R_2+\frac{\log\alpha_{12}}{n})},
\end{align}
where $E_{p_1}(\cdot),E_{p_2}(\cdot)$ and $E_{p_{12}}(\cdot)$ are Gallager's error exponents for the input distributions $P_{X_1}, P_{X_2}$ and $P_X=P_{X_1}P_{X_2}$, defined using 
\begin{align}
E_{p_1}(R) & \defeq  \max_{0\leq \rho\leq 1} [E_0^{1}(\rho,P_{X_1})-\rho R], \label{eqn:Error_exp_2mac_a1}\\
E_{p_2}(R) & \defeq  \max_{0\leq \rho\leq 1} [E_0^{2}(\rho,P_{X_2})-\rho R], \label{eqn:Error_exp_2mac_a2}\\
E_{p_{12}}(R) & \defeq  \max_{0\leq \rho\leq 1} [E_0^{12}(\rho,P_X)-\rho R], \label{eqn:Error_exp_2mac_a12}\\
E_0^1(\rho,P_{X_1}) & \defeq  -\log\sum_y\sum_{x_2 \in \cX_2} P_{X_2}(x_2) \nonumber \\
&\hphantom{=}\left[\sum_{x_1\in\cX_1}P_{X_1}(x_1)P_{Y|X_1,X_2}(y|x_1,x_2)^{\frac{1}{1+\rho}}\right]^{1+\rho}, \label{eqn:Error_exp_2mac_a1_0}\\
E_0^2(\rho,P_{X_2}) & \defeq  -\log\sum_y\sum_{x_1 \in \cX_1} P_{X_1}(x_1) \nonumber \\
&\hphantom{=}\left[\sum_{x_2\in\cX_2}P_{X_2}(x_2)P_{Y|X_1,X_2}(y|x_1,x_2)^{\frac{1}{1+\rho}}\right]^{1+\rho},\label{eqn:Error_exp_2mac_a2_)} \\
E_0^{12}(\rho,P_X) & \defeq  -\log\sum_y \left[ \sum_{x_1 \in \cX_1}\sum_{x_2\in\cX_2} \right.\nonumber \\
&\hphantom{=}\left.  P_{X_1}(x_1)P_{X_2}(x_2)P_{Y|X_1,X_2}(y|x_1,x_2)^{\frac{1}{1+\rho}}\right]^{1+\rho},\label{eqn:Error_exp_2mac_a12_0}
\end{align}
and 
\begin{align}
\alpha_1 &=  \max_{\bt\in \cT_q^n\setminus\{\cT_q^n(\bzero)\}}\frac{\oS^n_{1}(\bt)}{(M_1-1)B(n,\bt)q^{-n}}, \label{eqn:alp1_2mac}\\
\alpha_2 &=  \max_{\bt\in \cT_q^n\setminus\{\cT_q^n(\bzero)\}}\frac{\oS^n_{2}(\bt)}{(M_2-1)B(n,\bt)q^{-n}},\label{eqn:alp2_2mac}\\
\alpha_{12} &=  \alpha_1\alpha_2. \label{eqn:alp12_2mac}
\end{align}
Here $\cT_q^n$ is the set of possible types for $n$ elements from alphabet $\GF(q)$, $\cT_q^n(\bzero)$ is the type of the all-zero codeword, $\oS^n_{i}(\bt)$ is the ${\rm LDPC}(\lambda_i,\rho_i;n)$ ensemble-average number of type-$\bt$ vectors, and $M_i=q^{nR_i}$ for $i\in\{1,2\}$.
\end{thm}

{\em Proof:}  See Appendix~\ref{app:Pe_2mac}.

Theorem~\ref{thm:Pe_2mac} presents an upper bound, which is valid for any blocklength $n$, on the ensemble-average error probability for an arbitrary DM-$2$-MAC when each transmitter employs a random code from the ${\rm LDPC}(\lambda_i,\rho_i,\delta_i;n)$ ensemble. The final expression is a function of three error exponents. Note from \cite{Liu:96} that all error exponents, $E_{p_1}(R_1),E_{p_2}(R_2)$, and $E_{p_{12}}(R_1+R_2)$ are positive when the rate pair $$(R_1,R_2) \in \cR(P_{X_1},P_{X_2}).$$ 
Note that the quantizers $\delta_i(\cdot), i\in\{1,2\}$ restrict achievable input distributions $P_{X_i},i\in\{1,2\}$ to be integer multiples of $\frac{1}{q}$, rate pairs $(R_1,R_2)$ that require irrational input distributions or rational input distributions with non-integer multiples of $\frac{1}{q}$ may require large alphabet size $q$ to closely approximate the desired input distributions.

However, restricting the ensemble from standard i.i.d. random codes to LDPC codes incurs the rate offset penalties $\frac{\log \alpha_1}{n},\frac{\log \alpha_2}{n}$, and $\frac{\log \alpha_{12}}{n}$.

To eliminate these rate offsets, one can apply the expurgation technique from Lemma~\ref{lem:dMin} in Appendix~\ref{app:dMin} to remove codes with small minimum distances for both ${\rm LDPC}(\lambda_1,\rho_1,\delta_1;n)$ and ${\rm LDPC}(\lambda_2,\rho_2,\delta_2;n)$ ensembles. The same argument in Theorem~\ref{thm:ePe} can then be used to prove these rate offsets can be made arbitrarily small, with large enough blocklength $n$, $\rho_1$, and $\rho_2$. More precisely, when $\rho_1 = \kappa_1 n$ and $\rho_2 = \kappa_2 n$ for some $\kappa_1$ and $\kappa_2$ that decay no more quickly than $\bThe(\frac{\log n}{n})$, these rate offsets decay as $O(\frac{\log n}{n})$ provided that (see Appendix~\ref{app:ePe} for details). Therefore, the proposed quantized coset-shifted LDPC MAC codes are capable of achieving any rate pair $(R_1,R_2) \in \cR(P_{X_1},P_{X_2})$.

The true capacity region for the DM-$2$-MAC is the convex closure of the set $$\cR \defeq \bigcup_{P_{X_1}P_{X_2}}\cR(P_{X_1},P_{X_2}),$$
for all $P_{X_1}P_{X_2}$. To justify any rate pair in the capacity region $\cR$ is achievable with the proposed quantized coset-shifted LDPC MAC codes, one can apply the standard time sharing technique \cite{Han:1981} to introduce an auxiliary random variable $W\in \cW$ with $|\cW|\leq 2$. The two quantizers are then defined to be dependent on the auxiliary random variable, giving the distribution $P_W(w)P_{X_1|W}(x_1|w)P_{X_2|W}(x_2|w)$.

\subsection{Finite-Blocklength Bound via Error Exponent}\label{sec:exp}
We next seek to relate Gallager's error exponent bound \cite{Gallager:68} to the dispersion-style bound \cite{Polyanskiy:10}, which accurately approximates the maximal achievable rate in the non-asymptotic regime.

We begin with a short overview of both results. In \cite{Polyanskiy:10}, Polyanskiy et al. bound the maximal code size $M^*(n,\epsilon)$ achievable with error probability $\epsilon$ and blocklength $n$ as a function of the channel capacity $C$, the channel dispersion $V$, and the inverse complementary Gaussian CDF $Q^{-1}(\cdot)$. The resulting bound is reproduced as Theorem \ref{thm:Pe_Pol} below. 

% $$\frac{1}{n}\log_2 M^*(n,\epsilon) \approx C- \sqrt{\frac{V}{n}}Q^{-1}(\epsilon),$$ where $n$ is the blocklength, $C$ is the capacity, $V$ is the channel dispersion, $\epsilon$ is the targeted error probability, $Q$ is the complementary Gaussian cumulative distribution function, and $M^*(n,\epsilon)$ is the maximal code size achievable with error probability $\epsilon$ and blocklength $n$, we note the following results by Polyanskiy and Gallager.

\begin{thm} (\cite[Cor. 51]{Polyanskiy:10}). \label{thm:Pe_Pol} %average error probability by Polyanskiy
For a DM-PPC, if $0 < \epsilon \leq \frac{1}{2}$, then
\begin{align}
     \frac{\log_2 M^*(n,\epsilon)}{n} \geq C  - \sqrt{\frac{ V_{\min}}{n}}Q^{-1}(\epsilon) + O\left(\frac{1}{n}\right), \label{eqn:Pe_Pol}
\end{align}
where $V_{\min}$ is the minimal channel dispersion over all capacity-achieving channel input distributions.  
\end{thm}

The same paper also bounds the dispersion $V$ for DM-PPCs.

\begin{thm} (\cite[Th.~50]{Polyanskiy:10}). \label{thm:Vbd} %Dispersion bound
Consider a DM-PPC  with input alphabet $\cX$ and output alphabet $\cY$ such that $\min\{|\cX|,|\cY|\} > 2$. Then
\begin{align}
    V\leq 2\log_2^2(\min\{|\cX|,|\cY|\}) -C^2. \label{eqn:Pe_Pol2}
\end{align}
For DM-PPCs with $\min\{|\cX|,|\cY|\} = 2$, the upper bound becomes
\begin{align}
    V\leq 1.2 \log_2^2e -C^2. \label{eqn:Pe_Pol3}
\end{align}
\end{thm}

While Theorems \ref{thm:Pe_Pol} and \ref{thm:Vbd} together bound the maximal code size, and therefore rate, as a function of the blocklength $n$ and error probability $\epsilon$, Gallager's error exponent bounds error probability as a function of the blocklength $n$ and rate $R$, as described in Theorem \ref{thm:Pe_Gal}.

\begin{thm} (\cite[Th. 5.6.2., Corollary 1]{Gallager:68}). \label{thm:Pe_Gal} %average error probability by Gallager
% Given a DMC with transition probability $P_{Y|X}$, for any positive integer $n$ and positive number $R$, consider the ensemble of length-$n$ block codes, in which each letter of each $1\leq m \leq \lceil e^{nR} \rceil$ codeword is independently selected with probability assignment $P_X$. The ensemble-average probability of decoding error using maximum likelihood decoding satisfies
Given a DM-PPC with transition probability $P_{Y|X}$, for any positive integer $n$ and positive number $R$, consider the ensemble of length-$n$ block codes, in which each symbol of each codeword $m$, $m \in \left[\lceil e^{nR}\rceil \right]$, is independently drawn according to $P_X$. The ensemble-average probability of decoding error using ML decoding satisfies
\begin{align}
    \oP_e \leq e^{-nE_p(R)}, \label{eqn:Pe_Gal}
\end{align}
where $E_p(R) = \max_{0\leq \rho\leq 1} [E_0(\rho,P_X)-\rho R]$ is Gallager's random coding error exponent for input distribution $P_X$ defined in Theorem \ref{thm:Pe}.
\end{thm}
\begin{remark}
Note that the bound \eqref{eqn:Pe_Gal} also applies to an ensemble of random linear codes, see \cite[Section 6.2]{Gallager:68}. Therefore, there is no loss in performance for using only linear codes in Gallager's approach.
\end{remark}

\begin{thm} (\cite[Exercise~5.23]{Gallager:68}). \label{thm:ebd} %error exponent bound
Given a DM-PPC with transition probability $P_{Y|X}$, \eqref{eqn:Pe_Gal} can be bound as
\begin{align}
    \oP_e \leq e^{\left[-n \frac{(C-R)^2}{8/e^2 + 4 (\log_e |\cY|)^2} \right]}, \forall R \in [0,C],~ \label{eqn:Pe_Gal2}
\end{align}
where $|\cY|$ is the size of the output alphabet.

This bound results from a power series expansion of $E_p(R)$ evaluated at the capacity achieving input distribution $P_X$. Bounding the second derivative $E_0''(\rho,P_X)$ with respect to $\rho$ from below yields the given lower bound on $E_p(R)$.
\end{thm}
{\em Proof:}  An outline is shown in Appendix~\ref{app:ebd}. 

Note that a stronger bound can be proved by following the approach outlined in \cite[Exercise~5.23]{Gallager:68}, as shown in Corollary \ref{cor:ebd} below.
\begin{cor} \label{cor:ebd} Given a DM-PPC with transition probability $P_{Y|X}$, \eqref{eqn:Pe_Gal} can be bound as 
\begin{align}
    \oP_e \leq e^{\left[-n \frac{(C-R)^2}{8/e^2 + 2 (\log_e |\cY|)^2 - 2R_{cr}^2} \right]}, \label{eqn:Pe_Gal2.5}
\end{align}
for $R \in [\max\{0,C- (\frac{4}{e^2} +\log_e^2|\cY| -R_{cr}^2)\},C]$. Here $R_{cr}\defeq E_0'(1,P_X)$ is the critical rate \cite[Eq. (5.6.30)]{Gallager:68}.
\end{cor}
\begin{remark}
Gallager's error exponent $E_p(R)$ is a lower (achievability) bound on the true error exponent (known as the reliability function \cite[eq. 5.8.8]{Gallager:68}) for a given $R$.
A key property of the critical rate $R_{cr}$ is that for rates $R \in (R_{cr},C)$, Gallager's error exponent $E_p(R)$ equals the sphere-packing upper bound (converse) of the true error exponent \cite[Section 5.8]{Gallager:68}).
\end{remark}

% {\em Proof:}  An outline is shown in Appendix~\ref{app:ebd}. In fact, a stronger bound can be proved by following the approach outlined in \cite[Exercise~5.23]{Gallager:68}
% \begin{align}
%     E_p(R) \geq \frac{(C-R)^2}{8/e^2 + 2 (\log_e \cY)^2 - [E_0'(\rho,Q)]^2},
% \end{align}
% for $R \in [\max\{0,C- (\frac{4}{e^2} +\log_e^2|\cY| -[E'_0(\rho,Q)]^2)\},C]$.

Let the ensemble-average error probability $\oP_e$ be the targeted error probability $\epsilon$. The stronger bound \eqref{eqn:Pe_Gal2.5} can be rearranged as
\begin{align}
    R \geq C - \sqrt{\frac{8/e^2 + 2 (\log_e|\cY|)^2 -2R_{cr}^2}{n} \log_e \frac{1}{\epsilon}}. \label{eqn:Pe_Gal3}
\end{align}

Polyanskiy's and Gallager's strategies yield random coding achievability bounds. For Polyanskiy's approach, Theorem~\ref{thm:Pe_Pol} bounds the rate as a function of the channel's capacity and dispersion, while Theorem~\ref{thm:Vbd} bounds the dispersion of a DM-PPC in terms of the input and output alphabet sizes of the channel. In Gallager's approach, Theorem~\ref{thm:ebd} bounds the error probability using the capacity and (only) the output alphabet size. Comparing these two approaches yield the following observations.

% For Polyanskiy's approach, Theorem~\ref{thm:Pe_Pol} gives an achievability bound using the capacity and dispersion, and Theorem~\ref{thm:Vbd} bounds the dispersion of a DM-PPC in terms of the input/output alphabet size of the channel. On the other hand, Theorem~\ref{thm:ebd} in Gallager's approach bounds the error probability using the capacity and only the output alphabet size. Comparing these two approaches, we note
\begin{itemize}
    \item The first order term in \eqref{eqn:Pe_Pol} (Polyanskiy's approach) and \eqref{eqn:Pe_Gal3} (Gallager's approach) are both the channel capacity $C$.
    \item Polyanskiy's approach yields a tighter coefficient in the second-order term. Precisely, the second-order terms in \eqref{eqn:Pe_Pol} and \eqref{eqn:Pe_Gal3} are both $O(\sqrt{1/n})$. However, the upper bound on the coefficient in Polyanskiy's approach \eqref{eqn:Pe_Pol} is $$ 2\log_2^2(\min\{|\cX|,|\cY|\}) -C^2 \mbox{ ($C$ in bits)},$$ which is tighter than the coefficient in Gallager's approach \eqref{eqn:Pe_Gal3} 
    % (after multiplying by a factor of $\log_2^2(e)$ due to different bases) 
    \begin{align*}
        &\hphantom{.=}8/e^2 + 2 (\log_e|\cY|)^2 -2R_{cr}^2 \hspace{18pt}\mbox{ ($R_{cr}$ in nats)}\\
        &=\frac{8\log_2^2(e)}{e^2} + 2 (\log_2|\cY|)^2 -2R_{cr}^2 \mbox{ ($R_{cr}$ in bits)}.
    \end{align*} % From \eqref{eqn:E0_1} to \eqref{eqn:E0_3}, we know $E_0'(\rho,Q)$ is maximized at $\rho = 0$, which equals to the $C$ at the capacity achieving distribution $Q$ (\cite[Eq. 5.6.30]{Gallager:68}).
    Therefore, we conclude that the error-exponent approach yields a sub-optimal coefficient in the $\sqrt{1/n}$ term.
    \item Gallager's approach yields a better scaling at small error probability $\epsilon$. More precisely, for a given targeted error probability $\epsilon$, the $\sqrt{1/n}$ term in \eqref{eqn:Pe_Pol} (Polyanskiy's approach) scales as $Q^{-1}(\epsilon)$, while the corresponding term in \eqref{eqn:Pe_Gal3} (Gallager's approach) scales as $\sqrt{\log_e \frac{1}{\epsilon}}$. A comparison between these scaling terms is shown in Figure~\ref{fig:Qlog}, which confirms the advantage of the error-exponent approach (originally designed for analyzing exponentially small error) at small $\epsilon$.
    
    \begin{figure} 
        \includegraphics[width=0.45\textwidth]{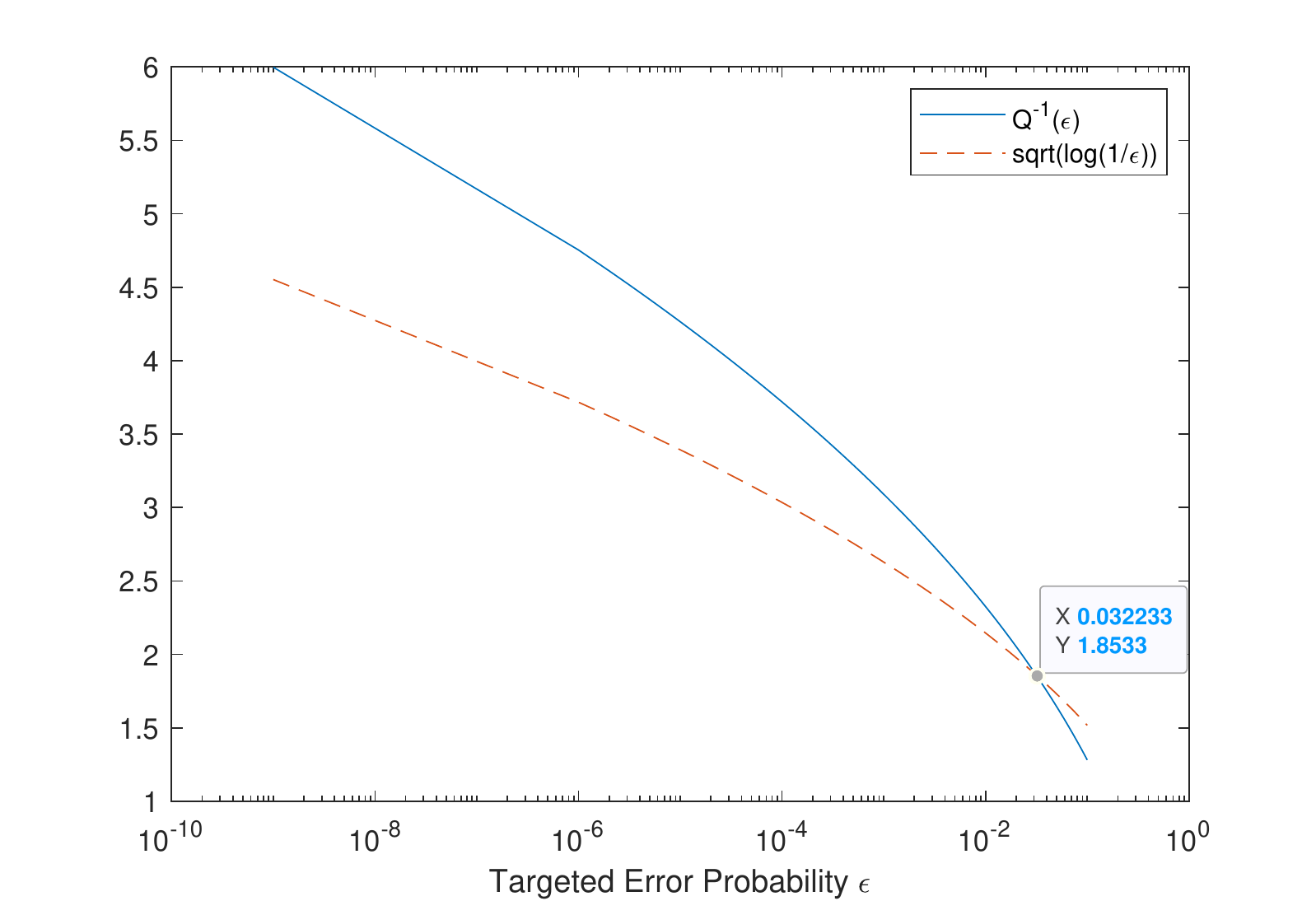} 
      \caption{Comparison between $Q^{-1}(\epsilon)$ with
      $\sqrt{\log_e\frac{1}{\epsilon}}$.}
      \label{fig:Qlog}
    \end{figure}
\end{itemize}

Applying the outcome of the error-exponent approach to Theorem~\ref{thm:ePe}, we obtain the following achievability result.
\begin{thm} \label{thm:exp_achi}
    Let $P_{Y|X}$ be the transition probability
for a symmetrical DM-$K$-MAC 
with input alphabet $\cX=\cU^K$ 
and output alphabet $\cY$.  
Let $C, P_U$, and $\delta(\cdot)$ be defined as in Theorem \ref{thm:Pe}. Then there exist LDPC parameters $(\lambda,\rho)$ for which the expurgated ensemble ${\rm LDPC}_K$-${\rm Ex}_\sigma(\lambda,\rho,\delta;n)$ contains at least one code with average error probability less than $\epsilon$ under ML decoding and 
\begin{align}
\lefteqn{R \geq} \nonumber \\
&\frac{1}{K}\left[KC - \sqrt{\frac{8\log^2(e)/e^2 + 2 (\log|\cY|)^2}{n} \log \frac{1}{\epsilon}} - \frac{\log \alpha_{\rm ex}}{n}\right],
\end{align}
where
\begin{align}
    \alpha_{\rm ex} = \max_{\bthe \in J_\sigma} \frac{\oS_{{\rm ex},\sigma}^{n}(n\bthe)}{(M^K-1)B(n,n\bthe)q^{-nK}},
\end{align}
and $\oS_{{\rm ex},\sigma}^{n}(n\bthe)$ is the average spectrum of the expurgated ensemble. If $\rho = \kappa n$ and $\kappa$ approaches zero no more quickly than $\bThe\left(\frac{\log n}{n}\right)$, then $\frac{\log \alpha_{\rm ex}}{n}=O\left(\frac{\log n}{n}\right)$.
\end{thm}
\begin{remark}
The error-exponent approach imposes a sub-optimal $\sqrt{1/n}$ second-order term even for codes drawn i.i.d. from $P_X$. The additional penalty for using LDPC codes instead of i.i.d. $P_X$ codes is $\frac{\log \alpha_{\rm ex}}{n}$, which is $O\left(\frac{\log n }{n}\right)$ for large enough $\rho$ as shown in the proof of Theorem~\ref{thm:ePe}. This observation raises the question of whether the dispersion-style approach can be applied to LDPC code, and, if so, whether the LDPC code can achieve the optimal second-order term. To answer this question, we first review the derivation of dispersion-style bound and tighten a prior result for the PPCs.
\end{remark}

\section{RCU Bounds for I.I.D. Codes}
\subsection{RCU Bound for I.I.D. Code on the DM-PPC}\label{sec:rcu_p2p_px}
In \cite{Polyanskiy:10}, Polyanskiy, Poor, and Verd\'u  study the PPC using techniques including the RCU bound, the dependency-testing (DT) bound, and the $\kappa\beta$ bound. We here build on the RCU bound, which employs the optimal ML decoder.

Theorem \ref{thm:RCU}, below, presents a slightly more general version of the non-asymptotic RCU bound from \cite[Th. 16]{Polyanskiy:10}.
The key difference between Theorem~\ref{thm:RCU} and \cite[Th. 16]{Polyanskiy:10} is that the RCU bound in \cite{Polyanskiy:10} requires all codewords to be drawn i.i.d. according to $P_X$ while Theorem~\ref{thm:RCU} requires only that the marginal distribution on each codeword equals $P_X$. For example, Theorem~\ref{thm:RCU} can be applied to codes whose codewords are dependent, in which case the joint distribution $P_{X\brX}(a,b)$ on a pair of codewords $X$ and $\brX$ is not equal to $P_X(a)P_X(b)$ for some $(a,b)\in \cX^2$.

\begin{thm} (RCU bound, modified from \cite[Th. 16]{Polyanskiy:10}) \label{thm:RCU} 
Consider an ensemble of codes with $M$ codewords drawn according to some $P_{X{(1)}X{(2)}\ldots X{(M)}}$ such that 
\begin{align}
    % P_{X(1)} &= P_X,~ \forall i \in [M], \label{eqn:rcu_con1}\\
    P_{X{(\cA)}}&= P_{X{(\cB)}},~ \forall \cA,\cB \subseteq [M] \mbox{ s.t. }|\cA| = |\cB| \label{eqn:rcu_con2}.
\end{align}
Under ML decoding, the ensemble-average error probability $\epsilon$ satisfies
\begin{align}
    \epsilon \leq \bbE\left[\min\{1, (M-1)\Pr[i(\brX;Y)\geq i(X;Y)|X,Y]\} \right], \label{eqn:RCU}
\end{align}
where 
\begin{align}
    P_{X\brX Y}(a,b,c) &= P_{X\brX}(a,b)P_{Y|X}(c|a) \\
    P_{X\brX}(a,b) &= P_{X(1)X(2)}(a,b).
\end{align}
\end{thm}
{\em Proof:} Denote the conditional error probability given that the $j$-th codeword is sent by $\epsilon_j$, then the average error probability is
\begin{align}
    \epsilon_{\text{avg}} = \frac{1}{M}\sum_{j=1}^M \epsilon_j.
\end{align}
By the symmetry of both the code design (implied by \eqref{eqn:rcu_con2}) and the ML decoder
\begin{align}
    \bbE[\epsilon_{\text{avg}}] = \bbE[\epsilon_1], 
\end{align}
where the expectation is taken over the random codebook design. 

The ML decoder $g(\cdot)$ gives
\begin{align}
    g(y) &=\argmax_{j\in [M]} ~P_{Y|X}(y|X(j)) \\
    &=\argmax_{j\in [M]} ~\frac{P_{Y|X}(y|X(j))}{P_Y(y)}\\
    &=\argmax_{j\in [M]} ~i(X(j);y).
\end{align}
For the case of a tie, the decoder chooses uniformly at random among the most probable codewords. %For the worst case scenario, we assume ties are included in the error event. 

Given that the first codeword $X(1)$ is transmitted, an error or tie occurs when the channel output is some value $y$ for which
\begin{align}
    \exists ~j\in [M]\setminus \{1\}, \text{ s.t. } i(X(j);y) \geq i(X(1);y).
\end{align}
Therefore, $\bbE[\epsilon_1]$ can be bounded from above as
\begin{align}
    \lefteqn{\bbE[\epsilon_1]} \nonumber\\
    &\leq \Pr\left[\bigcup_{j=2}^M \{i(X(j);Y)\geq i(X(1);Y)\} \right] \label{eqn:RCU_1}\\
    &= \bbE\left[\Pr\left[\bigcup_{j=2}^M \{i(X(j);Y)\geq i(X(1);Y)\}|X(1),Y \right] \right]\label{eqn:RCU_2} \\
    &\leq \bbE\left[\min\{1,(M-1) \right. \nonumber \\
    &\hphantom{=====}\left. \Pr\left[ \{i(X(2);Y)\geq i(X(1);Y)\}|X(1),Y \right] \}\right] \label{eqn:RCU_3},
\end{align}
where \eqref{eqn:RCU_1} is an inequality as the decoder might resolve some ties correctly, \eqref{eqn:RCU_2} follows from the law of iterated expectation, and \eqref{eqn:RCU_3} holds by union bound and the bounded nature of probability. Note that \eqref{eqn:RCU_3}  follows since all terms in the union bound are equal as the conditional distribution of all $X(2), \ldots, X(M)$ given the transmitted $X(1)$ are the same by the symmetry of code design.
\hfill$\blacksquare$

It is useful to notice that the bound in Theorem~\ref{thm:RCU} equation \eqref{eqn:RCU} takes the same form for all choices of $P_{X{(1)}X{(2)}\ldots X{(M)}}$ satisfying \eqref{eqn:rcu_con2}, but that the evaluation of that bound varies with the precise dependence or independence of $X_1$ and $X_2$ or, equivalently, $X$ and $\brX$ under the chosen code distribution. For example, the value of $\Pr[i(\brX;Y)\geq i(X;Y)|X,Y]$ is exactly one when $\brX = X$ with probability one, but it is less than or equal to one for other choices of $P_{X{(1)}X{(2)}}$. While we begin by evaluating Theorem~\ref{thm:RCU} under the case of independent codewords ($P_{X{(1)}X{(2)}\ldots X{(M)}} = (P_X)^M$), we require the more general form for evaluating LDPC codes, where codewords are not drawn i.i.d. but instead result from a shared Tanner graph.

% \begin{remark}
% The key difference in Theorem~\ref{thm:RCU} and \cite[Th. 16]{Polyanskiy:10}) is we allow a more general code design. The RCU bound in \cite{Polyanskiy:10} requires all codewords to be drawn i.i.d. according to $P_X$, while Theorem~\ref{thm:RCU} only requires codeword to have the same marginal distribution $P_X$ and the code design to be symmetric. For example, Theorem~\ref{thm:RCU} can be applied to codes with these properties, but $P_{X\brX}(a,b)$ may not equal to $P_X(a)P_X(b), \forall (a,b)\in \cX^2$ .
% \end{remark}
We now follow the approach in \cite[Th. 5]{Chen:19_ARXIV} to apply Theorem~\ref{thm:RCU} and two other important theorems to prove the achievability bound in Theorem~\ref{thm:rcu_p2p_px}. The given analysis tightens the achievability result from a third-order term $O(\log n)$ in \cite[Th. 49]{Polyanskiy:10} to $\frac{1}{2}\log n - O(1)$, yielding a result that matches the corresponding converse bound \cite[Th. 48]{Polyanskiy:10} up to the third order.

% \begin{remark} \label{rmk:why_rcu_ldpc}
% Theorem~\ref{thm:rcu_p2p_px}, stated below, does not hold for LDPC codes, since the derivation of \eqref{eqn:iid_px_property} relies on the independence between codewords, which is absent in LDPC code. % Note \eqref{eqn:iid_px_property} relies on the independence of $X^n$ and $\brX^n$ of i.i.d. codes, which is absent in LDPC codes. In LDPC code, when one codeword is known, one can eliminate 
% \end{remark}

\begin{thm} (Random coding finite-blocklength bound and asymptotic third-order-optimal achievability for the PPC). \label{thm:rcu_p2p_px}
Consider a DM-PPC with channel transition probability $P_{Y|X}$ and capacity achieving distribution $P_X$. If each symbol of each codeword is drawn i.i.d. according to $P_X$, then there exists a blocklength-$n$ code with $M$ codewords and average error probability $\epsilon$ such that for any blocklength $n$
\begin{align}
    \epsilon &\leq \bbE\left[\min\left\{1, M\frac{A(P_X)}{\sqrt{n}}\exp(-i(X^n;Y^n))\right\} \right],
\end{align}
and for large enough $n$ 
\begin{align}
    \frac{\log M}{n} \geq C - \sqrt{\frac{V(P_X)}{n}}Q^{-1}(\epsilon) + \frac{\log n}{2n} - O\left(\frac{1}{n}\right), \label{eqn:rcu_p2p_px_result}
\end{align}
provided the following moment assumptions are satisfied when $X\sim P_X$
\begin{align}
    I(P_X) &>0, \\
    V(P_X) &>0 \label{eqn:rcu_p2p_a1},\\
     V^Y(P_X) &>0 \label{eqn:rcu_p2p_a1_5},\\
    T(P_X) &< \infty, \label{eqn:rcu_p2p_a2}
\end{align}
where
\begin{align}
    I(P_X) &= \bbE[i(X;Y)], \label{eqn:rcu_p2p_I}\\
    V(P_X) &= \text{Var}[i(X;Y)], \label{eqn:rcu_p2p_V}\\
    V^Y(P_X) &= \text{Var}[i(X;Y)|Y], \label{eqn:rcu_p2p_Vr}\\
    T(P_X) &= \bbE[|i(X;Y) - \bbE[i(X;Y)]|^3], \label{eqn:rcu_p2p_T} \\
    B(P_X) &\defeq \frac{C_0 T(P_X)}{V(P_X)^{3/2}} \\
     A(P_X) &\defeq 2\left(\frac{\log 2}{\sqrt{2\pi V(P_X)}} +2 B(P_X)\right).
\end{align}
\end{thm}

The proof of Theorem~\ref{thm:rcu_p2p_px} relies on the Berry-Ess\'een inequality and \cite[Lemma 47]{Polyanskiy:10}, as stated in Theorem~\ref{thm:BET} and Lemma~\ref{lem:Pol47}, respectively.
\begin{thm} (Berry-Ess\'een Theorem, \cite[Chapter XVI.5]{Feller:71}). \label{thm:BET}
Let $Z_1,\ldots, Z_n$ be a sequence of independent random variables with distribution $Z_j \sim P_{Z_j}$. Assume that 
\begin{align}
    \bbE[Z_j] &= \mu_j, \forall j\in\{1,\ldots,n\}, \label{eqn:BE_A1} \\ % Berry-Ess\'een Assumption 1
    \mu &= \frac{1}{n}\sum_{j=1}^n \mu_j, \label{eqn:BE_A2} \\
    V &= \frac{1}{n}\sum_{j=1}^n \text{Var}[Z_j] > 0, \label{eqn:BE_A3}\\
    T &= \frac{1}{n}\sum_{j=1}^n\bbE[|Z_j-\mu_j|^3] <\infty. \label{eqn:BE_A4}
\end{align}
Then for any $-\infty < \lambda < \infty$ and $n\geq1$
\begin{align}
    \left| \Pr\left[\sum_{j=1}^n Z_j \geq n\left(\mu + \lambda\sqrt{\frac{V}{n}}\right) \right] -Q(\lambda) \right| \leq \frac{C_0 T}{V^{3/2}}\frac{1}{\sqrt{n}},
\end{align}
where $C_0 \leq 0.5583$ for independent random variables, and $C_0 \leq 0.4690$ for i.i.d. random variables \cite{Shevtsova:14}. 

The exact value of $C_0$ does not affect the results in this paper. We employ $C_0 = 0.5583$ even for the i.i.d. case.
\end{thm}

\begin{lem} (\cite[Lemma 47]{Polyanskiy:10}). \label{lem:Pol47}
% Let $Z_1,\ldots, Z_n$ be a sequence of independent random variables with distribution $Z_j \sim P_{Z_j}$. Employ the notations from Theorem~\ref{thm:BET} and assume that \eqref{eqn:BE_A1} to \eqref{eqn:BE_A4} are satisfied, then for any $A$
Let $Z_1,\ldots, Z_n$ be a sequence of independent random variables with distribution $Z_j \sim P_{Z_j}$. Assume
\begin{align}
     V &= \frac{1}{n}\sum_{j=1}^n \text{Var}[Z_j] > 0,\\
    T &= \frac{1}{n}\sum_{j=1}^n\bbE[|Z_j-\mu_j|^3] <\infty.
\end{align}
Then for any constant $\zeta$
\begin{align}
    \lefteqn{\bbE\left[\exp\left\{-\sum_{j=1}^n Z_j\right\} \ind\left\{\sum_{j=1}^n Z_j \geq \zeta\right\} \right]} \\
    &\leq 2\left(\frac{\log 2}{\sqrt{2\pi V}}+2 \frac{C_0T}{V^{3/2}}\right)\frac{1}{\sqrt{n}}\exp(-\zeta).
\end{align}
\end{lem}

The proof of Theorem~\ref{thm:rcu_p2p_px} follows the proof of a similar source coding argument in \cite[Th. 5]{Chen:19_ARXIV}.

{\em Proof of Theorem~\ref{thm:rcu_p2p_px}:} Setting $X = X^n, \brX = \brX^n, Y =Y^n$ in Theorem~\ref{thm:RCU}, we note that the ensemble-average error probability $\epsilon'$ satisfies
\begin{align}
    \epsilon' \leq \bbE\left[\min\{1, M\Pr[i(\brX^n;Y^n)\geq i(X^n;Y^n)|X^n,Y^n]\} \right],
\end{align}
where 
\begin{align*} 
P_{X^n\brX^nY^n}(x^n,\brx^n,y^n)&= P_{X^n,\brX^n}(x^n,\brx^n)P_{Y^n|X^n}(y^n|x^n)\\ &=P_{X^n}(x^n)P_{X^n}(\brx^n)P_{Y^n|X^n}(y^n|x^n),
\end{align*}
as the codewords are drawn i.i.d. according to $P_{X^n} = P_{X}^n$.

Denote for brevity
\begin{align}
    I_n &\defeq i(X^n;Y^n) = \sum_{j=1}^n i(X_j;Y_j) \\
    \brI_n &\defeq i(\brX^n;Y^n) = \sum_{j=1}^n i(\brX_j;Y_j),
\end{align}
where $V(P_X)$ and $T(P_X)$ are the second-order moment and third-order central moment of the information density, respectively as defined in \eqref{eqn:rcu_p2p_V} and \eqref{eqn:rcu_p2p_T}, and $ B(P_X)$ and $ A(P_X)$ are positive and finite by the moment assumptions \eqref{eqn:rcu_p2p_a1}-\eqref{eqn:rcu_p2p_a2}.

Since the codewords are drawn i.i.d. according to $P_{X^n} = P_X^n$, $\brX^n$ is independent of $(X^n,Y^n)$, and if $P_{Y^n|X^n}(Y^n|\brx^n)>0$, then
\begin{align}
    \lefteqn{\Pr[\brX^n = \brx^n|X^n,Y^n]}\\
    &= \Pr[\brX^n = \brx^n] \label{eqn:iid_px_property}\\
    &= P_{X^n}(\brx^n) \frac{P_{Y^n|X^n}(Y^n|\brx^n)}{P_{Y^n}(Y^n)} \frac{P_{Y^n}(Y^n)} {P_{Y^n|X^n}(Y^n|\brx^n)} \\
    &= \Pr[X^n = \brx^n |Y^n]\exp\left\{-i(\brx^n;Y^n)\right\}. \label{eqn:Pr_xn_bar}
\end{align}
If $P_{Y^n|X^n}(Y^n|\brx^n)=0$, then $\Pr[\brX^n = \brx^n|X^n,Y^n]=\Pr[\brX^n = \brx^n]$. However, since $P_{Y^n|X^n}(Y^n|\brx^n)=0$ implies $\brI_n=-\infty$, we only sum over $\brx^n$ such that $P_{Y^n|X^n}(Y^n|\brx^n)>0$ in the following derivation.

Fix some constant $\zeta$. Using \eqref{eqn:Pr_xn_bar} and summing over all $\brx^n$ such that $\brI_n \geq \zeta$ gives
\begin{align}
    \Pr[\brI_n \geq \zeta|Y^n] &= \bbE[\exp\{-I_n\} \ind\{I_n\geq \zeta\}|Y^n]. \label{eqn:px_ldpc_start}
\end{align}
Given $Y^n$, $I_n$ is a sum of independent random variables. Note from \eqref{eqn:rcu_p2p_a1_5} that $V^Y(P_X)>0$. Taking $Z_j = -i(X_j;Y_j)$, Lemma~\ref{lem:Pol47} implies %not necessarily identical By the assumption that $V^Y(P_X)>0$ and note $V^Y(P_X)\leq V(P_X)$, 
\begin{align}
    \Pr[\brI^n \geq \zeta|Y^n] \leq \frac{A(P_X)}{\sqrt{n}}\exp(-\zeta).
\end{align}

Therefore,
\begin{align}
    \epsilon' &\leq \bbE\left[\min\left\{1, M\frac{A(P_X)}{\sqrt{n}}\exp(-I_n)\right\} \right] \label{eqn:rcu_p2p_px_0} \\
    &= 1 \cdot \Pr\left[I_n < \log \frac{MA(P_X)}{\sqrt{n}} \right]  \nonumber \\
    &\mathrel{\phantom{=}} + \bbE\left[\frac{MA(P_X)}{\sqrt{n}}\exp(-I_n)\ind\left\{ I_n \geq \log \frac{MA(P_X)}{\sqrt{n}} \right\} \right]  \label{eqn:rcu_p2p_px_1} \\
    &= \Pr\left[I_n < \log \frac{MA(P_X)}{\sqrt{n}} \right]  \nonumber \\
    &\mathrel{\phantom{=}} + \frac{MA(P_X)}{\sqrt{n}}\bbE\left[\exp(-I_n)\ind\left\{ I_n \geq \log \frac{MA(P_X)}{\sqrt{n}} \right\} \right]\label{eqn:rcu_p2p_px_2} \\
    &\leq \Pr\left[I_n < \log M + \log A(P_X) - \frac{1}{2}\log n\right]\nonumber \\
    &\mathrel{\phantom{=}} +\frac{MA(P_X)}{\sqrt{n}}\cdot \frac{A(P_X)}{\sqrt{n}}\cdot \exp\left(-\log \frac{MA(P_X)}{\sqrt{n}}\right) \label{eqn:rcu_p2p_px_3} \\
    &= \Pr\left[I_n < \log M + \log A(P_X) - \frac{1}{2}\log n\right] + \frac{A(P_X)}{\sqrt{n}} \label{eqn:rcu_p2p_px_r1},
\end{align}
where $\eqref{eqn:rcu_p2p_px_1}$ separates the two possible outcomes of the minimization in \eqref{eqn:rcu_p2p_px_0}, and \eqref{eqn:rcu_p2p_px_3} applies Lemma~\ref{lem:Pol47} to the last term in \eqref{eqn:rcu_p2p_px_2} with $\zeta =\log \frac{MA(P_X)}{\sqrt{n}}$.

Recall from \eqref{eqn:rcu_p2p_I} that $I(P_X) = \bbE[i(X;Y)]$ and that, under our i.i.d. codeword design, $\bbE[I^n] = nI(P_X)$. Therefore, setting
\begin{align}
    \log M &= nI(P_X) + \frac{1}{2}\log n - \log A(P_X) \nonumber \\
    &\mathrel{\phantom{}} +\sqrt{nV(P_X)}Q^{-1}\left(1 - \left(\epsilon - \frac{B(P_X)+A(P_X)}{\sqrt{n}}\right)\right),
\end{align}
we have
\begin{align}
    \lefteqn{\Pr\left[I_n < \log M + \log A(P_X) - \frac{1}{2}\log n\right]} \\
    &= 1- \Pr\bigg[I_n \geq nI(P_X) \nonumber \\
    &\mathrel{\phantom{=}} +\sqrt{nV(P_X)}Q^{-1}\left(1 - \left(\epsilon - \frac{B(P_X)+A(P_X)}{\sqrt{n}}\right)\right)\bigg]  \label{eqn:rcu_p2p_px_4}\\
    &\leq 1 - \bigg(-\frac{B(P_X)}{\sqrt{n}} \nonumber
    \\
    &\mathrel{\phantom{=}} +Q\left(Q^{-1}\left(1 - \left(\epsilon - \frac{B(P_X)+A(P_X)}{\sqrt{n}}\right)\right)\right)\bigg) \label{eqn:rcu_p2p_px_5}\\
    & = 1 - \left(1- \epsilon + \frac{A(P_X)}{\sqrt{n}}\right) \\
    &= \epsilon - \frac{A(P_X)}{\sqrt{n}} \label{eqn:rcu_p2p_px_r2},
\end{align}
where \eqref{eqn:rcu_p2p_px_4} follows from $\Pr[X<a] = 1- \Pr[X\geq a]$, and \eqref{eqn:rcu_p2p_px_5} holds by applying the Berry-Ess\'een Theorem (Theorem~\ref{thm:BET}) to the last term in \eqref{eqn:rcu_p2p_px_4} with  $\sum_{j=1}^n Z_j = I_n$ and $\lambda =Q^{-1}\left(1 - \left(\epsilon - \frac{B(P_X)+A(P_X)}{\sqrt{n}}\right)\right) $. Note that the Berry-Ess\'een Theorem is given in the form $|a-b|\leq c$, and \eqref{eqn:rcu_p2p_px_5} applies the lower bound, i.e., $a-b \geq -c$.

Plugging \eqref{eqn:rcu_p2p_px_r2} into \eqref{eqn:rcu_p2p_px_r1} gives
\begin{align}
    \epsilon' \leq \epsilon- \frac{A(P_X)}{\sqrt{n}} + \frac{A(P_X)}{\sqrt{n}} =\epsilon, \label{eqn:rcu_p2p_px_ldpc_symbol_end}
\end{align}
which gives an achievability bound
\begin{align}
    \log M &\geq nI(P_X) + \frac{1}{2}\log n - \log A(P_X) \nonumber\\
    &\mathrel{\phantom{=}} +\sqrt{nV(P_X)}Q^{-1}\left(1 - \left(\epsilon - \frac{B(P_X)+A(P_X)}{\sqrt{n}}\right)\right) \\
    &=  nI(P_X) + \frac{1}{2}\log n - \log A(P_X) \nonumber\\
    &\mathrel{\phantom{=}} - \sqrt{nV(P_X)}Q^{-1}\left(\epsilon - \frac{B(P_X)+A(P_X)}{\sqrt{n}}\right) \label{eqn:rcu_p2p_px_6},
\end{align}
where \eqref{eqn:rcu_p2p_px_6} follows from the property of inverse $Q$ function, $Q^{-1}(1-\epsilon) = - Q^{-1}(\epsilon)$ for all $0<\epsilon<1$.

Finally, we use the 1st-order Taylor bound and the inverse function theorem as in \cite[Eq. (65)-(69)]{Chen:19_ARXIV} to derive the bounds
\begin{align}
    \lefteqn{Q^{-1}\left(\epsilon - \frac{B(P_X)+A(P_X)}{\sqrt{n}}\right)} \nonumber\\
    &\leq Q^{-1}(\epsilon)+\frac{B(P_X)+A(P_X)}{\sqrt{n}\phi\left(\Phi^{-1}\left(\Phi(Q^{-1}(\epsilon)) + \frac{B(P_X)+A(P_X)}{\sqrt{n}}\right)  \right)} \label{eqn:q_Inv_bound_1},
\end{align}
when $\epsilon \leq \frac{1}{2}$ and $n > \big(\frac{B(P_X)+A(P_X)}{\epsilon}\big)^2$, and
\begin{align}
    \MoveEqLeft[9] Q^{-1}\left(\epsilon - \frac{B(P_X)+A(P_X)}{\sqrt{n}}\right) \nonumber \\
    &\leq Q^{-1}(\epsilon)+\frac{B(P_X)+A(P_X)}{\sqrt{n}\phi(Q^{-1}(\epsilon))}\label{eqn:q_Inv_bound_2},
\end{align}
when $\epsilon >\frac{1}{2}$ and $n > \big(\frac{B(P_X)+A(P_X)}{\epsilon - 1/2}\big)^2$. Recall here that $\Phi(\cdot)$ and $\phi(\cdot)$ are the CDF and PDF for the standard Gaussian distribution.

By choosing $P_X$ to be the capacity achieving distribution, we obtain the existence of an $M(n,\epsilon)$ code with
\begin{align}
    \frac{\log M}{n} \geq C - \sqrt{\frac{V(P_X)}{n}}Q^{-1}(\epsilon) + \frac{\log n}{2n} - O\left(\frac{1}{n}\right).
\end{align}
\hfill$\blacksquare$
% \begin{remark}
% The achievability result in Theorem~\ref{thm:rcu_p2p_px} is optimal up to the third order. See the corresponding converse bound in \cite[Th. 48]{Polyanskiy:10}).
% \end{remark}

% \begin{remark} \label{rmk:why_rcu_ldpc}
% Theorem~\ref{thm:rcu_p2p_px} does not hold for LDPC codes, since the derivation of \eqref{eqn:iid_px_property} relies on the independence between codewords, which is absent in LDPC code. % Note \eqref{eqn:iid_px_property} relies on the independence of $X^n$ and $\brX^n$ of i.i.d. codes, which is absent in LDPC codes. In LDPC code, when one codeword is known, one can eliminate 
% \end{remark}
\subsection{RCU Bound for i.i.d. Code on the DM-$2$-MAC}\label{sec:rcu_2mac_px}
In this section, we first extend the RCU bound from the PPC to the MAC with two transmitters. We then present an asymptotic achievability result based on the two-user RCU bound. Our argument follows the multiple access source coding proof in \cite[Th. 11]{Chen:19_ARXIV} and is similar to \cite[Th. 1]{Recep:20_ARXIV}. The results generalize to MACs with more than two transmitters. We then present an asymptotic achievability result based on the two-user RCU bound. The bound improves the third-order MAC achievability bound $-O(\frac{\log n}{n})\bone$ in \cite{Huang:12}, and the best prior MAC achievability bound $-\nu\frac{\log n}{n}\bone$ in \cite{Tan:14}, with $\nu \geq 2|\cX_1||\cX_2||\cY|$, to $+\frac{\log n}{2n} \bone - O(\frac{1}{n})\bone$.
% n-dimensional verison
% Consider a discrete memoryless two-user MAC (DM2-MAC), $(\cX_1 \times \cX_2, P_{Y|X_1,X_2}, \cY)$, an $(n, M_1,M_2,\epsilon)$ code is defined by
% \begin{itemize}
%     \item two encoding functions: $f_1: [M_1] \rightarrow \cX_1^n $, $f_2:  [M_2]\rightarrow\cX_2^n $, and 
%     \item one decoding function: $g: \cY^n \rightarrow [M_1] \times [M_2]$,
% \end{itemize}
% such that the decoding constraint is satisfied 
% \begin{align}
% \frac{1}{M_1M_2}\sum_{\substack{
%     (W_1,W_2) \\
%     \in [M_1]\times [M_2]}}\Pr(g(Y^n)\neq (W_1,W_2)) \leq \epsilon,
% \end{align}
% where $(W_1,W_2)$ are the transmitters' (independent, equiprobable) messages over $[M_1]\times[M_2]$, and $X_1^n = f_1(W_1), X_2^n=f_2(W_2), P_{Y^n|X_1^n,X_2^n} = P_{Y|X_1,X_2}^n$. 

Consider a two-user MAC, $(\cX_1 \times \cX_2, P_{Y|X_1,X_2}, \cY)$. An $( M_1,M_2,\epsilon)$ code is defined by two encoding functions
\begin{eqnarray*}
f_1&:& [M_1] \rightarrow \cX_1  \\
f_2&:& [M_2]\rightarrow\cX_2 
\end{eqnarray*}
and one decoding function
\[
g: \cY^n \rightarrow [M_1] \times [M_2]
\]
such that the average error probability is bounded by $\epsilon$ 
\begin{align}
 \frac{1}{M_1M_2}\mkern-18mu\sum_{\substack{
    (w_1,w_2) \\
    \in [M_1]\times [M_2]}} &\Pr[g(Y)\neq (w_1,w_2)| \nonumber \\
   & X_1=f_1(w_1),X_2=f_2(w_2)] \leq \epsilon.
\end{align}

Similarly, given a two-user MAC, $(\cX_1 \times \cX_2, P_{Y|X_1,X_2}, \cY)$, a blocklength-$n$ $(M_1,M_2,\epsilon)$ code for the two-user MAC, denoted as $(n,M_1,M_2,\epsilon)$, is defined by two encoding functions
\begin{align*}
f_1\colon & [M_1] \rightarrow \cX_1^n  \\
f_2\colon & [M_2]\rightarrow\cX_2^n
\end{align*}
and one decoding function
\[
g:\cY^n \rightarrow [M_1] \times [M_2]
\]
such that the average error probability is bounded by $\epsilon$ 
\begin{align}
 \frac{1}{M_1M_2}\mkern-18mu\sum_{\substack{
    (w_1,w_2) \\
    \in [M_1]\times [M_2]}} &\Pr[g(Y^n)\neq (w_1,w_2)| \nonumber \\
   & X_1^n=f_1(w_1),X_2^n=f_2(w_2)] \leq \epsilon,
\end{align}

The corresponding (finite-blocklength) rate pair for an $(n,M_1,M_2,\epsilon)$ is defined as
\begin{align}
    R_1 &= \frac{1}{n}\log M_1, \\
    R_2 &= \frac{1}{n}\log M_2.
\end{align}
A rate pair $(R_1,R_2)$ is said to be $(n,\epsilon)$-achievable if there exists an $(n,M_1,M_2,\epsilon)$ code. The closure of the set of all $(n,\epsilon)$-achievable rate pairs is called the $(n,\epsilon)$-achievable rate region, denoted as $\cR_{n,\epsilon}$.

\begin{remark}
The definition of an $(n,M_1,M_2,\epsilon)$ code and the corresponding rate region $\cR_{n,\epsilon}$ apply to general two-user MACs and are not restricted to the discrete or memoryless case. In this paper, we focus on the subclass of DM-$2$-MACs; in this case, $P_{Y^n|X_1^n,X_2^n} = P_{Y|X_1,X_2}^n$ and $\cX_1,\cX_2$, and $\cY$ are all discrete.
\end{remark}
\begin{thm} (Two-user RCU bound, extended from \cite[Th. 16]{Polyanskiy:10}) \label{thm:RCU_2MAC} 
% For an arbitrary $P_{X_1}P_{X_2}$, there exists an $(M_1,M_2,\epsilon)$ code with symmetric code design (no codeword is treating any better or worse than any other codeword in each of the codebook), where each codeword in the codebook of transmitter $i$ has the same marginal distribution $P_{X_i}, i \in \{1,2\}$, and average error probability $\epsilon$ such that
% For an arbitrary $P_{X_1}P_{X_2}$, 
Consider an ensemble of MAC codes with $M_1 \times M_2$ codeword pairs drawn according to some $P_{X_{1}(1) \ldots X_1(M_1)}P_{X_2(1) \ldots X_2(M_2)}$ such that
\begin{align}
    % P_{X_{1}(i)} &= P_{X_1}, ~\forall i \in [M_1], \\
    % P_{X_{2}(j)} &= P_{X_2}, ~\forall j \in [M_2], \\
    P_{X_{1}(\cA)} &=P_{X_{1}(\cB)}, ~\forall \cA,\cB \subseteq [M_1] \mbox{ s.t. } |\cA| = |\cB|,\\
    P_{X_{2}(\cA)} &=P_{X_{2}(\cB)}, ~\forall \cA,\cB \subseteq [M_2] \mbox{ s.t. } |\cA| = |\cB|,
\end{align}
 Under ML decoding, the ensemble-average error probability $\epsilon$ satisfies
%%%%%%% Multi-line Expression %%%%%%%%
% \begin{align}
%     \lefteqn{\epsilon \leq \bbE[\min\{1,} \nonumber\\  &\mathrel{\phantom{==}} (M_1-1)\Pr[i(\brX_1;Y|X_2)\geq i(X_1;Y|X_2)|X_1,X_2,Y] \nonumber \\
%      &\mathrel{\phantom{==}} +(M_2-1)\Pr[i(\brX_2;Y|X_1)\geq i(X_2;Y|X_1)|X_1,X_2,Y]\} \nonumber \\
%      &\mathrel{\phantom{==}} +(M_1-1)(M_2-1)\nonumber \\
%      &\mathrel{\phantom{====}}\Pr[i(\brX_1,\brX_2;Y)\geq i(X_1,X_2;Y)|X_1,X_2,Y]\},  \label{eqn:RCU_MAC2}
% \end{align}
%%%%%%% Single Expression %%%%%%%%
\begin{align}
    \epsilon \leq \bbE[\min\{1, V_1 + V_2 + V_{12}\}] \label{eqn:RCU_MAC2},
\end{align}
where 
\begin{align} 
    V_1 &=(M_1-1)\Pr[i(\brX_1;Y|X_2)\geq i(X_1;Y|X_2)|X_1,X_2,Y], \\
    V_2 &= (M_2-1)\Pr[i(\brX_2;Y|X_1)\geq i(X_2;Y|X_1)|X_1,X_2,Y], \\
    V_{12} &= (M_1-1)(M_2-1) \nonumber \\
    &\mathrel{\phantom{==}}\Pr[i(\brX_1,\brX_2;Y)\geq i(X_1,X_2;Y)|X_1,X_2,Y]\},
\end{align}
and
\begin{align}
% \MoveEqLeft[6] P_{X_1X_2\brX_1\brX_2 Y}(a,b,c,d,e) = \nonumber\\ &P_{X_1X_2\brX_1\brX_2}(a,b,c,d)P_{Y|X_1,X_2}(e|a,b).
\MoveEqLeft[5] P_{X_1X_2\brX_1\brX_2 Y}(a,b,c,d,e) = \nonumber\\ &P_{X_1\brX_1}(a,c)P_{X_2\brX_2}(b,d)P_{Y|X_1,X_2}(e|a,b),
\\
P_{X_1\brX_1}(a,c) &= P_{X_1(1)X_1(2)}(a,c) \nonumber\\
P_{X_2\brX_2}(b,d) &= P_{X_2(1)X_2(2)}(b,d) \nonumber.
\end{align}
\end{thm}
{\em Proof:} Denote the random MAC codebook as $$(X_{1}(1),\ldots, X_{1}(M_1)) \times(X_{2}(1),\ldots, X_{2}(M_2)),$$
where codewords $(X_{k}(1),\ldots, X_{k}(M_k))$ are chosen according to $P_{X_{k}(1),\ldots,X_{k}(M_k)}$ for $k\in\{1,2\}$. 

Denote the conditional error probability given the codeword pair $(X_{1}(i),X_{2}(j))$ is sent as $\epsilon_{i,j}$.

The average error probability is
\begin{align}
    \epsilon_{\text{avg}} = \frac{1}{M_1M_2}\sum_{(i,j) \in [M_1]\times [M_2]} \epsilon_{i,j}.
\end{align}
By the symmetry of code design
\begin{align}
    \bbE[\epsilon_{\text{avg}}] = \bbE[\epsilon_{1,1}], 
\end{align}
where the expectation is taken over the random codebook design. 

The ML decoder $g(\cdot)$ gives
\begin{align}
    g(y) &=\argmax_{(i,j)\in [M_1] \times [M_2]} P_{Y|X_1,X_2}(y|X_{1}(i),X_{2}(j)) \\
    &=\argmax_{(i,j)\in [M_1] \times [M_2]} \frac{P_{Y|X_1,X_2}(y|X_{1}(i),X_{2}(j))}{P_Y(y)}\\
    &=\argmax_{(i,j)\in [M_1] \times [M_2]} i(X_{1}(i),X_{2}(j);y),
\end{align}
and ties are broken uniformly at random. %The error probability bound includes all ties in the error event.

Given that codeword pair $(X_{1}(1),X_{2}(1))$ is transmitted, an error or tie occurs if
\begin{align}
   \MoveEqLeft[6] \exists ~(i,j) \in [M_1]\times[M_2] \setminus \{1,1\}, \nonumber
    \\&\text{ s.t. } i(X_{1}(i),X_{2}(j);y) \geq i(X_{1}(1),X_{2}(1);y). \label{eqn:rcu_2mac_px_c1} %rcu bound for 2-user mac error condition 1
\end{align}

Note that condition \eqref{eqn:rcu_2mac_px_c1} can be equivalently written as the union of the following events
\begin{enumerate}
    \item $ \begin{aligned}[t] \exists ~i \in [M_1]& \setminus \{1\}, \text{s.t. }\\& i(X_{1}(i),X_{2}(1);y) \geq i(X_{1}(1),X_{2}(1);y); \end{aligned}$
    \item $\begin{aligned}[t] \exists ~j \in [M_2] &\setminus \{1\}, \text{s.t. }\\& i(X_{1}(1),X_{2}(j);y) \geq i(X_{1}(1),X_{2}(1);y); \end{aligned}$
    \item $ \begin{aligned}[t] \exists ~i \in [M_1] &\setminus \{1\}, ~j \in [M_2] \setminus \{1\},\text{s.t. } \\ & i(X_{1}(i),X_{2}(j);y) \geq i(X_{1}(1),X_{2}(1);y). \end{aligned} $
\end{enumerate}

Therefore, $\bbE[\epsilon_{1,1}]$ can be bounded from above as
\begin{align}
    \lefteqn{\bbE[\epsilon_{1,1}]} \nonumber \\
    &\leq \Pr\left[\left\{\bigcup_{i=2}^{M_1} \left\{i(X_{1}(i),X_{2}(1);Y) \geq i(X_{1}(1),X_{2}(1);Y)\right\}\right\}   \nonumber \right.\\
    &\mathrel{\phantom{==}} \cup \left\{\bigcup_{j=2}^{M_2} \left\{i(X_{1}(1),X_{2}(j);Y) \geq i(X_{1}(1),X_{2}(1);Y)\right\}\right\} \nonumber\\
    & \left. \cup \left\{\bigcup_{\substack{
    i \in [M_1]\setminus\{1\} \\
    j \in [M_2]\setminus\{1\}}} \mkern-18mu\left\{i(X_{1}(i),X_{2}(j);Y) \geq i(X_{1}(1),X_{2}(1);Y)\right\}\right\} \right] \label{eqn:RCU_2MAC_1}\\
    &=\Pr\left[\left\{\bigcup_{i=2}^{M_1} \left\{i(X_{1}(i);Y|X_{2}(1)) \geq i(X_{1}(1);Y|X_{2}(1))\right\}\right\}   \nonumber \right.\\
    &\mathrel{\phantom{==}} \cup \left\{\bigcup_{j=2}^{M_2} \left\{i(X_{2}(j);Y|X_{1}(1)) \geq i(X_{2}(1);Y|X_{1}(1))\right\}\right\} \nonumber\\
    & \left. \cup \left\{\bigcup_{\substack{
    i \in [M_1]\setminus\{1\} \\
    j \in [M_2]\setminus\{1\}}} \mkern-18mu\left\{i(X_{1}(i),X_{2}(j);Y) \geq i(X_{1}(1),X_{2}(1);Y)\right\}\right\} \right] \label{eqn:RCU_2MAC_2},
    \end{align}
where \eqref{eqn:RCU_2MAC_1} is an inequality instead of an equality since the decoder might resolve some ties correctly, and \eqref{eqn:RCU_2MAC_2} removes common terms from the first two terms of \eqref{eqn:RCU_2MAC_1}, thereby replacing information density by conditional information density.

Let $W=(X_{1}(1),X_{2}(1),Y)$. Then
    \begin{align}
    \lefteqn{\bbE[\epsilon_{1,1}]} \nonumber \\
    &=\bbE\left[\Pr\left[  \right. \right.\nonumber\\
    &\mathrel{\phantom{}}\left\{\bigcup_{i=2}^{M_1} \left\{i(X_{1}(i);Y|X_{2}(1)) \geq i(X_{1}(1);Y|X_{2}(1))\right\}\right\} \cup  \nonumber  \\
    &\mathrel{\phantom{}}  \left\{\bigcup_{j=2}^{M_2} \left\{i(X_{2}(j);Y|X_{1}(1)) \geq i(X_{2}(1);Y|X_{1}(1))\right\}\right\} \cup\nonumber\\
    & \left. \left. \left. \left\{\bigcup_{\substack{
    i \in [M_1]\setminus\{1\} \\
    j \in [M_2]\setminus\{1\}}} \mkern-28mu\left\{i(X_{1}(i),X_{2}(j);Y)\geq i(X_{1}(1),X_{2}(1);Y) \right\}\right\} \right\vert T\right]\right] \label{eqn:RCU_2MAC_3}\\
    &\leq \bbE\left[ \min\left\{1,\right. \right. \nonumber \\
    &\mathrel{\phantom{==}}  (M_1-1)\Pr[\left\{i(\brX_1;Y|X_{2}) \geq i(X_{1};Y|X_{2})\right\}|T]\  \nonumber \\ 
    &\mathrel{\phantom{=}} +(M_2-1) \Pr[\left\{i(\brX_2;Y|X_{1}) \geq i(X_{2};Y|X_{1})\right\}|T]\nonumber\\
    &\mathrel{\phantom{=}} +(M_1-1)(M_2-1) \nonumber \\
    &\mathrel{\phantom{==}} \left. \Pr\left[\{i(\brX_1,\brX_2;Y)\geq i(X_{1},X_{2};Y)\} \vert T\right]\} \right], \label{eqn:RCU_2MAC_4}
\end{align}
where \eqref{eqn:RCU_2MAC_3} follows from the law of iterated expectation, and \eqref{eqn:RCU_2MAC_4} holds by the bounded nature of probability and symmetry of our code design. 
\hfill$\blacksquare$

\begin{remark}
The authors in \cite{Recep:20_ARXIV} achieve lower decoder complexity in the symmetrical rate case by replacing the three events in \eqref{eqn:rcu_2mac_px_c1} by one. While we do not assume the symmetrical rate point, we note that only events corresponding to constraints that are active at a given rate point have a non-negligible impact in \eqref{eqn:RCU_2MAC_4}. This observation enables decoder simplification for most rate points.
\end{remark}
% We now give a definition of achievable rate region for block-$n$ MAC code, and then derive an asymptotic third order achievability rate region.

% The definition of $(M_1,M_2,\epsilon)$ can be extended to the case where the input and output alphabet are defined with Cartesian product structure. Formally, consider a two-user MAC, $(\cX_1 \times \cX_2, P_{Y|X_1,X_2}, \cY)$, an block-$n$ $( M_1,M_2,\epsilon)$ code, denoted as $(n,M_1,M_2,\epsilon)$, is defined by

 Prior to stating the achievability theorem, we generalize the inverse complementary CDF $Q^{-1}(\cdot)$ to higher dimension. Let $\bZ$ be a Gaussian random vector in $\mathbb{R}^d$ with mean zero and covariance matrix $\bK_{\bZ\bZ}$, denote the set $Q_{\text{inv}}(\bK_{\bZ\bZ},\epsilon)$ as
 \begin{align}
     Q_{\text{inv}}(\bK_{\bZ\bZ},\epsilon) \defeq \left\{\bz \in \mathbb{R}^d: \Pr[\bZ\leq \bz] \geq 1-\epsilon\right\}. \label{eqn:mul_Q}
 \end{align}
\begin{thm} (Random coding finite-blocklength bound and third-order achievability bound on the DM-$2$-MAC). \label{thm:rcu_mac_px}
Consider a DM-$2$-MAC $(\cX_1 \times \cX_2, P_{Y|X_1,X_2}, \cY)$. Let each symbol of each codeword for transmitter $i$ be drawn i.i.d. according to $P_{X_i}$, for $i\in\{1,2\}$. Then there exists an $(n,M_1,M_2,\epsilon)$ code such that for any blocklength $n$
\begin{align}
        \epsilon&\leq \bbE\left[\min\left\{1, E_1+E_2+E_{12}\right\}\right],
\end{align}
and for large enough blocklength $n$
\begin{align}
    \bbrR \in \bbrI - \frac{Q_{\text{inv}}(\rmV,\epsilon)}{\sqrt{n}} + \frac{\log n}{2n}\bone - O\left(\frac{1}{n}\right) \bone,
\end{align}
providing the following moment assumptions are satisfied
\begin{align}
    V^Y(P_{X_1}|P_{X_2}) &>0, &V^Y(P_{X_2}|P_{X_1}) >0, \label{eqn:rcu_2mac_a1}~\\
      V^Y(P_{X_1},P_{X_2}) &>0,  &T(P_{X_1}|P_{X_2})  < \infty,\label{eqn:rcu_2mac_a4} \\
    T(P_{X_2}|P_{X_1}) &< \infty,  &T(P_{X_1},P_{X_2}) < \infty,
 \label{eqn:rcu_2mac_a6}
\end{align}
% \begin{align}
%     V(P_{X_1}|P_{X_2}) &>0, \label{eqn:rcu_2mac_a1}\\
%      V(P_{X_2}|P_{X_1}) &>0, \label{eqn:rcu_2mac_a2}\\
%       V(P_{X_1},P_{X_2}) &>0, \label{eqn:rcu_2mac_a3}\\
%     T(P_{X_1}|P_{X_2}) & < \infty,
%  \label{eqn:rcu_2mac_a4} \\
%     T(P_{X_2}|P_{X_1}) &< \infty,
%  \label{eqn:rcu_2mac_a5} \\
%     T(P_{X_1},P_{X_2}) &< \infty,
%  \label{eqn:rcu_2mac_a6}
% \end{align}
where
% where $W_1, W_2, E_{12}$ are defined in \eqref{eqn:e1} to \eqref{eqn:e12},
% \begin{align}
%     \bbrR \defeq \begin{bmatrix} R_1 \\ R_2 \\ R_1+R_2\end{bmatrix}, ~\bbrI \defeq \begin{bmatrix} \bbE[i(X_1^n;Y^n|X_2^n)] \\ \bbE[i(X_2^n;Y^n|X_1^n)] \\ \bbE[i(X_1^n,X_2^n;Y^n)]\end{bmatrix},
% \end{align}
\begin{align}
    F_1 &\defeq 2\left(\frac{\log 2}{\sqrt{2\pi V(P_{X_1}|P_{X_2})}} +2 \frac{C_0 T(P_{X_1}|P_{X_2})}{V(P_{X_1}|P_{X_2})^{3/2}}\right) \label{eqn:rcu_mac_f1}\\
    F_2 &\defeq 2\left(\frac{\log 2}{\sqrt{2\pi V(P_{X_2}|P_{X_1})}} +2 \frac{C_0 T(P_{X_2}|P_{X_1})}{V(P_{X_2}|P_{X_1})^{3/2}}\right)\label{eqn:rcu_mac_f2}\\
    F_{12} &\defeq 2\left(\frac{\log 2}{\sqrt{2\pi V(P_{X_1},P_{X_2})}} +2 \frac{C_0 T(P_{X_1},P_{X_2})}{V(P_{X_1},P_{X_2})^{3/2}}\right)\label{eqn:rcu_mac_f} \\
    E_1 &\defeq M_1\frac{F_1}{\sqrt{n}}\exp(-i(X_1^n;Y^n|X_2^n)) \label{eqn:e1}\\
    E_2 &\defeq M_2\frac{F_2}{\sqrt{n}}\exp(-i(X_{2j};Y_j|X_{1j}))\label{eqn:e2}\\
    E_{12} &\defeq M_1M_2\frac{F_{12}}{\sqrt{n}}\exp(-i(X_1^n,X_2^n;Y^n)) \label{eqn:e12} \\
    \bbrR &\defeq \begin{bmatrix} R_1 \\ R_2 \\ R_1+R_2\end{bmatrix}, ~\bbrI \defeq \begin{bmatrix} \bbE[i(X_1;Y|X_2)] \\ \bbE[i(X_2;Y|X_1)] \\ \bbE[i(X_1,X_2;Y)]\end{bmatrix},
\end{align}
$\rmV$ is the covariance matrix of %vector $\bbri(P_{X_1},P_{X_2})$
\begin{align}
    \bbri(P_{X_1},P_{X_2}) \defeq \begin{bmatrix} i(X_1;Y|X_2) \\ i(X_2;Y|X_1) \\ i(X_1,X_2;Y)\end{bmatrix},
\end{align}
and $Q_{\text{inv}}$ is defined in \eqref{eqn:mul_Q}.
\end{thm}

The proof of Theorem~\ref{thm:rcu_mac_px} requires a multi-dimensional version of Berry Ess\'een theorem, shown as Lemma~\ref{thm:BET_2MAC} below.
\begin{lem}
(Multi-dimensional Berry-Ess\'een Theorem, \cite[Lemma 15]{Chen:19_ARXIV}, \cite[Cor. 8]{Tan:14}). \label{thm:BET_2MAC}
Let $\bU_1,\ldots, \bU_n \in \mathbb{R}^d$ be a sequence of i.i.d. random vectors with mean zero and covariance matrix $\bSgm$ of rank $r \defeq \text{rank}(\bSgm)$. Let $\bZ \in \mathbb{R}^d$ be a Gaussian vector with mean zero and the same covariance matrix  $\bSgm$. Let $\rmT$ be a $d\times r$ matrix, where the columns of $\rmT$ are the $r$ normalized eigenvectors of $\bSgm$ with non-zero eigenvalues. Define $\bW_1,\ldots,\bW_n \in \mathbb{R}^r$ to be a sequence of i.i.d. random vectors, such that $\bU_i = \rmT\bW_i$ for all $i\in[n]$. If $r\geq 1$, then for all $n$,
\begin{align}
\sup_{\bz\in \mathbb{R}^d} \left|\Pr\left[\frac{1}{\sqrt{n}}\sum_{j=1}^n \bU_i \leq \bz \right] - \Pr[\bZ\leq \bz] \right|\leq \frac{400d^{\frac{1}{4}}\beta_r}{\lambda_{\min}^{\frac{3}{2}}}\frac{1}{\sqrt{n}},
\end{align}
where $\bSgm_r$ is the covariance matrix of $\bW_1$, $\beta_r \defeq \bbE[\|\bW_1 \|^3_2]$ ($\|\cdot\|$ is the $\ell^2$ norm), and $\lambda_{\min}$ is the minimum eigenvalue of $\bSgm_r$.
\end{lem}

{\em Proof of Theorem~\ref{thm:rcu_mac_px}:} Setting $X_1 = X_1^n, \brX_1 = \brX_1^n,X_2 = X_2^n, \brX_2 = \brX_2^n, Y =Y^n$ in Theorem~\ref{thm:RCU_2MAC}, we note that there exists an $(n,M_1,M_2,\epsilon')$ code with
\begin{align}
    \epsilon' \leq \bbE[\min\{1, V_1 + V_2 + V_{12}\}] \label{eqn:RCU_MAC2_iid},
\end{align}
where 
\begin{align} 
    V_1 &=(M_1-1)\nonumber \\
    &\mathrel{\phantom{==}}\Pr[i(\brX_1^n;Y^n|X_2^n)\geq i(X_1^n;Y^n|X_2^n)|X_1^n,X_2^n,Y^n], \label{eqn:rcu_2mac_iid_v1}\\
    V_2 &= (M_2-1)\nonumber \\
    &\mathrel{\phantom{==}}\Pr[i(\brX_2^n;Y^n|X_1^n)\geq i(X_2^n;Y^n|X_1^n)|X_1^n,X_2^n,Y^n], \\
    V_{12} &= (M_1-1)(M_2-1) \nonumber \\
    &\mathrel{\phantom{==}}\Pr[i(\brX_1^n,\brX_2^n;Y^n)\geq i(X_1^n,X_2^n;Y)^n|X_1^n,X_2^n,Y^n]\},
\end{align}
and 
\begin{align*} 
\lefteqn{P_{X_1^nX_2^n\brX_1^n\brX_2^nY^n}(x_1^n,x_2^n,\brx_1^n,\brx_2^n,y^n)} \\
&= P_{X_1^n\brX_1^n}(x_1^n,\brx_1^n)P_{X_2^n\brX_2^n}(x_2^n,\brx_2^n)P_{Y^n|X_1^nX_2^n}(y^n|x_1^n,x_2^n)\\ &=P_{X_1^n}(x_1^n)P_{X_1^n}(\brx_1^n)P_{X_2^n}(x_2^n)P_{X_2^n}(\brx_2^n)P_{Y^n|X^n}(y^n|x^n),
\end{align*}
as the codewords for transmitter $i\in\{1,2\}$ are drawn i.i.d. according to $P_{X_i^n} = P_{X_i}^n$.

Denote for brevity
\begin{align}
    I_{1n} &\defeq i(X_1^n;Y^n|X_2^n) = \sum_{j=1}^n i(X_{1j};Y_j|X_{2j}), \\
    I_{2n} &\defeq i(X_2^n;Y^n|X_1^n) = \sum_{j=1}^n i(X_{2j};Y_j|X_{1j}), \\
    I_n &\defeq i(X_1^n,X_2^n;Y^n) = \sum_{j=1}^n i(X_{1j},X_{2j};Y_j), \\
    \brI_{1n} &\defeq i(\brX_1^n;Y^n|X_2^n) = \sum_{j=1}^n i(\brX_{1j};Y_j|X_{2j}), \\
    \brI_{2n} &\defeq i(\brX_2^n;Y^n|X_1^n) = \sum_{j=1}^n i(\brX_{2j};Y_j|X_{1j}), \\
    \brI_n &\defeq i(\brX_1^n,\brX_2^n;Y^n) = \sum_{j=1}^n i(\brX_{1j},\brX_{2j};Y_j), 
\end{align}
where $(X_{1j},X_{2j})$ and $(\brX_{1j},\brX_{2j})$ are the $j$-th symbols of the transmitted codeword pair and an untransmitted codeword pair, respectively. 

Note that since the codewords are drawn i.i.d. by assumption, $\brX_1^n$ is independent of $X_1^n, X_2^n$, and $Y^n$. If $P_{Y^n|X_1^n,X_2^n}(Y^n|\brx_1^n,X_2^n)>0$, then
\begin{align}
    \lefteqn{\Pr[\brX_1^n = \brx_1^n|X_1^n,X_2^n,Y^n]}\\
    &= \Pr[\brX_1^n = \brx_1^n|X_2^n] \\
    &= \Pr[\brX_1^n = \brx_1^n|X_2^n] \label{eqn:2MAC_Pr_Y_x1bx2} \frac{P_{Y^n|X_1^n,X_2^n}(Y^n|\brx_1^n,X_2^n)}{P_{Y^n|X_2^n}(Y^n|X_2^n)} \nonumber \\
    &\mathrel{\phantom{\qquad\qquad\qquad\qquad\qquad}} \cdot\frac{P_{Y^n|X_2^n}(Y^n|X_2^n)} {P_{Y^n|X_1^n,X_2^n}(Y^n|\brx_1^n,X_2^n)} \\
    &= \Pr[X_1^n = \brx_1^n |Y^n,X_2^n]\exp\left\{-i(\brx_1^n;Y^n|X_2^n)\right\}.
\end{align}

If $P_{Y^n|X_1^n,X_2^n}(Y^n|\brx_1^n,X_2^n)=0$, then we stop at \eqref{eqn:2MAC_Pr_Y_x1bx2}. Note that the following derivation only sums over $\brx_1^n$ such that $P_{Y^n|X_1^n,X_2^n}(Y^n|\brx_1^n,X_2^n)>0$ as $P_{Y^n|X_1^n,X_2^n}(Y^n|\brx_1^n,X_2^n)=0$ implies $\brI_{1n}=-\infty$.

Summing over all $\brx_1^n$ such that $\brI_{1n} \geq \zeta$ gives
\begin{align}
    \Pr[\brI_{1n} \geq \zeta|Y^n,X_2^n] &= \bbE[\exp\{-I_{1n}\} \ind\{I_{1n}\geq \zeta\}|Y^n,X_2^n] \nonumber \\
    &\leq \frac{F_1}{\sqrt{n}}\exp(-\zeta) \label{eqn:rcu_mac_iid_1},
\end{align}
where \eqref{eqn:rcu_mac_iid_1} follows from Lemma~\ref{lem:Pol47}.

Plugging \eqref{eqn:rcu_mac_iid_1} into \eqref{eqn:rcu_2mac_iid_v1}, we obtain
\begin{align}
V_1 \leq M_1\frac{F_1}{\sqrt{n}}\exp(-I_{1n}) =E_1\label{eqn:rcu_mac_iid_2}.
\end{align}
A similar approach yields
\begin{align}
    V_2 &\leq M_2\frac{F_2}{\sqrt{n}}\exp(-I_{2n})= E_2 \label{eqn:rcu_mac_iid_3}, \\
    V_{12} &\leq M_1M_2\frac{F_{12}}{\sqrt{n}}\exp(-I_{n}) =E_{12} \label{eqn:rcu_mac_iid_4}.
\end{align}

Therefore
\begin{align}
    \epsilon'&\leq \bbE\left[\min\left\{1, E_1+E_2+E_{12}\right\}\right]\\
    &=\Pr[E_1+E_2+E_{12} >1] \nonumber \\
    &\mathrel{\phantom{=}}+ \bbE[(E_1+E_2+E_{12})\ind\{(E_1+E_2+E_{12})\leq 1\}] \label{eqn:rcu_mac_iid_5} \\
    &\leq \Pr[E_1+E_2+E_{12} >1]+ \bbE[E_1\ind\{E_1\leq 1\}] \nonumber \\
    &\mathrel{\phantom{=}}+\bbE[E_2\ind\{E_2\leq 1\}] +\bbE[E_{12}\ind\{E_{12}\leq 1\}] \label{eqn:rcu_mac_iid_6} \\
    &\leq \Pr[E_1+E_2+E_{12} >1] +\frac{F_1}{\sqrt{n}}+\frac{F_2}{\sqrt{n}}+\frac{F_{12}}{\sqrt{n}} \label{eqn:rcu_mac_iid_7} \\
    &= 1 - \Pr[E_1+E_2+E_{12} \leq1]+\frac{F_1}{\sqrt{n}}+\frac{F_2}{\sqrt{n}}+\frac{F_{12}}{\sqrt{n}} \\
    &\leq 1 - \Pr\left[ \left\{E_1 \leq \frac{1}{3}\right \} \cap \left\{E_2 \leq \frac{1}{3}\right \} \cap \left\{E_{12} \leq \frac{1}{3}\right \}\right] \nonumber \\
    &\mathrel{\phantom{=}}+\frac{F_1}{\sqrt{n}}+\frac{F_2}{\sqrt{n}}+\frac{F_{12}}{\sqrt{n}} \label{eqn:rcu_mac_iid_8},
\end{align}
where \eqref{eqn:rcu_mac_iid_5} holds by separating the cases based on whether $E_1+E_2+E_{12}<1$ or not, \eqref{eqn:rcu_mac_iid_6} follows from linearity of expectation and weakening the indicator function threshold, applying Lemma~\ref{lem:Pol47} to the each of the last three terms in \eqref{eqn:rcu_mac_iid_6} yields \eqref{eqn:rcu_mac_iid_7}, and \eqref{eqn:rcu_mac_iid_8} holds since the event $\{\left\{E_1 \leq \frac{1}{3}\right \} \cap \left\{E_2 \leq \frac{1}{3}\right \} \cap \left\{E_{12} \leq \frac{1}{3}\right \}\}$ is a subset of the event $\{E_1+E_2+E_{12} \leq1\}$.

Denote
\begin{align}
    \bU_j &\defeq  \begin{bmatrix} i(X_{1j};Y_j|X_{2j}) \\ i(X_{2j};Y_j|X_{1j}) \\ i(X_{1j},X_{2j};Y_j)\end{bmatrix}-\bbrI, ~\forall j\in [n],\\
    \bS_n &\defeq \frac{1}{\sqrt{n}}\sum_{j=1}^n \bU_j =  \frac{1}{\sqrt{n}} \begin{bmatrix} I_{1n} \\ I_{2n}\\ I_n \end{bmatrix}-\sqrt{n\bbrI},
\end{align}
where each $\bU_j, j\in [n]$, is a random vector with  mean zero and covariance matrix \rmV. Note $\bbE[\|\bU_1 \|^3_2]$ is finite by the moment assumptions \eqref{eqn:rcu_2mac_a4}-\eqref{eqn:rcu_2mac_a6}; hence Lemma~\ref{thm:BET_2MAC} is applicable.

Therefore,
\begin{align}
    &\mathrel{\phantom{=}}\Pr\left[ \left\{E_1 \leq \frac{1}{3}\right \} \cap \left\{E_2 \leq \frac{1}{3}\right \} \cap \left\{E_{12} \leq \frac{1}{3}\right \}\right] \nonumber \\
    &=\Pr\left[ \left\{I_{1n}\geq \log M_1 + \log 3F_1 - \frac{1}{2}\log n \right\} \cap \right. \nonumber \\
    &\mathrel{\phantom{===}}\left\{I_{2n}\geq \log M_2 + \log 3F_2 - \frac{1}{2}\log n \right\} \cap \nonumber \\
    &\left.  \mathrel{\phantom{===}}\left\{I_{n}\geq \log M_1 +\log M_2 + \log 3F_{12} - \frac{1}{2}\log n \right\}   \right] \label{eqn:rcu_mac_iid_9} \\
    &=\Pr\left[\bS_n \geq \sqrt{n}\left(\bbrR - \bbrI - \frac{\log n}{2n}\bone + O\left(\frac{1}{n}\right) \bone\right)\right] \label{eqn:rcu_mac_iid_10}\\
    &=1 - \Pr\left[\bS_n < \sqrt{n}\left(\bbrR - \bbrI - \frac{\log n}{2n}\bone + O\left(\frac{1}{n}\right) \bone\right)\right] \\
    &\geq 1 - \Pr\left[\bS_n \leq \sqrt{n}\left(\bbrR - \bbrI - \frac{\log n}{2n}\bone + O\left(\frac{1}{n}\right)\bone \right)\right] \\
    &\geq 1 - \Pr\left[\bZ \leq \sqrt{n}\left(\bbrR - \bbrI - \frac{\log n}{2n}\bone + O\left(\frac{1}{n}\right) \bone\right)\right] \nonumber \\
     &\mathrel{\phantom{\qquad\qquad\qquad\qquad\qquad\qquad\qquad\qquad}} -  O\left(\frac{1}{\sqrt{n}}\right) \label{eqn:rcu_mac_iid_11},
\end{align}
where \eqref{eqn:rcu_mac_iid_9} follows by expanding $E_1,E_2,$ and $E_{12}$ using \eqref{eqn:e1}-\eqref{eqn:e12}, \eqref{eqn:rcu_mac_iid_10} rewrites \eqref{eqn:rcu_mac_iid_9} using the definition of $\bS_n, \bbrR$ and $\bbrI$, and \eqref{eqn:rcu_mac_iid_11} follows from the multi-dimensional Berry-Ess\'een Theorem, Lemma~\ref{thm:BET_2MAC}.

For any rate $(R_1,R_2)$ satisfying
\begin{align}
       \bbrR \in \bbrI - \frac{Q_{\text{inv}}(\rmV,\epsilon - \frac{c}{\sqrt{n}})}{\sqrt{n}} + \frac{\log n}{2n}\bone - O\left(\frac{1}{n}\right) \bone,
\end{align}
by the definition of $Q_{\text{inv}}$ in \eqref{eqn:mul_Q}, we have
\begin{align}
\Pr\left[\bZ \leq \sqrt{n}\left(\bbrR - \bbrI - \frac{\log n}{2n}\bone + O\left(\frac{1}{n}\right) \bone\right)\right] \leq  \epsilon -\frac{c}{\sqrt{n}}.
\end{align}

Therefore, \eqref{eqn:rcu_mac_iid_11} becomes
\begin{align}
    &\mathrel{\phantom{=}}\Pr\left[ \left\{E_1 \leq \frac{1}{3}\right \} \cap \left\{E_2 \leq \frac{1}{3}\right \} \cap \left\{E_{12} \leq \frac{1}{3}\right \}\right] \nonumber \\
    &\geq 1 -  \epsilon + \frac{c}{\sqrt{n}} -  O\left(\frac{1}{\sqrt{n}}\right)\label{eqn:rcu_mac_iid_12}.
\end{align}

Substituting \eqref{eqn:rcu_mac_iid_12} into \eqref{eqn:rcu_mac_iid_8} gives
\begin{align}
    \epsilon' &\leq  1 - \left(1 -  \epsilon + \frac{c}{\sqrt{n}} -  O\left(\frac{1}{\sqrt{n}}\right)\right) +\frac{F_1+F_2+F_{12}}{\sqrt{n}}\\
    &=\epsilon + \frac{F_1+F_2+F_{12}-c}{\sqrt{n}} +  O\left(\frac{1}{\sqrt{n}}\right).
\end{align}

Recall that the constants $F_1, F_2$, and $F_{12}$ are positive and finite by the moment assumptions \eqref{eqn:rcu_2mac_a1}-\eqref{eqn:rcu_2mac_a6}. Therefore, there exists some constant $c$ and $N$ such that $\epsilon' \leq \epsilon$ for all $n \geq N$. Finally, we apply part 1) of \cite[Lemma 16]{Chen:19_ARXIV} to conclude the existence of an $(n,M_1,M_2,\epsilon)$ code when
\begin{align}
        \bbrR \in \bbrI - \frac{Q_{\text{inv}}(\rmV,\epsilon)}{\sqrt{n}} + \frac{\log n}{2n}\bone - O\left(\frac{1}{n}\right) \bone.
\end{align}
\hfill$\blacksquare$
% \begin{remark} \label{rmk:why_rcu_ldpc_mac}
% Theorem~\ref{thm:rcu_mac_px} does not hold for LDPC codes by the same argument as Remark~\ref{rmk:why_rcu_ldpc}. For the MAC case, the problem is more complicated. Apart from the dependence between the untransmitted codeword $\brX_1^n$ for user 1 and the transmitted codeword $X_1^n$ for user 1, $\brX_1^n$ might be dependent on the transmitted codeword $X_2^n$ for user 2, if user 1 and user 2 are using the same LDPC codebook.
% \end{remark}

\section{RCU Bounds for LDPC Codes}
\subsection{RCU Bound for LDPC Code on the DM-PPC}\label{sec:rcu_p2p_ldpc}
In this section, we apply the generalized RCU bound, Theorem~\ref{thm:RCU}, to the ${\rm LDPC}(\lambda,\rho,\delta;n)$ ensemble to prove an achievability result for LDPC codes. The LDPC achievability result matches the optimal achievable performance of an unrestricted point-to-point code in its first- and second-order terms. The penalty incurred for using the LDPC  code ensemble is $\frac{\log \alpha}{n}$, where $\alpha = \alpha_1\Big|_{(\lambda_1,\rho_1)=(\lambda,\rho)}$ and $\alpha_1$ is from \eqref{eqn:alp1_2mac}. We show that $\frac{\log \alpha}{n}$ is $O(\frac{\log n }{n})$ if $\rho = \kappa n$ and $\kappa$ approaches zero no more quickly than $\bThe(\frac{\log n}{n})$, provided that we first expurgate codes with low minimal distance, as shown in Appendix~\ref{app:ePe}. Whether the penalty in the third-order term results from the LDPC structure or the bounding technique remains an open problem.

The PPC and MAC achievability results for i.i.d. $P_X$ codes (see Theorem~\ref{thm:rcu_p2p_px} and Theorem~\ref{thm:rcu_mac_px}) do not apply for the LDPC code ensemble. The challenges in applying the proof techniques in Theorem~\ref{thm:rcu_p2p_px} and Theorem~\ref{thm:rcu_mac_px} to LDPC codes are as follows.
\begin{enumerate}
    \item The codewords in our LDPC  code ensembles, ${\rm LDPC}(\lambda,\rho;n)$ and ${\rm LDPC}(\lambda,\rho,\delta;n)$, are not independent of each other. For example, if a particular vector is known to be in a random codebook, then it must be true that the underlying Tanner graph describes a family of parity-check equations that are consistent with the given codeword. Further, all other codewords in the codebook must satisfy the same parity checks. Thus, both the parity-check matrix and the other codewords are dependent on the given codeword. For example, when $q=2$ and the check node degree $\rho$ is odd, if $x^n$ is a codeword, then $x^n+ 1^n$ cannot be a codeword, and vice versa.
    \item The symbols within a codeword for the ${\rm LDPC}(\lambda,\rho;n)$ ensemble are not independent. In fact, the symbols within a codeword must be dependent to fulfill the set of parity-check equations.
\end{enumerate}

Nonetheless, the code design of the ${\rm LDPC}(\lambda,\rho,\delta;n)$ ensemble meets the condition of our generalized RCU bound, Theorem~\ref{thm:RCU}.

We here present two results for the ${\rm LDPC}(\lambda,\rho,\delta;n)$ ensemble. The first one is a finite-blocklength error probability bound, which holds for any blocklength $n$. The second one is an asymptotic achievability expansion.

\begin{thm} (LDPC code finite-blocklength bound and second-order-optimal achievability for the DM-PPC). \label{thm:rcu_p2p_ldpc}
Consider a DM-PPC with channel transition probability $P_{Y|X}$ and rational input distribution $P_X$, chosen to approximate the optimal input distribution $P_X^*$. Then there exist LDPC parameters $(\lambda,\rho)$ for which the ${\rm LDPC}(\lambda,\rho,\delta;n)$ ensemble, with $\delta(\cdot)$ chosen to approximate $P_X$, contains at least one code with average error probability less than $\epsilon$ such that for any blocklength $n$
\begin{align}
        \epsilon \leq \bbE\left[\min\left\{1, \alpha M\frac{A(P_X)}{\sqrt{n}}\exp(-I_n)\right\} \right],
\end{align}
and for large enough blocklength $n$
\begin{align}
    \lefteqn{R = 1 - \frac{\lambda}{\rho} = \frac{\log M}{n} \geq C(P_X)} \nonumber \\
   &- \sqrt{\frac{V(P_X)}{n}}Q^{-1}(\epsilon) + \frac{\log n}{2n} - \frac{\log \alpha}{n} - O\left(\frac{1}{n}\right),
\end{align}
% provided the same moment assumptions in \eqref{eqn:rcu_p2p_a1} and \eqref{eqn:rcu_p2p_a2} are satisfied when $X\sim P_X$.
providing the following moment assumptions are satisfied when $X\sim P_X$
\begin{align}
    I(P_X) &>0, \\
    V^Y(P_X) &>0 ,\\
    T(P_X) &< \infty. 
\end{align}
Here
\begin{align}
     A(P_X) &= 2\left(\frac{\log 2}{\sqrt{2\pi V(P_X)}} +2 \frac{C_0 T(P_X)}{V(P_X)^{3/2}}\right), \\
    \alpha &=  \max_{\bt\in \cT_q^n\setminus \{\cT_{q}^n(\bzero)\}}\frac{\oS^n(\bt)}{(M-1)B(n,\bt)q^{-n}},
\end{align}
 $\cT_q^n$ is the set of all possibles types for a list of $n$ elements in $\GF(q)$, $\cT_q^n(\bzero)$ is the type of the all-zero vector, $B(n,\bt)$ is the number of length-$n$ vectors with type $\bt$ (the multinomial coefficient), $M=q^{nR}$, $\oS^n(\bt)$ is the ${\rm LDPC}(\lambda,\rho;n)$ ensemble-average number of type-$\bt$ vectors, and $C(P_X)$ is the mutual information $i(X^n;Y^n)$ evaluated at input distribution $P_X$.
\end{thm}
\begin{remark}
Due to the nature of the quantizer $\delta(\cdot)$, we are only able to achieve rational input distributions that are integer multiples of $\frac{1}{q}$. When the optimal input distribution $P_X^*$ is irrational or not an integer multiple of $\frac{1}{q}$, then a large alphabet size $q$ may be required to closely approximate $P_X^*$.
\end{remark}

{\em Proof of Theorem~\ref{thm:rcu_p2p_ldpc}:}
Since the codeword distribution under the LDPC design meets the constraint of Theorem~\ref{thm:RCU}, the generalized RCU bound is applicable. Setting $X = X^n, \brX = \brX^n, Y =Y^n$ in Theorem~\ref{thm:RCU}, we note that there exists at least one code in this ensemble with average error probability $\epsilon'$ satisfying
\begin{align}
    \epsilon' \leq \bbE\left[\min\{1, M\Pr[i(\brX^n;Y^n)\geq i(X^n;Y^n)|X^n,Y^n]\} \right],
\end{align}
where 
\begin{align*} 
P_{X^n\brX^nY^n}(x^n,\brx^n,y^n)&= P_{X^n,\brX^n}(x^n,\brx^n)P_{Y^n|X^n}(y^n|x^n).
\end{align*}
Here $P_{X^n,\brX^n}(x^n,\brx^n) \neq P_{X^n}(x^n)P_{X^n}(\brx^n)$ in general due to codeword dependence in the ${\rm LDPC}(\lambda,\rho,\delta;n)$ ensemble.

For the ${\rm LDPC}(\lambda,\rho,\delta;n)$ ensemble,  $Y^n$ depends on $\brX^n$ only through its dependence on $X^n$ and therefore $\brX^n \rightarrow X^n \rightarrow Y^n$ forms a Markov chain. Thus,
\begin{align}
   \Pr[\brX^n = \brx^n|X^n,Y^n]  &= \Pr[\brX^n = \brx^n|X^n] \\
    &= \frac{\Pr[\brX^n = \brx^n,X^n=x^n]}{\Pr[X^n=x^n]}.
    \end{align}
Recall from Appendix~\ref{app:Pe} equation \eqref{eqn:ldpc_ub} that
\begin{align}
    \Pr[\bC_1+\bv= \ba,C_{m'}+\bv = \ba'] \leq q^{-n}\alpha q^{-n},
\end{align}
where $\bC_1$ is the first codeword in a random LDPC codebook $\bC$ from the ${\rm LDPC}(\lambda,\rho;n)$ ensemble, $\bC_{m'}$, $m'\neq 1$, is another codeword in $\bC$, and $\bv$ is the random coset vector. Thus
\begin{align}
\lefteqn{\Pr[\brX^n = \brx^n,X^n=x^n]} \label{eqn:ldpc_alpha_start} \\
    & =\sum_{a,a':\delta(a)=x^n,\delta(a')=\brx^n}  \Pr[\bC_1+\bv = \ba, \bC_{m'}+\bv = \ba'] \\
    &\leq \sum_{a,a':\delta(a)=x^n,\delta(a')=\brx^n} q^{-n}\alpha q^{-n} \\
    &= \alpha \sum_{a:\delta(a)=x^n} q^{-n}\cdot \sum_{a:\delta(a')=\brx^n} q^{-n} \\
    &= \alpha \Pr[X^n=x^n]\Pr[\brX^n = \brx^n] \label{eqn:ldpc_alpha_end},
\end{align}
giving
\begin{align}
    \lefteqn{\Pr[\brX^n = \brx^n|X^n,Y^n]} \\
    &\leq \frac{\alpha \Pr[X^n=x^n]\Pr[\brX^n = \brx^n]}{\Pr[X^n=x^n]}\\
    &= \alpha\Pr[X^n=x^n] \\
    &= \alpha P_{X^n}(\brx^n) \frac{P_{Y^n|X^n}(Y^n|\brx^n)}{P_{Y^n}(Y^n)} \frac{P_{Y^n}(Y^n)} {P_{Y^n|X^n}(Y^n|\brx^n)} \\
    &= \alpha\Pr[X^n = \brx^n |Y^n]\exp\left\{-i(\brx^n;Y^n)\right\}.
\end{align}  

Next, we follow the approach from the proof of Theorem~\ref{thm:rcu_p2p_px} to show
\begin{align}
    \epsilon' \leq \bbE\left[\min\left\{1, \alpha M\frac{A(P_X)}{\sqrt{n}}\exp(-I_n)\right\} \right]. \label{eqn:ldpc_p2p_rcu_1}
\end{align}
% We then express $I_n$ as a sum using the memoryless property of the channel
% \begin{align*}
%     I_n = i(X^n;Y^n) = \sum_{j=1}^n i(X_j;Y_j).
% \end{align*}
% We would like to bound this sum using the Berry-Ess\'een Theorem (Lemma~\ref{thm:BET}), but first need to check whether $I_n$ is a sum of independent random variables under the ${\rm LDPC}(\lambda,\rho,\delta;n)$ ensemble.
% To apply Lemma~\ref{thm:BET}, we first need to show that
% \begin{align}
%     \log \frac{P_{Y^n|X^n}(y^n|x^n)}{P_{Y^n}(y^n)} = \sum_{j=1}^n\log \frac{P_{Y|X}(y_j|x_j)}{P_{Y}(y_j)}.
% \end{align}
To bound $I_n$ using the Berry-Ess\'een Theorem (Lemma~\ref{thm:BET}), we first need to check whether $I_n$ is a sum of independent random variables under the ${\rm LDPC}(\lambda,\rho,\delta;n)$ ensemble. That is, we need to show
\begin{align}
    \log \frac{P_{Y^n|X^n}(y^n|x^n)}{P_{Y^n}(y^n)} = \sum_{j=1}^n\log \frac{P_{Y|X}(y_j|x_j)}{P_{Y}(y_j)}.
\end{align}
The given equality holds due to the uniform distribution of coset vector $\bv$. Formally,
\begin{align}
   &\mathrel{\phantom{=}}\Pr[\delta((\bc_1+\bv)[j])=X_j|\delta((\bc_1+\bv)[1:j-1])=X^{j-1}] \nonumber \\
    &=\Pr[\delta((\bc_1+\bv)[j])=X_j], ~\forall j\in \{2,\ldots,n\}.
\end{align}

Setting 
\begin{align}
    \log M &=  nI(P_X) + \frac{1}{2}\log n - \log A(P_X)- \frac{\log \alpha}{n} \nonumber\\
    &\mathrel{\phantom{=}} - \sqrt{nV(P_X)}Q^{-1}\left(\epsilon - \frac{B(P_X)+A(P_X)}{\sqrt{n}}\right)
\end{align}
and following the derivation from \eqref{eqn:rcu_p2p_px_1} to \eqref{eqn:rcu_p2p_px_ldpc_symbol_end}, we can show \eqref{eqn:ldpc_p2p_rcu_1} is bounded by $\epsilon$.

Therefore, setting $P_X$ to the capacity achieving distribution, and using \eqref{eqn:q_Inv_bound_1} and \eqref{eqn:q_Inv_bound_2} to bound $Q^{-1}\left(\epsilon - \frac{B(P_X)+A(P_X)}{\sqrt{n}}\right)$ yields the desired achievability bound.
\hfill$\blacksquare$

\begin{remark}
Theorem~\ref{thm:rcu_p2p_ldpc} provides an achievability bound for the LDPC code ensemble. The result is optimal in its first- and second-order terms. The third-order term exceeds the optimal third-order term for i.i.d. codeword design in \eqref{eqn:rcu_p2p_px_result}, providing an upper bound on the effect of LDPC codeword dependence. By this result, the penalty incurred for using the LDPC code ensemble is at most $\frac{\log \alpha}{n}$, which we show to be $O\left(\frac{\log n }{n} \right)$ if $\rho = \kappa n$ and $\kappa$ approaches zero no more quickly than $\bThe(\frac{\log n}{n})$, provided and we first expurgate codes with low minimal distance as shown in Appendix~\ref{app:ePe}.
\end{remark}

\subsection{RCU Bound for LDPC Code on the DM-$2$-MAC}\label{sec:rcu_2mac_ldpc}
Just as Theorem~\ref{thm:rcu_p2p_ldpc} extends the proof of Theorem~\ref{thm:rcu_p2p_px} from i.i.d. code design to LDPC code design in the PPC, Theorem~\ref{thm:rcu_mac_ldpc}, below, extends Theorem~\ref{thm:rcu_mac_px} from i.i.d. code design to LDPC code design in the MAC.

\begin{thm} (LDPC code finite-blocklength bound, and second-order best-prior achievability on the DM-$2$-MAC). \label{thm:rcu_mac_ldpc}
Consider a DM-$2$-MAC $(\cX_1 \times \cX_2, P_{Y|X_1,X_2}, \cY)$. Assume transmitter $i$ employs the ${\rm LDPC}(\lambda_i,\rho_i,\delta_i;n)$ ensemble with coset vector $\bv_i$, and quantizer $\delta_i(\cdot)$ chosen to approximate $P_{X_i}$ for $i\in\{1,2\}$. Then there exist LDPC parameters $(\lambda_1,\rho_1)$ and $(\lambda_2,\rho_2)$ for which the ${\rm LDPC}(\lambda_1,\rho_1,\delta_1;n)\times {\rm LDPC}(\lambda_2,\rho_2,\delta_2;n)$ ensemble contains at least one MAC code with average error bounded by $\epsilon$ such that for any blocklength $n$
\begin{align}
    \epsilon&\leq \bbE\left[\min\left\{1, \alpha_1 E_1+\alpha_2 E_2+\alpha_1\alpha_2E_{12}\right\}\right],
\end{align}
and for large enough $n$
\begin{align}
    \bbrR \in \bbrI - \frac{Q_{\text{inv}}(\rmV,\epsilon)}{\sqrt{n}} + \frac{\log n}{2n}\bone - \frac{\log \bar{\boldsymbol{\alpha}}}{n} \bone- O\left(\frac{1}{n}\right) \bone,
\end{align}
provided the moment assumptions \eqref{eqn:rcu_2mac_a1}-\eqref{eqn:rcu_2mac_a6} are satisfied. The definitions of $\bbrI$ and $\rmV$ are the same as those in Theorem~\ref{thm:rcu_mac_px}, $E_1, E_2$ and $E_{12}$ are defined in \eqref{eqn:e1}-\eqref{eqn:e12}, and $\alpha_1$ and $\alpha_2$ are the same as those in Theorem~\ref{thm:Pe_2mac}. The remaining terms are defined as
\begin{align}
\bbrR &\defeq \begin{bmatrix} R_1 \\ R_2 \\ R_1+R_2\end{bmatrix}=\begin{bmatrix} 1-\frac{\lambda_1}{\rho_1} \\ 1-\frac{\lambda_2}{\rho_2} \\ 1-\frac{\lambda_1}{\rho_1} + 1-\frac{\lambda_2}{\rho_2}\end{bmatrix}, \\ 
    \bar{\boldsymbol{\alpha}} &= \begin{bmatrix} \alpha_1 \\\alpha_2 \\\alpha_1\alpha_2 \end{bmatrix}. \\
    % \alpha_1 &=  \max_{\bt\in \cT_q^n\setminus \{\cT_{q}^n(\bzero)\}}\frac{\oS_1^n(\bt)}{(M_1-1)B(n,\bt)q^{-n}},\\
    % \alpha_2 &=  \max_{\bt\in \cT_q^n\setminus \{\cT_{q}^n(\bzero)\}}\frac{\oS_2^n(\bt)}{(M_2-1)B(n,\bt)q^{-n}},
\end{align}
%  $\cT_q^n$ is the set of all possibles types for a list of $n$ elements in $\GF(q)$, $\cT_q^n(\bzero)$ is the type of the all zero vector, $B(n,\bt)$ is the number of length $n$ vectors with type $\bt$ (multinomial coefficient), $M_i = q^{nR_i}$, $\oS_i^n(\bt)$ is the ${\rm LDPC}(\lambda_i,\rho_i;n)$ ensemble-average (before applying coset vector and quantization) number of type-$\bt$ vectors for $i\in \{1,2\}$, and .
\end{thm}

\begin{remark}
Since the quantizers $\delta_i(\cdot), i\in\{1,2\}$ restrict achievable input distributions $P_{X_i},i\in\{1,2\}$ to be integer multiples of $\frac{1}{q}$, rate pairs $(R_1,R_2)$ that require irrational input distributions or rational input distributions with non-integer multiples of $\frac{1}{q}$ may require large alphabet size $q$ to closely approximate the desired input distributions.
\end{remark}

{\em Proof of Theorem~\ref{thm:rcu_mac_ldpc}:}
Notice that the ${\rm LDPC}(\lambda_i,\rho_i,\delta_i;n)$ ensemble meets the codeword distribution constraint of Theorem~\ref{thm:RCU_2MAC}.

Setting $X_1 = X_1^n, \brX_1 = \brX_1^n,X_2 = X_2^n, \brX_2 = \brX_2^n, Y =Y^n$ in Theorem~\ref{thm:RCU_2MAC}, we note that there exists at least one code in the joint ensemble ${\rm LDPC}(\lambda_1,\rho_1,\delta_1;n) \times {\rm LDPC}(\lambda_2,\rho_2,\delta_2;n)$ such that the average error probability $\epsilon'$ satisfies
\begin{align}
    \epsilon' \leq \bbE[\min\{1, V_1 + V_2 + V_{12}\}],
\end{align}
where 
\begin{align} 
    V_1 &=(M_1-1)\nonumber \\
    &\mathrel{\phantom{==}}\Pr[i(\brX_1^n;Y^n|X_2^n)\geq i(X_1^n;Y^n|X_2^n)|X_1^n,X_2^n,Y^n], \\
    V_2 &= (M_2-1)\nonumber \\
    &\mathrel{\phantom{==}}\Pr[i(\brX_2^n;Y^n|X_1^n)\geq i(X_2^n;Y^n|X_1^n)|X_1^n,X_2^n,Y^n], \\
    V_{12} &= (M_1-1)(M_2-1) \nonumber \\
    &\mathrel{\phantom{==}}\Pr[i(\brX_1^n,\brX_2^n;Y^n)\geq i(X_1^n,X_2^n;Y)^n|X_1^n,X_2^n,Y^n]\}.
\end{align}
Since the codebooks for transmitter $1$ and transmitter $2$ are independently designed, but the codewords in each are dependent under LDPC design
\begin{align*} 
\lefteqn{P_{X_1^nX_2^n\brX_1^n\brX_2^nY^n}(x_1^n,x_2^n,\brx_1^n,\brx_2^n,y^n)} \\
&= P_{X_1^n\brX_1^n}(x_1^n,\brx_1^n)P_{X_2^n\brX_2^n}(x_2^n,\brx_2^n)P_{Y^n|X_1^nX_2^n}(y^n|x_1^n,x_2^n).
\end{align*}

% , we cannot claim that $P_{X_i^n\brX_i^n}(x_i^n,\brx_i^n)$ equals $P_{X_i^n}(x_i^n)P_{X_i^n}(\brx_i^n)$ for $i\in\{1,2\}$.

% Denote for $i \in \{1,2\}$,
% \begin{align}
%     \alpha_i =  \max_{\bt\in \cT_q^n\setminus \{\cT_{q}^n(\bzero)\}}\frac{\oS_i^n(\bt)}{(M_i-1)B(n,\bt)q^{-n}},
% \end{align}
% where $\cT_q^n$ is the set of all possibles types in $a^n \in \GF(q)^n$ $\cT_q^n(\bzero)$ is the type of the all zero vector, $B(n,\bt)$ is the number of length $n$ vectors with type $\bt$ (multinomial coefficient), and $\oS^n_i(\bt)$ is the ${\rm LDPC}(\lambda_i,\rho_i,\delta_i;n)$ ensemble-average number of type $\bt$ vectors.

From Appendix~\ref{app:Pe}, we know that for each $i \in \{1,2\}$
\begin{align}
    \Pr[\bC_{i,1}+\bv_i= \ba,\bC_{i,2}+\bv_i= \ba'] \leq q^{-n}\alpha_i q^{-n},
\end{align}
where $\bC_{i,1}$ and $\bC_{i,2}$ are the codewords for messages $1$ and $2$ from a random code in the ${\rm LDPC}(\lambda_i,\rho_i,\delta_i;n)$ ensemble for transmitter $i$, and $\bv_i$ is the coset vector for transmitter $i$.

For each of the LDPC  code ensembles, we note that for $i \in \{1,2\}$, $\brX_i^n \rightarrow X_i^n \rightarrow (X_{3-i}^n,Y^n)$ forms a Markov chain, as the dependence of $Y^n$ or $X_{3-i}^n$ on $\brX_i^n$ is through $X_i^n$. A given $Y^n$ affects the conditional distribution on $X_i^n$ through the structure of the channel, and thus affects the conditional distribution of $\brX_i^n$ through the dependence between $X_i^n$ and $\brX_i^n$. By the assumption of independent coset vectors $\bv_1$ and $\bv_2$, $X_{3-i}^n$ is independent of $\brX_i^n$.

Therefore,
\begin{align}
 \Pr[\brX_1^n = \brx_1^n|X_1^n,X_2^n,Y^n] &= \Pr[\brX_1^n = \brx_1^n|X_1^n] \\
    &= \frac{\Pr[\brX^n = \brx^n,X^n=x^n]}{\Pr[X^n=x^n]}.
    \end{align}
    
By an argument similar to \eqref{eqn:ldpc_alpha_start}-\eqref{eqn:ldpc_alpha_end}, we have
\begin{align}
    \lefteqn{\Pr[\brX_1^n = \brx_1^n|X_1^n,X_2^n,Y^n]} \nonumber \\
    &\leq \alpha_1    \Pr[\brX_1^n = \brx_1^n] \\
    &= \alpha_1 \Pr[X_1^n = \brx_1^n |Y^n,X_2^n]\exp\left\{-i(\brx_1^n;Y^n|X_2^n)\right\}.
\end{align}

Summing over all $\brx_1^n=i(\brx_1^n;Y^n|X_2^n)$ such that $\brI_{1n}\geq \zeta$ gives
\begin{eqnarray}
    \lefteqn{\Pr[\brI_{1n} \geq \zeta|Y^n,X_2^n]} \nonumber \\
    &=& \alpha_1\bbE[\exp\{-\brI_{1n}\} \ind\{\brI_{1n}\geq \zeta\}|Y^n,X_2^n] \nonumber \\
    &\leq& \alpha_1\frac{F_1}{\sqrt{n}}\exp(-\zeta) \label{eqn:rcu_mac_ldpc_1},
\end{eqnarray}
where \eqref{eqn:rcu_mac_ldpc_1} follows from Lemma~\ref{lem:Pol47}, and $F_1$ is defined in \eqref{eqn:rcu_mac_f1}.

Therefore, 
\begin{align}
    V_1 \leq \alpha_1\frac{F_1}{\sqrt{n}}\exp(-i(X_1^n;Y^n|X_2^n)) \label{eqn:LDPC:V1_i}
\end{align}
Switching the role of transmitter 1 and transmitter 2 yields
\begin{align}
     \lefteqn{\Pr[\brX_2^n = \brx_2^n|X_1^n,X_2^n,Y^n]}\\
     &\leq \alpha_2 \Pr[X_2^n = \brx_2^n |Y^n,X_1^n]\exp\left\{-i(\brx_2^n;Y^n|X_1^n)\right\},
\end{align}
and therefore
\begin{align}
    V_2 \leq \alpha_1\frac{F_2}{\sqrt{n}}\exp(-i(X_2^n;Y^n|X_1^n)). \label{eqn:LDPC:V2_i}
\end{align}
Finally, for $\Pr[\brX_1^n = \brx_1^n, \brX_2^n = \brx_2^n|X_1^n,X_2^n,Y^n]$, we have
\begin{align}
    \lefteqn{\Pr[\brX_1^n = \brx_1^n, \brX_2^n = \brx_2^n|X_1^n,X_2^n,Y^n]} \\
    &=\Pr[\brX_2^n = \brx_2^n|X_1^n,X_2^n,Y^n] \nonumber \\
    &\mathrel{\phantom{\qquad\qquad\qquad}} \Pr[\brX_1^n = \brx_1^n, |X_1^n,X_2^n,Y^n,\brX_2^n] \\
    &=\Pr[\brX_2^n = \brx_2^n|X_2^n]\Pr[\brX_1^n = \brx_1^n, |X_1^n] \\
    &\leq \alpha_1\alpha_2 \Pr[X_1^n = \brx_1^n,X_2^n = \brx_2^n |Y^n]\exp\left\{-i(\brx_1^n,\brx_2^n;Y^n)\right\},
\end{align}
and 
\begin{align}
    V_{12} \leq \alpha_1\alpha_2\frac{F_{12}}{\sqrt{n}}\exp(-i(X_1^n,X_2^n;Y^n)). \label{eqn:LDPC:V12_i}
\end{align}

With the definitions of $E_1, E_2, E_{12}$ from \eqref{eqn:e1}-\eqref{eqn:e12}, combining the above three results on $V_1,V_2$, and $V_{12}$ gives the finite-blocklength error bound.

The rest of the proof follows from the proof of Theorem~\ref{thm:rcu_mac_px} by invoking Berry-Ess\'een Theorem (Lemma~\ref{thm:BET}) to bound each of the three terms in \eqref{eqn:LDPC:V1_i}, \eqref{eqn:LDPC:V2_i}, and \eqref{eqn:LDPC:V12_i}, which gives the following achievability result for large enough $n$
\begin{align}
        \bbrR \in \bbrI - \frac{Q_{\text{inv}}(\rmV,\epsilon)}{\sqrt{n}} + \frac{\log n}{2n}\bone - \frac{\log \bar{\boldsymbol{\alpha}}}{n} \bone - O\left(\frac{1}{n}\right) \bone.
\end{align}
\hfill$\blacksquare$

\begin{remark}
Theorem~\ref{thm:rcu_mac_ldpc} provides an achievability bound for the random LPDC ensemble that achieves the same second-order term as the best known bound for i.i.d. MAC codes. The penalty for the codeword dependence that result from using the LDPC code ensemble is the $\frac{\log \bar{\boldsymbol{\alpha}}}{n}$ term, which is $O\left(\frac{\log n }{n} \right)\bone$ if $\rho_1= \kappa_1 n,\rho_2=\kappa_2 n$ and $\kappa_1, \kappa_2$ approach zero no more quickly than $\bThe(\frac{\log n}{n})$, provided that we first expurgate codes with small minimal distance.
\end{remark}
\begin{remark}
The proof of Theorem~\ref{thm:rcu_mac_ldpc} uses an independent code ensemble for each transmitter and independent coset vectors. For many practical scenarios, it is useful to allow different transmitters to use the same LDPC code for simplicity. If the same code ensemble ${\rm LDPC}(\lambda,\rho;n)$ (before applying the coset vector and quantization) is used for both transmitters, then $\alpha_1=\alpha_2$. In addition, if the transmitters use the same coset vector, then both $X_1^n$ and $X_2^n$ have an impact on the distribution of $\brX_1^n$ (similar for $\brX_2^n$), as knowing both $X_1^n,X_2^n$ (assuming $X_1^n \neq X_2^n$) reveals two different codewords in the codebook. In this case, the penalty term $\log \bar{\boldsymbol{\alpha}}$ becomes $$\log \bar{\boldsymbol{\alpha}} = \begin{bmatrix} 2\log\alpha_1\\2\log\alpha_2\\2\log\alpha_1+2\log\alpha_2\end{bmatrix} =\begin{bmatrix} 2\log\alpha_1\\2\log\alpha_1\\4\log\alpha_1\end{bmatrix}. $$
Hence, different transmitters may use the same or different coset vectors depending on their sensitivity to the factor of 2 difference in the rate penalty bound.
% In the latter case, we rely on each user having a unique identity (such as phone number), which can be used to choose different coset vector.
\end{remark}

\section{Summary and Conclusions}
This paper studies the performance of quantized coset LDPC codes over the DM-PPC and the DM-MAC using finite-blocklength and error-exponent analyses. 

For the error-exponent analysis, we extend the result of \cite{BennatanB:04} from the DM-PPC to symmetrical rates in the symmetric DM-$K$-MAC and arbitrary rates in the general DM-$2$-MAC using Gallager's error exponent. A non-asymptotic expansion of Gallager's error exponent is provided using \cite[Exercise 5.23]{Gallager:68}.

For the dispersion-style approach, we derive finite-blocklength error bounds and asymptotic third-order achievability results for the DM-PPC and the DM-$2$-MAC for standard i.i.d. codes; the achievability result is optimal up to the third order for the DM-PPC (Theorem~\ref{thm:rcu_p2p_px}), and is the tightest bound available to date for the DM-$2$-MAC (Theorem~\ref{thm:rcu_mac_px}). Application of two generalized RCU bounds (Theorem~\ref{thm:RCU} for the DM-PPC and Theorem~\ref{thm:RCU_2MAC} for the DM-$2$-MAC) shows that quantized coset LDPC codes achieve first- and second-order performance 
that is optimal for the DM-PPC (Theorem~\ref{thm:rcu_p2p_ldpc})
and identical to the best-prior results for the DM-MAC (Theorem~\ref{thm:rcu_mac_ldpc}), provided that we first expurgate LDPC codes with small minimum distance, and the sparsity of LDPC codes ($\kappa = \frac{\rho}{n}$) decays no more quickly than $\bThe(\frac{\log n}{n})$. 

A comparison of both approaches (Section~\ref{sec:exp}) demonstrates that the error-exponent analysis achieves a sub-optimal second-order coefficient in blocklength $n$ but a superior bound when target error probability $\epsilon$ is small.

%\section{Message Passing Decoding}\label{sec:mp}

%The message passing algorithm transmits a description of a pmf on $\cX^k$.  
%This description may be the probability mass function (pmf) itself, 
%which we describe as a $|\cX|^k$-dimensional vector $\bp=(p_\bx: \bx\in\cX^k)$
%of non-negative real numbers satisfying $\sum_{\bx\in\cX^k}p_\bx=1$, 
%or it may be the log-likelihood ratio (LLR), 
%which we describe as a $|\cX|^k$-dimensional vector $\ell=(\ell(x^k):x^k\in\cX^k)$ 
%such that 
%\[
%\ell(x^k)\defeq\log(p(0^k)/p(x^k)),\ \ x^k\in\cX^k.
%\]
%A pmf chosen at random from some family of possible pmfs is denoted by $P$.  
%The corresponding LLR is $L$.  

\appendices
\section{Proof of Theorem~\ref{thm:Pe}}\label{app:Pe}

Recall that by the given code construction, 
all transmitters employ the same codebook, but each is 
offset by an independent random coset vector. Recall further that the codebook is restricted to include precisely $M = q^{nR}$ codewords, where $R = 1 - \frac{\lambda}{\rho} $ is the design rate. Here $\bc=\{\bc_1,\ldots,\bc_{M}\}$ describes the single-transmitter codebook and $\bd=\{\bd_\bm:\bm\in [M]^K\}$ describes the corresponding MAC codebook, where for any $\bm=(m(1),\ldots,m(K))$, $\bd_\bm=(\bc_{m(1)},\ldots,\bc_{m(K)})$.
Given a coset matrix $\bv$ and quantizer $\delta$, 
the resulting set of channel inputs is $\{\delta(\bd_\bm+\bv):\bm\in [M]^K\}$.

The expected value under our random code construction 
of the average error probability is
\begin{eqnarray*}
E[P_e^{(n)}] 
	& = & \sum_\bm\sum_\bd\sum_\bv P_\bM(\bm)P_\bD(\bd)P_\bV(\bv)P_{e|\bm,\bd,\bv}^{(n)} \\
	& = & E_{\bM\bD\bV}\left[P_{e|\bM,\bD,\bV}^{(n)}\right],
\end{eqnarray*}
where $P_{e|\bm,\bd,\bv}^{(n)}$ is the conditional error probability 
under fixed values of the message vector $\bm$, codebook $\bd$, and coset matrix $\bv$, 
$P_\bM(\bm)$, $P_\bD(\bd)$, and $P_\bV(\bv)$ capture the (independent, uniform) 
distributions on the vectors of possible messages, set of possible codebooks, and cosets, respectively, and 
$E_{\bM\bD\bV}[\cdot]$ is the resulting expectation. 

We begin by bounding the conditional error probability 
$P_{e|\bm,\bd,\bv}^{(N)}$.  
Let 
\begin{eqnarray*}
\bcY_{\bm,\bd,\bv}  & =  & \left\{\by:\exists\ \bm'\in [M]^K\setminus\{\bm\} \mbox{ s.t. }\right. \\
&& \left.\Pr\left[\by|\delta(\bd_{\bm'}+\bv)\right]\geq\Pr\left[\by|\delta(\bd_\bm+\bv)\right]\right\},
\end{eqnarray*}
denote the set of channel outputs for which message vector $\bm$ is not the unique most likely explanation. Then 
\[
P_{e|\bm,\bd,\bv}^{(n)} \leq \Pr\left[\left.\bcY_{\bm,\bd,\bv}\right|\bm,\bd,\bv\right], 
\]
which is an inequality rather than an equality since an error is not guaranteed 
when $\Pr[\by|\delta(\bd_{\bm'}+\bv)]=\Pr[\by|\delta(\bd_\bm+\bv)]$.  
For any set $\Tau\subseteq\cT^n_\cQ$, define $\bcY^\Tau_{\bm,\bd,\bv}$ as 
\begin{eqnarray*}
\bcY^\Tau_{\bm,\bd,\bv} = \left\{\by:\exists\ \bm'\in [M]^K\setminus\{\bm\} \mbox{ s.t. }\cT^n_\cQ(\Delta^\bd_{\bm,\bm'})\in \Tau\right. \\
 \left. \mbox{ and }\Pr\left[\by|\delta(\bd_{\bm'}+\bv)\right]\geq\Pr\left[\by|\delta(\bd_\bm+\bv)\right]\right\}, %\wedge\
\end{eqnarray*}
where 
\[
\Delta^\bd_{\bm',\bm}=\bd_\bm-\bd_{\bm'}.
\]
Recall that $\Tau^c=\cT_\cQ^n\setminus \Tau\setminus\{\cT_\cQ^n(\bzero)\}$, 
where $\bzero$ is the all-zeros  codematrix.
Then
\[
\bcY_{\bm,\bd,\bv}=	\bcY^\Tau_{\bm,\bd,\bv}\cup
				\bcY^{\Tau^c}_{\bm,\bd,\bv}\cup
				\bcY^{\{\cT_\bQ^n(\bzero)\}}_{\bm,\bd,\bv}, 
\]
and we have 
\begin{eqnarray*}
P_{e|\bm,\bd,\bv}^{(n)}  & \leq & \Pr[\bcY^\Tau_{\bm,\bd,\bv}|\bm,\bd,\bv]+\Pr[\bcY^{\Tau^c}_{\bm,\bd,\bv}|\bm,\bd,\bv] \\
	&& +\Pr[\bcY^{\{\cT_\cQ^n(\bzero)\}}_{\bm,\bd,\bv}|\bm,\bd,\bv].
\end{eqnarray*}
Since all codewords in the single-transmitter codebook are distinct by definition 
($\bc_m\neq\bc_{m'}$ for all $m'\in [M]\setminus\{m\}$), 
all codematrices are also distinct 
($\bd_\bm\neq\bd_{\bm'}$ for all $\bd$ and all $\bm'\in [M]^K\setminus\{\bm\}$),
set $\bcY^{\{\cT_\cQ^n(\bzero)\}}_{\bm,\bd,\bv}$ is always empty, 
and we can bound $P_{e|\bm,\bd,\bv}$ by bounding the remaining two terms.

Let $\bt_{\bm,\bm'}=\cT^n_\cQ(\Delta^\bd_{\bm,\bm'})$.
Then, for the first term, 
\begin{align*}
\lefteqn{\Pr[\bcY^\Tau_{\bm,\bd,\bv}|\bm,\bd,\bv]} \\
& =  \sum_{\by\in\bcY^\Tau_{\bm,\bd,\bv}}P_{\bY|\bX}(\by|\delta(\bd_\bm+\bv)) \\
& \leq  \sum_{\by\in\bcY^\Tau_{\bm,\bd,\bv}}\left[ \vphantom{\sum_{\bm':\bt_{\bm,\bm'}\in \Tau}} P_{\bY|\bX}(\by|\delta(\bd_\bm+\bv)) \right.\\
&\qquad\qquad\left. \sum_{\bm':\bt_{\bm,\bm'}\in \Tau}
\sqrt{\frac{P_{\bY|\bX}(\by|\delta(\bd_{\bm'}+\bv))}{P_{\bY|\bX}(\by|\delta(\bd_\bm+\bv))}} \right]\\
& \leq  \sum_{\by}\sum_{\bm':\bt_{\bm,\bm'}\in \Tau} \\
&\mkern80mu \sqrt{P_{\bY|\bX}(\by|\delta(\bd_{\bm'}+\bv))P_{\bY|\bX}(\by|\delta(\bd_\bm+\bv))} \\
& =  \sum_{\bm':\bt_{\bm,\bm'}\in \Tau}\sum_{\by}\prod_{i=1}^n \\
& \sqrt{P_{Y|X}(y_i|\delta((\bd_{\bm'}+\bv)[i,*]))P_{Y|X}(y_i|\delta((\bd_{\bm}+\bv)[i,*]))} \\
& =  \sum_{\bm':\bt_{\bm,\bm'}\in \Tau}\prod_{i=1}^n \sum_y\\
& \sqrt{P_{Y|X}(y|\delta((\bd_{\bm'}+\bv)[i,*]))P_{Y|X}(y|\delta((\bd_{\bm}+\bv)[i,*]))}.
\end{align*}
Taking the expectation over random cosets gives
\begin{align*}
\lefteqn{E_\bV[\Pr[\bcY^\Tau_{\bm,\bd,\bV}|\bm,\bd,\bV]]} \\
& \leq  E_\bV\left[\sum_{\bm':\bt_{\bm,\bm'}\in \Tau}\prod_{i=1}^n \sum_y\right.\\
&\left.\vphantom{\sum_{\bt_{\bm,\bm'}}}\sqrt{P_{Y|X}(y|\delta((\bd_{\bm'}+\bV)[i,*]))P_{Y|X}(y|\delta((\bd_{\bm}+\bV)[i,*]))}\right] \\
& =  \sum_{\bm':\bt_{\bm,\bm'}\in \Tau}\prod_{i=1}^n E_{\bV[i,*]}\left[\sum_y\right.\\
& \left.\vphantom{\sum_{\by}}\sqrt{P_{Y|X}(y|\delta((\bd_{\bm'}+\bV)[i,*]))P_{Y|X}(y|\delta((\bd_{\bm}+\bV)[i,*]))}\right] \\
& \stackrel{(a)}{=} 
\sum_{\bm':\bt_{\bm,\bm'}\in \Tau}\prod_{i=1}^n \left[\sum_{g'\in\cQ}\frac1{q^K}\sum_y\right.\\
& \hphantom{=}\left.\vphantom{\sum_{\bt_{\bm,\bm'}}}\sqrt{P_{Y|X}(y|\delta(g'+\Delta^\bd_{\bm,\bm'}[i,*]))P_{Y|X}(y|\delta(g'))}\right] \\
& \stackrel{(b)}{=} 
\sum_{\bm':\bt_{\bm,\bm'}\in \Tau}\prod_{i=1}^n \cD(\Delta^\bd_{\bm,\bm'}[i,*]) \\
& \stackrel{(c)}{=}  
\sum_{\bm':\bt_{\bm,\bm'}\in \Tau}\bcD^{\bt_{\bm,\bm'}} \\
& \stackrel{(d)}{\leq} 
\sum_{\bm^*:\cT^n_\cQ(\bd_{\bm^*})\in \Tau}\bcD^{\cT^n_\cQ(\bd_{\bm^*})} \\
& \stackrel{(e)}{=} 
\sum_{\bt\in \Tau}S^n_{\bd}(\bt)\bcD^{\bt}.
\end{align*}
Here $(a)$ follows since each row of $\bV$ is uniformly distributed over $\cQ$, 
which  implies that each row of $\bd_{\bm}+\bV$ is uniformly distributed over $\cQ$; 
$(b)$ and $(c)$ apply definitions~(\ref{eqn:defD}) and~(\ref{eqn:defDt}); $(d)$ uses the fact that the difference between two codewords is a codeword in any liner code, and therefore the given upper bound applies after our random selection of codewords; and $(e)$ applies definition~(\ref{eqn:defS}). 
Finally, taking the expectation 
with respect to the random choice of the codebook and message gives 
\[
E[\Pr[\bcY^\Tau_{\bM,\bD,\bV}|\bM,\bD,\bV]]
\leq 
\sum_{\bt\in \Tau}\obS^n(\bt)\bcD^{\bt}.
\]

For the second term, abbreviating $\bm \in [M]^K\setminus\{\bm\}$ to $\bm \neq \bm'$,
\begin{align*}
\lefteqn{E[\Pr[\bcY^{\Tau^c}_{\bM,\bD,\bV}|\bM,\bD,\bV]]} \\
& =  \sum_{\bm,\ba,\by}P_\bM(\bm)P_{\bD_\bm+\bV}(\ba)P_{\bY|\bX}(\by|\delta(\ba)) \\
& \hphantom{ \sum} \cdot\Pr[\exists\bm'\neq\bm: \cT_\cQ^n(\bD_{\bm'}-\bD_{\bm})\in \Tau^c,  \\
& \hphantom{\cdot\Pr[]}P_{\bY|\bX}(\by|\delta(\bD_{\bm'}+\bV))\geq P_{\bY|\bX}(\by|\delta(\bD_\bm+\bV)) \\
&\hphantom{\cdot\Pr[]}|\bD_\bm+\bV=\ba] \\
& = \sum_{\bm,\ba,\by}P_\bM(\bm)P_{\bD_\bm+\bV}(\ba)P_{\bY|\bX}(\by|\delta(\ba)) \\
&\hphantom{ \sum} \cdot\Pr[\exists\bm'\neq\bm: \bD_{\bm'}+\bV=\ba',\ \cT_\cQ^n(\ba'-\ba)\in \Tau^c,  \\
& \hphantom{\cdot\Pr[]}P_{\bY|\bX}(\by|\delta(\ba'))\geq P_{\bY|\bX}(\by|\delta(\ba))|\bD_\bm+\bV=\ba] \\
& \stackrel{(e)}{\leq}  \sum_{\bm,\ba,\by}P_\bM(\bm)P_{\bD_\bm+\bV}(\ba)P_{\bY|\bX}(\by|\delta(\ba))\min\left\{1,\sum_{\bm'\neq\bm}\right.\\
& \hphantom{ \sum} \left. \sum_{\substack{\ba':\cT_\cQ^n(\ba'-\ba)\in \Tau^c\\P_{\bY|\bX}(\by|\delta(\ba'))\geq P_{\bY|\bX}(\by|\delta(\ba))}}
	\mkern-60mu \Pr[\bD_{\bm'}+\bV=\ba'|\bD_\bm+\bV=\ba]\right\} \\
& \stackrel{(f)}{\leq}  \sum_{\bm,\ba,\by}P_\bM(\bm)P_{\bD_\bm+\bV}(\ba)P_{\bY|\bX}(\by|\delta(\ba))\left(\sum_{\bm'\neq\bm}\right. \\
&\hphantom{ \sum} \sum_{\substack{\ba':\cT_\cQ^n(\ba'-\ba)\in \Tau^c\\P_{\bY|\bX}(\by|\delta(\ba'))\geq P_{\bY|\bX}(\by|\delta(\ba))}}
	\mkern-60mu \Pr[\bD_{\bm'}+\bV=\ba'|\bD_\bm+\bV=\ba]\left.\vphantom{\sum_{\bm'\neq\bm}}\right)^\rho \\
& \stackrel{(g)}{=}  \sum_{\by,\ba} P_{\bD_\bone+\bV}(\ba) P_{\bY|\bX}(\by|\delta(\ba))\left(\sum_{\substack{\bm'\neq\bone}}\right.\\
&\hphantom{ \sum}\!\!\! \sum_{\substack{\ba: \cT^n_\cQ(\ba'-\ba)\in \Tau^c \\ P_{\bY|\bX}(\by|\delta(\ba'))\geq P_{\bY|\bX}(\by|\delta(\ba))}}
	\mkern-60mu\Pr[\bD_{\bm'}+\bV=\ba'|\bD_\bone+\bV=\ba]\left.\vphantom{\sum_{\bm'\neq\bm}}\right)^\rho,
\end{align*}
where $(e)$ follows from the union bound and the bounded nature of probabilities;
$(f)$ follows by a case analysis for any $\rho\in[0,1]$ ($\min\{1,a\} = 1 \leq a^\rho$ when $a\geq 1$, and $\min\{1,a\} = a \leq a^\rho$ when $0\leq a<1$); and 
$(g)$ follows by taking $\bm=\bone=(1,\ldots,1)$ 
by the symmetry of our random code design. 
Under our random code design and coset choice, 
for any $\bm'\neq\bone$ 
\begin{eqnarray}
\lefteqn{\Pr[\bD_\bone+\bV=\ba,\bD_{\bm'}+\bV=\ba']} \nonumber \\
& = & \sum_{\bv}\Pr[\bV=\bv,\bD_\bone=\ba-\bv,\bD_{\bm'}-\bD_\bone=\ba'-\ba] \nonumber \\
& = & q^{-nK}\sum_{\bv}\Pr[\bD_\bone=\ba-\bv,\bD_{\bm'}-\bD_\bone=\ba'-\ba] \nonumber \\
& = & q^{-nK}\Pr[\bD_{\bm'}-\bD_\bone=\ba'-\ba] \nonumber\\
& \stackrel{(h)}{\leq} & q^{-nK}\Pr[\ba'-\ba\in \bD] \nonumber \\
&& \cdot \Pr[\bD_{\bm'}-\bD_\bone=\ba'-\ba|\ba'-\ba\in\bD] \nonumber \\
& \stackrel{(i)}{=} & q^{-nK}\frac{\oS^n(\cT^n_\cQ(\ba'-\ba))}{B\left(n,\cT^n_\cQ(\ba'-\ba)\right)} \frac1{M^K-1} \nonumber \\
& \stackrel{(j)}{\leq} & q^{-nK}\left(\alpha_{\rm \scalebox{0.4}{MAC}} q^{-nK}\right) \label{eqn:ldpc_ub},
\end{eqnarray}
where $(h)$ follows since the difference between two codematrices 
is also a codematrix in any linear MAC code, and the upper bound continues to hold even when we select $M=q^{nR}$ codewords from the set of parity-check solutions; $(i)$ follows from the symmetry of our code design (since no variable node is treated any better or worse than any other variable node) and from our restriction to 
precisely $M = q^{nR}$ codewords in each single-transmitter codebook; 
and $(j)$ follows from the definition of $\alpha_{\rm \scalebox{0.4}{MAC}}$ in (\ref{eqn:alpha}).  
Since $P_{\bD_\bone+\bV}(\ba)=q^{-nK}$ 
by the uniformity of random matrix $\bV$, 
% $(i)$ follows since the number of codematrices $M_\bD=(M(\bC))^K$ 
% in random MAC codebook $\bD$ 
% with underlying random single-transmitter codebook $\bC$ 
% satisfies $M(\bC)\geq M = q^{nR}$ prior to our random restriction to precisely $M$ codewords, 
\[
\Pr[\bD_{\bm'}+\bV=\ba'|\bD_\bone+\bV=\ba]\leq\alpha_{\rm \scalebox{0.4}{MAC}} q^{-nK}.
\]
%Let $R_\bC$ denote the rate of the underlying LDPC code corresponds to MAC codebook $\bD$ (which can be related as $R_\bC =\log (M_\bD) / (nK)$), then breaking the error event based on whether $R_\bC > R+\epsilon$ (equivalently, $M_\bD > q^{n(R+\epsilon)K}$) or $R_\bC \leq R+\epsilon$ using law of total probability gives
Therefore
\begin{align*}
\lefteqn{E[\Pr[\bcY^{\Tau^c}_{\bM,\bD,\bV}|\bM,\bD,\bV]]} \\
%&=& E[\Pr(\bcY^{\Tau^c}_{\bM,\bD,\bV}|\bM,\bD: R_\bC \leq R +\epsilon,\bV)] \\
%&& \cdot \Pr(R_\bC \leq R+\epsilon) + \Pr(R_\bC > R+\epsilon) \\
%&& \cdot E[\Pr(\bcY^{\Tau^c}_{\bM,\bD,\bV}|\bM,\bD: R_\bC > R +\epsilon,\bV)] \\
%&=& \lefteqn{E[\Pr(\bcY^{\Tau^c}_{\bM,\bD,\bV}|\bM,\bD,\bV)]}\\
% &\leq &  \sum_{\by,\ba} q^{-nK}p(\by|\delta(\ba))\left(\sum_{\bm' \neq \bone}\right.
% 	\sum_{\substack{\ba'\in\cQ:\cT^n_\cQ(\ba'-\ba)\in \Tau^c \\ p(\by|\delta(\ba'))\geq p(\by|\delta(\ba))}}
% 	\!\!\!\!\!\!\!\!\!\!\!\!\!\!\!\!\!\alpha q^{-nK}\left.\vphantom{\sum_{\bm'\neq\bone}}\right)^\rho  \\
% &=&  \sum_{\by,\ba} q^{-nK}p(\by|\delta(\ba))\left(\sum_{\bm' =\btwo}^{M_\bD}\right.
% 	\sum_{\substack{\ba'\in\cQ:\cT^n_\cQ(\ba'-\ba)\in \Tau^c \\ p(\by|\delta(\ba'))\geq p(\by|\delta(\ba))}}
% 	\!\!\!\!\!\!\!\!\!\!\!\!\!\!\!\!\!\alpha q^{-nK}\left.\vphantom{\sum_{\bm'\neq\bone}}\right)^\rho \\
% &&\cdot \Pr(R_\bC > R+ \epsilon) \\
& \leq   \sum_{\by,\ba} q^{-nK}P_{\bY|\bX}(\by|\delta(\ba)) \\
& \hphantom{====} \cdot\left(\sum_{\bm'=\btwo}^{q^{nRK}}\right.
	\sum_{\substack{\ba'\in\cQ:\cT^n_\cQ(\ba'-\ba)\in \Tau^c \\
	P_{\bY|\bX}(\by|\delta(\ba'))\geq P_{\bY|\bX}(\by|\delta(\ba))}}
	\!\!\!\!\!\!\!\!\!\!\!\!\!\!\!\!\!\alpha_{\rm \scalebox{0.4}{MAC}} q^{-nK}\left.\vphantom{\sum_{\bm'\neq\bone}}\right)^\rho \\
%&& \cdot 1 +  q^{-n\epsilon /2}  \cdot 1\\
& \leq   \alpha_{\rm \scalebox{0.4}{MAC}}^\rho \sum_{\by,\ba}q^{-nK}  P_{\bY|\bX}(\by|\delta(\ba)) \\
&\hphantom{====} 	\cdot \left((q^{nRK}-1)\sum_{\ba':P_{\bY|\bX}(\by|\delta(\ba'))\geq  P_{\bY|\bX}(\by|\delta(\ba))}
	\!\!\!\!\!\!\!q^{-nK}\right)^\rho 
\\ 
& \leq   \alpha_{\rm \scalebox{0.4}{MAC}}^\rho q^{nR K\rho}
	\sum_{\bx,\by} P_{\bY|\bX}(\by|\bx)\sum_{\ba:\delta(\ba)=\bx}q^{-nK} \\
&\hphantom{====}	\cdot \left(\sum_{\bx': P_{\bY|\bX}(\by|\bx')\geq  P_{\bY|\bX}(\by|\bx)}\sum_{\ba':\delta(\ba')=\bx'}q^{-nK}\right)^\rho \\
& =  \alpha_{\rm \scalebox{0.4}{MAC}}^\rho q^{nR K\rho}\sum_{\bx,\by} P_{\bY|\bX}(\by|\bx) P_\bX(\bx)  \\
&\hphantom{====}
	\cdot \left(\sum_{\bx': P_{\bY|\bX}(\by|\bx')\geq  P_{\bY|\bX}(\by|\bx)}P_\bX(\bx')\right)^\rho \\
& \stackrel{(k)}{\leq}   \alpha_{\rm \scalebox{0.4}{MAC}}^\rho q^{nR K\rho}\sum_{\bx,\by} P_{\bY|\bX}(\by|\bx) P_\bX(\bx) \\
&\hphantom{===}
	\cdot \left(\sum_{\bx'}P_\bX(\bx')\left(\frac{P_{\bY|\bX}(\by|\bx')}{P_{\bY|\bX}(\by|\bx)}\right)^s\right)^\rho \\
& =   \alpha_{\rm \scalebox{0.4}{MAC}}^\rho q^{nRK\rho}\sum_\by\left(\sum_\bx P_{\bX}(\bx)P_{\bY|\bX}(\by|\bx)^{1-s\rho}\right) \\
&\hphantom{===}	\cdot \left(\sum_{\bx'}P_\bX(\bx')P_{\bY|\bX}(\by|\bx')^s\right)^\rho ,
\end{align*}
%where $(k)$ uses Theorem~\ref{thm:full_R} and bounded nature of probability, $(l)$ follows by setting $\epsilon = 2\log n /n$, and $(m)$ holds for any $s>0$.  
where  $(k)$ holds for any $s>0$.

When $s=1/(1+\rho)$, 
\begin{align}
\lefteqn{E[\Pr[\bcY^{\Tau^c}_{\bM,\bD,\bV}|\bM,\bD,\bV]]} \nonumber \\
&\leq  \alpha_{\rm \scalebox{0.4}{MAC}}^\rho q^{nR K\rho}
\sum_\by\left(\sum_\bx P_\bX(\bx)P_{\bY|\bX}(\by|\bx)^{1/(1+\rho)}\right)^{1+\rho}. \label{eqn:Pe_app_res}
\end{align}
Rewriting \eqref{eqn:Pe_app_res} in an exponential form using Gallager's error exponent gives the desired result .  
\hfill$\blacksquare$

% \section{Proof of Lemma~\ref{lem:Rcs}} \label{app:Rcs}
% In the code spectrum of a $K$-transmitter MAC $\GF(q)$ random code ensemble with rate  $\bR=(R,\ldots,R)$, with $R=1-\frac{r}{n}$, the average number of codematrices of type $n\bthe$ is 
% \begin{eqnarray*}
% \obS_r^n(n\bthe) = \frac{q^{(n-r)K}}{q^{nK}}B(n,n\bthe).
% \end{eqnarray*}
% Applying the definition of normalized ensemble spectrum gives
% \begin{eqnarray*}
% \oS_U(\bthe) &\defeq& \lim_{n \rightarrow \infty} \frac{1}{n}\log \obS_r^n(n\bthe) \\
% &\stackrel{(a)}{=}& \lim_{n \rightarrow \infty} \frac{1}{n} \log \left(q^{-rK} \right) +H(\bthe) \\
% & =&H(\bthe) - K\frac{r}{n} \\
% &=& H(\bthe) - K(1-R),
% \end{eqnarray*}
% where $(a)$ follows from Stirling's approximation on $B(n,n\bthe)$. Note that $H(\bthe)$ is taken with $\log$ base $q$.

% \section{Proof of Theorem~\ref{thm:BML}}\label{app:BML}

% R

\section{Tools used to bound $\log \alpha_{\rm {MAC} }/n$ in Theorem~\ref{thm:ePe}} \label{app:log_a}
% \subsection{Outline}
% The following definitions are useful in the discussion of $\frac{\log \alpha}{n}$.
To bound the rate offset $\frac{\log \alpha_{\rm \scalebox{0.4}{MAC}}}{n}$, we first seek to understand how the normalized ensemble spectra (see Definition~\ref{def:spectrum}) for the MAC under the uniform random ensemble and the random LDPC  code ensemble, here denoted by $\oS_U(\bthe)$ and $\oS_L(\bthe)$, respectively, differ. Lemma~\ref{lem:Rcs} first evaluates $\oS_U(\bthe)$. Theorem~\ref{thm:BML} then evaluates $\oS_L(\bthe)$. Theorem~\ref{thm:BRthe} relates $\oS_L(\bthe)$ to $\oS_U(\bthe)$ for a restricted family of pmfs $\bthe$, corresponding to codes in which the minimal distance is sufficiently large . Lemma~\ref{lem:dMin} then paves the way for expurgation to remove codes with small minimal distance by showing that the probability of all codes with small minimum distance approaches zero as $n$ grows without bound under the proposed LDPC  code ensemble. 

% Finally, Theorem~\ref{thm:ePe} bounds $\frac{\log \alpha}{n}$ using the results from Theorem~\ref{thm:BML} and Lemma~\ref{lem:dMin}.
\begin{defin} (Normalized ensemble spectrum) \label{def:spectrum} Consider any ensemble of codes with ensemble-average spectrum $\obS^n=(\oS^n(\bt):\bt \in \cT_\cQ^n)$. Given any rational pmf $\bthe=(\theta(g):g\in\cQ)$, let $\{n_i\}$ be a series of all indices $j$ such that $j\bthe \in \cT_\cQ^j$, the asymptotic exponent for $\bthe$ is defined by 
% Let $\bthe = (\theta(g):g\in \cQ)$ be a rational pmf for which $n\bthe$ is a type $\bt \in \cT_\cQ^n$, the normalized ensemble spectrum for $\bthe$ is defined by
\begin{align}
    \oS(\bthe)= \lim_{i\rightarrow\infty} \frac{1}{n_i}\log \oS^{n_i}(n_i\bthe),
\end{align}
and the normalized ensemble spectrum for this ensemble is the collection of all asymptotic exponents $\oS^n(\bthe)$ for all pmfs $\bthe$.  
\end{defin}
\begin{remark}
For notational simplicity, we omit the index $i$ to write $\oS^{n_i}(n_i\bthe)$ as $\oS^{n}(n\bthe)$ with the implicit assumption that $n\bthe \in \cT_\cQ^n$.
\end{remark}

We consider the normalized ensemble spectrum for two ensembles, each with the same fixed rate $R = 1 - \frac{\lambda}{\rho}$ $q$-ary symbols per channel use for each transmitter.
\begin{enumerate}
    \item The first ensemble is an ensemble of uniform random $\GF(q)$ $K$-transmitter MAC codes, where each transmitter employs a distinct blocklength-$n$ codebook with $q^{nR}$ codewords, $R = 1 - \frac{\lambda}{\rho}$, chosen uniformly at random from $\GF(q)^{n}$. We denote the normalized ensemble spectrum for this (uniform) random ensemble by
    \begin{align}
        \oS_U(\bthe) &\defeq \lim_{n \rightarrow \infty} \frac{1}{n}\log \oS_U^n(n\bthe),
    \end{align}
    where $\obS_U$ ($U$ stands for uniform) represents the ensemble-average spectrum under the $K$-MAC with independent codewords distributed uniformly on $\GF(q)^n$.
    \item The second ensemble is the ${\rm LDPC}_K(\lambda,\rho;n)$ ensemble. This is an ensemble of $K$-transmitter MAC codes for which all transmitters employ the same random codebook from the ${\rm LDPC}(\lambda,\rho;n)$ ensemble. We denote the normalized ensemble spectrum for this LDPC  code ensemble by
    \begin{equation}\label{eqn:bthe_1}
        \oS_L(\bthe)\defeq\lim_{n\rightarrow\infty}\frac1n\log\oS_L^n(n\bthe),
    \end{equation}
    where $\obS_L$ ($L$ stands for LDPC) represents the ensemble-average spectrum under the ${\rm LDPC}_K(\lambda,\rho;n)$ ensemble.
\end{enumerate}

% Given a rational pmf $\bthe=(\theta(g):g\in\cQ)$, 
% we define the {\bf normalized ensemble spectrum} for the ${\rm LDPC}(\lambda,\rho,\delta;n)_q$ ensemble of type $n\bthe$ as 

% where $\obS^n(n\bthe)$ is the ensemble-average number of codematrices with type $n\bthe$ that we would observe under our random LDPC design.

% Similarly, we define $\obS_r^n(n\bthe)$ as the ensemble-average number of codematrices under the (uniform) $\GF(q)$ random code ensemble of a $K$-transmitter MAC, with symmetrical rate $R = 1 -\frac{r}{n}$, the normalized spectrum for this ensemble of type $n\bthe$ is
% \begin{align}
%     \oS_U(\bthe) &\defeq \lim_{n \rightarrow \infty} \frac{1}{n}\log \obS_r^n(n\bthe).
% \end{align}

% The following results evaluate $\oS_L(\bthe)$ and $\oS_U(\bthe)$.

% Roughly, $\oS^n(n\bthe)$ is the expected number of codematrices  in a random codebook 
% with ``type'' $n\bthe$; 
% taking the logarithm of that value and dividing by the number of channel uses $n$ 
% converts a number of codewords into something like a rate. 

 \subsection{Normalized Ensemble Spectrum for Uniform Random MAC Ensemble} \label{sec:spectrum_U}
We begin by evaluating $ \oS_U(\bthe)$.

\begin{lem} \label{lem:Rcs} %random coding spectrum
The normalized ensemble spectrum of the $K$-transmitter MAC uniform random ensemble is given by
\begin{equation*}
    \oS_U(\bthe) = H(\bthe) - K(1-R),
\end{equation*}
where \[
H(\bthe)=-\sum_{g\in \cQ}\theta(g)\log \theta(g)
\]
is the entropy of the pmf $\bthe$ in $q$-ary digits.
\end{lem}

The proof of Lemma~\ref{lem:Rcs} is based on the discussion of binary codes in \cite[Th. 1]{Shulman:99}. 

{\em Proof:}  
% Consider the (uniform) $\GF(q)$ random code ensemble of a $K$-transmitter MAC, where each codeword symbol is chosen uniformly and independently for each transmitter. If the code's rate vector is $\bR=(R,\ldots,R)$ with $R=1-\frac{r}{n}$, then the ensemble-average number of codematrices of type $n\bthe$, denoted by $\obS_r^n(n\bthe)$, is 
When each codeword is chosen uniformly at random from $\GF(q)^{n}$, the ensemble-average number $\oS_U^n(n\bthe)$ of codematrices of type $n\bthe$ is 
% \begin{eqnarray*}
% \oS_U^n(n\bthe) = \frac{q^{nRK}}{q^{nK}}B(n,n\bthe).
% \end{eqnarray*}
\begin{align*}
    \oS_U^n(n\bthe) &= E_U[S_\bD^n(n\bthe)] \\
    &= \sum_\bm E_U\left[\ind\{ \cT_\cQ^n(\bD_\bm) = n\bthe\}\right] \\
    &= q^{nRK}{\rm Pr}_U[\cT_\cQ^n(\bD_\bone)=n\bthe] \\
    &= q^{nRK}\frac{B(n,n\bthe)}{q^{nK}}.
\end{align*}
Applying the definition of the normalized ensemble spectrum gives
\begin{eqnarray*}
\oS_U(\bthe) &\defeq& \lim_{n \rightarrow \infty} \frac{1}{n}\log \oS_U^n(n\bthe) \\
&\stackrel{(a)}{=}& \lim_{n \rightarrow \infty} \frac{1}{n} \log \left(q^{-n(1-R)K} \right) +H(\bthe) \\
&=& H(\bthe) - K(1-R),
\end{eqnarray*}
where $(a)$ follows from applying Stirling's upper and lower bounds on the factorial to the multinomial coefficient $B(n,n\bthe)$. Note that the definition of $H(\bthe)$ employs the base-$q$ logarithm. 
\hfill$\blacksquare$

\subsection{Normalized Ensemble Spectrum for ${\rm LDPC}_K(\lambda,\rho;n)$ Ensemble} \label{sec:spectrum_L}
Before moving on to the evaluating of $\oS_L(\bthe)$, recall that for any type $\bt\in\cT^\rho_\cQ$, $B(\rho,\bt)$ is the number of type-$\bt$ $\rho\times K$ matrices. For any type-$\bt$ matrix $G^T=[g_1^T,g_2^T,\ldots,g_\rho^T]$, $g_i\in\GF(q)^K$, let $G_\bt$ be the corresponding matrix transpose, then 
\[
\cN_\bt = \left|\left\{\be\in\{\GF(q)\setminus \{0\}\}^\rho:G_\bt\be=\bzero\right\}\right|
\]
is the number of vectors $\be\in\{\GF(q)\setminus \{0\}\}^\rho$ in the nullspace of $G_\bt$. Notice that $\cN_\bt$ is constant across all matrices $G_\bt$ with type $\bt$.  
Theorem~\ref{thm:BML} employs this definition of $\cN_\bt$ as well as the following notation.
Given $x\in \mathbb{R}$, 
\begin{eqnarray*}
\sgn(x) & \defeq & \left\{\begin{array}{cl}
	1 & \mbox{if }x>0 \\
	0 & \mbox{if }x=0 \\
	-1 & \mbox{if }x<0.
	\end{array}\right.
\end{eqnarray*}
% For any type $\bt\in\cT^\rho_\cQ$, $B(\rho,\bt)$ is the number of type $\bt$ matrices $[g_1;g_2;\ldots;g_d]$ with $K$-dimensional rows  $g_i\in\GF(q)^K$. % For any type-$\bt$ string $(g_1,\ldots,g_\rho)\in\cQ^\rho$, 
% Let $G_\bt=[g_1,\ldots,g_\rho]$ be a matrix 
% with $K$-dimensional columns,
% then 
% \[
% \cN_\bt = \left|\left\{\be\in[q-1]^\rho:G_\bt\be=\bzero\right\}\right|
% \]
% is the number of vectors $\be\in[q-1]^\rho$ 
% in the nullspace of $G_\bt$.  
Note that the calculation of $\obS_L^n$ in Theorem~\ref{thm:BML} is for the ${\rm LDPC}_K(\lambda,\rho;n)$ ensemble before codeword removal. The true spectrum $\obS_L^n$ is smaller, and $\alpha_{\rm \scalebox{0.4}{MAC}}$ in \eqref{eqn:alpha} is a valid upper bound for the ${\rm LDPC}_K(\lambda,\rho;n)$ ensemble (with codeword removal). %Appendix~\ref{sec:spectrum_L}
\begin{thm}\label{thm:BML}
% The asymptotic normalized spectrum 
% of a $(\lambda,\rho)$-regular LDPC code over $\GF(q)$ for a $K$-transmitter MAC is 
The normalized ensemble spectrum of the ${\rm LDPC}_K(\lambda,\rho;n)$ ensemble is given by 
% of the $K$-transmitter MAC LDPC ensemble, for which all transmitters employ the same random codebook from ${\rm LDPC}(\lambda,\rho;n)$ with independent random coset vector and same quantizer, is given by  
\begin{equation}
\oS_L(\bthe) 
= (1-\lambda)H(\bthe)-\lambda\log(q-1) 
+\frac{\lambda}{\rho}\log\inf_{\substack{\bx:\sgn(\bx)\\=\sgn(\bthe)}}
\frac{A(\bx)}{\bx^{\rho\bthe}}, \label{eqn:oSL}
\end{equation}
where for any pmf $\bx=(x_g: g\in\cQ)$ on $\cQ$, 
\begin{eqnarray*}
\bx^{\rho\bthe} & = & \prod_{g\in\cQ}x_g^{d\theta(g)} \\
A(\bx) & = & \sum_{\bt\in\cT^\rho_\cQ}\cN_\bt B(\rho,\bt)\bx^\bt.
\end{eqnarray*}
\end{thm}

{\em Proof:} Recall that when $n\bthe$ is a type, 
$\oS_L(\bthe)$ is the expected number of codematrices of type $n\bthe$ 
in a randomly drawn MAC codebook $\bD$, corresponding to underlying single-transmitter $(\lambda,\rho;n)$ LDPC code $\bC$.  
Recall further that when $\bD$ 
is the codebook of an LDPC MAC in $\GF(q)$, 
then $\obS((n\theta_g:g\in\cQ))=\obS((n\theta_{\pi(g)}:g\in\cQ))$ 
for any permutation $\pi$ on $[|\cQ|]$.
% since the code design is symmetric across variable nodes.  

Let $\bcM(n\bthe)$ denote all possible codematrices of type $n\bthe$. Then 
\begin{eqnarray*}
\oS_L(n\bthe) 
& = & E_L\left[\sum_{\bd_0\in\bcM(n\bthe)}\ind\{\bd_0\in\bD\}\right] \\
& = & \sum_{\bd_0\in\bcM(n\bthe)}E_L[\ind\{\bd_0\in\bD\}] \\
& = & \sum_{\bd_0\in\bcM(n\bthe)}\Pr[\bd_0\in\bD] \\
& = & B(n,n\bthe)\Pr[\bd_{n\bthe}\in\bD],
\end{eqnarray*} 
%=n!/(\prod_g(n\theta_g)!)$ (the multinomial coefficient)
where $B(n,n\bthe)$ is the size of $\bcM(n\bthe)$, $\bd_{n\bthe}$ is any fixed codematrix in $\bcM(n\bthe)$,
and the final equality follows from the symmetry of the code design. 
By the definition of $\oS_L(\bthe)$ in (\ref{eqn:bthe_1}) 
and Stirling's upper and lower bounds on the factorial,
\begin{equation}\label{eqn:Bthe_2}
\oS_L(\bthe) = H(\bthe)+\lim_{n\rightarrow\infty}\frac1n\log\Pr[\bd_{n\bthe}\in\bD].
\end{equation}

\begin{remark}
In the preceding characterization of $\obS_L(n\bthe)$, $\Pr[\bd_{n\bthe}\in\bD]$ refers to the probability that $\bd_{n\bthe}$ is in the codebook of a randomly drawn code $\bD$ from the ${\rm LDPC}_K(\lambda,\rho;n)$ ensemble. The calculation below evaluates this quantity by assuming that $\bD$ is from the ${\rm LDPC}_K(\lambda,\rho;n)$ ensemble without codeword removal, i.e., from the ${\rm LDPC}_K({\rm Full}, \lambda,\rho;n)$ ensemble. The true spectrum is smaller. Therefore, the resulting spectrum in \eqref{eqn:oSL} is a valid upper bound for ${\rm LDPC}_K(\lambda,\rho;n)$ ensemble (with codeword removal), 
\end{remark}

To find $\Pr[\bd_{n\bthe}\in\bD]$,  
note that the random choice of edge connections and labels 
associates with each check node socket 
a socket value equal to the product of 
the edge value and 
the variable node value.
There are $B(n\lambda,n\lambda\bthe)$
equally likely assignments 
of variable node values to sockets 
that are consistent with PDF $\bthe$.  
Combining this with the 
$q-1$ possible labels for each edge, 
we find that there are 
\[
t(\bthe,n)=B(n\lambda,n\lambda\bthe)(q-1)^{n\lambda}
\]
equally likely outcomes 
for the choice of connections and edge values 
under a fixed codematrix $\bd_{n\bthe}$.  
It is useful to note that some of these pairs 
yield the same socket values; 
for example, when a variable node holds value $\bzero$, 
the socket value is identical for all $q-1$ values of the edge.  
Since our probability calculation relies on a counting argument, 
the above value counts separately all events that yield the same output.  
This is different from the prior work \cite[Eq. (49)]{BennatanB:04}, 
which counts the number of distinct outcomes 
rather than the number of distinct events leading to these outcomes in its probability calculation.

For $\bd_{n\bthe}$ to be a codematrix, 
summing the $\rho$ (randomly chosen) socket values 
at each of the $n\lambda/\rho$ check nodes 
must give the value $\bzero\in\cQ$.  
The following strategy and notation from~\cite[Sect. III.B]{BurshteinM:04} 
are useful in calculating the number of assignments 
that yield this outcome.
First, for each fixed vector of edge values $\be=[e_1,\ldots,e_\rho]\in[q-1]^\rho$, 
we work to build a multinomial $f(\bx)$ in $\bx=(x_g:g\in\cQ)$ 
such that for any type $\bt=(t(g):g\in\cQ)\in\cT_\cQ^{n\lambda}$, 
the coefficient of the term $\bx^\bt=\prod_{g\in\cQ}x_g^{t(g)}$ 
equals the number of socket assignment and edge value pairs 
for which the socket assignment carries variable node values of type $\bt$, 
and the socket values satisfy all check node constraints.\footnote{The 
type is with respect to vectors of length $n\lambda$ 
since each of $n$ variable nodes is employed in $\lambda$ sockets, 
giving a total of $n\lambda$ socket values.} 
Then, using notation $\lfloor f(\bx)\rfloor_{\bt}$ 
to designate a function that maps multinomial $f(\bx)$ 
to the coefficient of element $\bx^\bt$, 
we extract the number of socket and edge value assignments 
that are consistent with the fixed codematrix $\bd_{n\bthe}$ 
and satisfy all constraint nodes; 
this is the number of randomly designed codes 
for which $\bd_{n\bthe}$ is a codematrix.  

To begin, consider a single check node.  
Let $g_1,\ldots,g_\rho$ denote the values 
at the $\rho$ variable nodes connected to that check node, 
and let $e_1,\ldots,e_\rho$ be the corresponding edge values.  
% Let $[g_1,\ldots,g_\rho]$ and $\be=[e_1,\ldots,e_\rho]$.  
We seek to build a multinomial $A(\bx)$ 
in which the coefficient of each term $\bx^\bt$ 
is the number of distinct edge value and socket assignments 
for which the variable-node inputs have type $\bt$ 
and the check node is satisfied.
That is, 
\begin{eqnarray*}
\lefteqn{A(\bx)} \\
& = & \sum_{g_1,\ldots,g_\rho\in\cQ}\sum_{e_1,\ldots,e_\rho\in[q-1]}
	\ind\left\{\sum_{i=1}^\rho e_ig_i=\bzero\right\}\left(\prod_{i=1}^\rho x_{g_i}\right) \\
& = & \sum_{e_1,\ldots,e_\rho\in[q-1]}\sum_{\hg_1,\ldots,\hg_\rho\in\cQ}
 	\ind\left\{\sum_{i=1}^\rho\hg_i=\bzero\right\}\left(\prod_{i=1}^\rho x_{\hg_i/e_i}\right).
\end{eqnarray*}
Note that the above expression implements the multiplication $e_ig_i$ by viewing $g_i$ as length-$K$ vector over $\GF(q)$, and similarly for the division $\hg_i/e_i$.
Recall that $q$ is a prime power, say $q=p^m$, and that $\cQ=\GF(q)^K$.  
We can therefore view each element $g\in\cQ$ 
as a corresponding vector $\bh\in\{0,\ldots,p-1\}^{mK}$ 
and implement addition in $\cQ$ as component-wise addition modulo-$p$.  
Thus, following the argument of~\cite[Theorem 8]{BennatanB:04},
for each fixed value of $(e_1,\ldots,e_\rho)$, 
the given sum equals a $\rho$-fold, $Km$-dimensional 
cyclic convolution evaluated at $\bzero$, 
giving 
\begin{eqnarray}
\lefteqn{A(\bx)} \nonumber \\
& = & \sum_{\be_1,\ldots,\be_\rho\in \GF(p)^m\setminus\{\bzero\}}
	  \sum_{\substack{\bh_1,\ldots,\bh_\rho\in\GF(p)^{mK} \\ \sum_{i=1}^\rho\bh_i=\bzero}}
	\left(\prod_{i=1}^\rho x_{\frac{\bh_i}{\be_i}}\right) \nonumber \\
& = & \sum_{\be_1,\ldots,\be_\rho\in\GF(p)^m\setminus\{\bzero\}}
	\left[x\left[\frac{\bh}{\be_1}\right]*\cdots*x\left[\frac{\bh}{\be_\rho}\right]\right]_{\bh=\bzero} \nonumber \\
& = & \sum_{\substack{\be_1,\ldots,\be_\rho\in\\ \GF(p)^m\setminus\{\bzero\}}}
	\left[{\rm IDFT}\left[
	\prod_{j=1}^\rho{\rm DFT}\left[x\left[\frac{\bh}{\be_j}\right]\right]\right]\right]_{\bh=\bzero} \nonumber \\
& = & \sum_{\substack{\be_1,\ldots,\be_\rho\in\\ \GF(p)^m\setminus\{\bzero\}}}
	\frac1{q^K}\sum_{\bk\in\GF(p)^{mK}} \left[\prod_{i=1}^\rho \right. \nonumber \\
&& 	\left. \left(\sum_{\bh\in\GF(p)^{mK}}e^{-j\frac{2\pi}{p}\sum_{\ell=0}^{mK}k_\ell h_\ell}
	x\left[\frac{\bh}{\be_i}\right]\right)
	\right], \label{eqn:A_x} 
\end{eqnarray}
where for any $\bh\in\{0,\ldots,p-1\}^{mK}$, $x[\bh]$ equals $x_g$ for the corresponding $g\in\cQ$.

Combining $n\lambda/\rho$ such multinomials,
corresponding to our $n\lambda/\rho$ check nodes,
gives multinomial $(A(\bx))^{n\lambda/\rho}$.
The coefficient of the term $\bx^{n\lambda\bthe}$
in this multinomial describes the number of edge and socket assignments
for which $\bd_{n\theta}$ is a codematrix.
We denote this number by
\[
e(\bthe,n)=\lfloor((A(\bx))^{n\lambda/\rho})\rfloor_{n\lambda\bthe}.
\]
Applying this definition, we have
\begin{eqnarray}
\lefteqn{\lim_{n\rightarrow\infty}
\frac1n\log\Pr[\bd_{n\bthe}\in\bD]} \nonumber \\
& = & \lim_{n\rightarrow\infty}\frac1n\log\frac{e(\bthe,n)}{t(\bthe,n)}  \nonumber\\
& = & \lim_{n\rightarrow\infty}\frac1n\log\frac{\lfloor((A(\bx))^{n\lambda/\rho})\rfloor_{n\lambda\bthe}}{B(n\lambda,n\lambda\bthe)(q-1)^{n\lambda}} \label{eqn:d_nbthe}\\
& = & -\lambda H(\bthe)-\lambda\log(q-1) \nonumber  \nonumber\\
&& +\frac{\lambda}{\rho}\lim_{\frac{\lambda}{\rho}n\rightarrow\infty}\frac1{\frac{\lambda}{\rho}n}
        \log\left\lfloor (A(\bx))^{\frac{\lambda}{\rho}n}\right\rfloor_{(\frac{\lambda}{\rho}n)\rho\bthe } \nonumber \\
& = & -\lambda H(\bthe)-\lambda\log(q-1) \nonumber \\
&& +\frac{\lambda}{\rho}\lim_{n\rightarrow\infty}\frac1n
        \log\left\lfloor (A(\bx))^n\right\rfloor_{n\rho\bthe } \nonumber \\
& \stackrel{(a)}{=} & -\lambda H(\bthe)-\lambda \log(q-1) +\frac{\lambda}{\rho}\log\inf_{\substack{\bx:\sgn(\bx)\\=\sgn(\bthe)}}\frac{A(\bx)}{\prod_{g}x_g^{\rho\theta_g}}, \nonumber
\end{eqnarray}
where $(a)$ follows from the definition of $\left\lfloor A(\bx)\right\rfloor_{\bt }$ as the coefficient of element $\bx^\bt$ in multinomial $A(\bx)$ and from the second equation in~\cite[Theorem~10]{BennatanB:04} (included below for reference), which gives an expression for evaluating the limit of multinomial coefficient exponent $\frac1n
        \log\left\lfloor (A(\bx))^n\right\rfloor_{n\rho\bthe }$.
Combining the given limit with (\ref{eqn:Bthe_2}) gives
\[
\oS_L(\bthe)
= (1-\lambda)H(\bthe)-\lambda\log(q-1)
+\frac{\lambda}{\rho}\log\inf_{\substack{\bx:\sgn(\bx)\\=\sgn(\bthe)}}
\frac{A(\bx)}{\bx^{\rho\bthe}},
\]
which is the desired result.
% For the purpose of completeness,~\cite[Theorem~10]{BennatanB:04} is restated below.
\hfill$\blacksquare$
\begin{lem} (\cite[Th.~10]{BennatanB:04}). \label{lem:MN} %multinomial bound
Let $\gamma >0$ be some rational number and $p(x,y)^\gamma$ be a multinomial with non-negative coefficient. Let $\alpha >0$ and $\beta >0$ be rational numbers, and $\{n_i\}$ be a series of all indices $j$ such that $j/\gamma \in \mathbb{Z}, \lfloor p(x,y)^j\rfloor_{\alpha j,\beta j}\neq 0$, then
\begin{equation}
    \lfloor p(x,y)^{n_i}\rfloor_{\alpha n_i,\beta n_i} \leq \inf_{x>0,y>0} \frac{p(x,y)^{n_i}}{x^{\alpha n_i}y^{\beta n_i}} \label{eqn:MN_1}
\end{equation}
and
\begin{equation}
    \lim_{i\rightarrow \infty} \frac{1}{n_i}\log \lfloor p(x,y)^{n_i}\rfloor_{\alpha n_i,\beta n_i} = \log \inf_{x>0,y>0} \frac{p(x,y)
    }{x^{\alpha }y^{\beta}}. \label{eqn:MN_2}
\end{equation}
\end{lem}

\subsection{Relationship between $\oS_U(\bthe)$ and $\oS_L(\bthe)$} \label{sec:BRthe} 
Rather than comparing $\oS_U(\bthe)$ and $\oS_L(\bthe)$ for all possible values of $\bthe$, Theorem~\ref{thm:BRthe}, below, makes this comparison only for pmfs $\bthe$ that lie in a restricted family of pmfs $J_\sigma$ on alphabet $\cQ$. We begin by defining this family. For any $\sigma \in (0,1)$, $J_\sigma$ eliminates all pmfs with $\theta(\bzero)$ above $1-\sigma$; precisely,
\begin{align}
J_\sigma & = \left\{ (\bthe = (\theta(g):g\in \cQ):  \sum_{g \in \cQ} \theta(g)=1, \right.\nonumber \\
& \hphantom{=\{}\left. \vphantom{\sum_{g \in \cQ}} 0\leq \theta(\bzero)\leq 1-\sigma,  \theta(g) \geq 0 ~\forall g \in \cQ \setminus \{\bzero\}\right\}, \label{eqn:J_sigma}
\end{align}
where $\bzero$ is the all zero vector.

By an argument similar to that used for LDPC codes on the PPC in \cite{BennatanB:04}, Theorem~\ref{thm:BRthe} shows uniform convergence of $\oS_L(\bthe)$ to $\oS_U(\bthe)$ for the subset of values of $\bthe \in J_\sigma$. 
\begin{thm}\label{thm:BRthe}
For any positive rational number $R<1$, any $\sigma \in (0,1)$, and any $\epsilon>0$, there exists a constant $\rho_0>0$ such that for all $\bthe \in J_\sigma$, and all $\lambda,\rho$ for which $R = 1- \frac{\lambda}{\rho}$ and $\rho >\rho_0$,
\begin{align}
    \oS_L(\bthe) < \oS_U(\bthe) + \epsilon.
\end{align}

% Fix a rational positive number $R<1$ and let $\oS_U(\bthe)$ be the MAC random coding normalized spectrum corresponding to the rate $R$, then for any $\epsilon >0$, there exists $d_0>0$  such that
% \begin{eqnarray*}
% \oS_L(\bthe) < \oS_U(\bthe) + \epsilon,
% \end{eqnarray*}
% for all $\bthe \in J_\sigma$, and all $d >d_0$ with $R = 1- \frac{\lambda}{\rho}$.
\end{thm}
 
{\em Proof:}  To prove $\oS_L(\bthe)<\oS_U(\bthe)+\epsilon$ for large enough $\rho$ and $\bthe \in J_\sigma$, we derive an upper bound on the limit of $\oS_L(\bthe)$ as $\rho$ approaches $n$ and show that the upper bound equals $\oS_U(\bthe)$. 

We start with the expression of $\oS_L(\bthe)$ from Theorem~\ref{thm:BML}, 
\begin{align}
\oS_L(\bthe) 
& =   (1-\lambda)H(\bthe)-\lambda\log(q-1) 
	+\frac{\lambda}{\rho}\log\inf_{\substack{\bx:\sgn(\bx)\\=\sgn(\bthe)}}
	\frac{A(\bx)}{\bx^{\rho\bthe}} \nonumber \\
& \leq   (1-\lambda)H(\bthe)-\lambda\log(q-1) 
	+\frac{\lambda}{\rho}\log\frac{A(\bthe)}{\bthe^{\rho\bthe}} \nonumber \\
& =   H(\bthe)-\lambda\log(q-1)+\frac{\lambda}{\rho}\log A(\bthe). \label{eqn:bthe_3}
\end{align}
We next focus on $A(\bthe)$ 
to bound the last term in this equation. 

Given our equation for $A(\bx)$ from \eqref{eqn:A_x}, it follows that
% for $\bx \in J_\sigma$
% \begin{align}
% A(\bx) \nonumber
% & =  \sum_{\substack{\be_1,\ldots,\be_\rho\in  \nonumber\\
% 	\GF(p)^m\setminus\{\bzero\}}}\frac1{q^K}\sum_{\bk\in\GF(p)^{mK}}  \nonumber\\
% &	\prod_{i=1}^\rho \left(\sum_{\bh\in\GF(p)^{mK}}
% 	e^{-j\frac{2\pi}{p}\sum_{\ell=0}^{mK}k_\ell 
% 	h_\ell}x\left[\frac\bh{\be_i}\right]\right) \nonumber \\
% & \stackrel{(a)}{=}  \frac1{q^K}\sum_{\bk\in\GF(p)^{mK}} 
% 	\left(\sum_{\be\in\GF(p)^m\setminus\{\bzero\}} \right.  \nonumber\\
% &	\left.\left(\sum_{\bh\in\GF(p)^{mK}}
% 	e^{-j\frac{2\pi}{p}\sum_{\ell=0}^{mK}k_\ell h_\ell}
% 	x\left[\frac\bh{\be}\right]\right)\right)^\rho  \nonumber\\
% & \stackrel{(b)}{=}  \frac1{q^K}\sum_{\bk\in\GF(p)^{mK}} 
% 	\left(\sum_{\bh\in\GF(p)^{mK}}
% 	e^{-j\frac{2\pi}{p}\sum_{\ell=0}^{mK}k_\ell h_\ell}\right.  \nonumber\\
% &	\left.\cdot\left(\sum_{\be\in\GF(p)^m\setminus\{\bzero\}}
% 	x\left[\frac\bh{\be}\right]\right)\right)^\rho  \nonumber \\
% & \stackrel{(c)}{=}  \frac1{q^K}\sum_{\bk\in\GF(p)^{mK}}\left(x[\bzero](q-1)
% 	\vphantom{\sum_{\bh\in\GF(p)^{mK}}}\right.  \nonumber\\
% &	\left.+(1-x[\bzero])\sum_{\substack{\bh\in\\\GF(p)^{mK}\setminus\{\bzero\}}}
% 	e^{-j\frac{2\pi}{p}\sum_{\ell=0}^{mK}k_\ell h_\ell}\right)^\rho   \nonumber \\
% & \stackrel{(d)}{=}  \frac{(q-1)^\rho}{q^K}\left(1+\sum_{\bk\neq\bzero}\left(x[\bzero]
% 	\vphantom{\frac1{(q-1)^\rho}}	+(1-x[\bzero])\right.\right.   \nonumber\\
% &	\left.\left.\hphantom{\frac{(q-1)^\rho}{q^K}}
% 	\cdot \frac{\sum_{\bh\neq\bzero}e^{-j\frac{2\pi}{p}\sum_{\ell=0}^{mK}k_\ell h_\ell}}{q-1}
% 	\right)^\rho\right), \label{eqn:Ax_BR_proof}
% \end{align}
\begin{align}
A(\bx) \nonumber
& =  \sum_{\substack{\be_1,\ldots,\be_\rho\in  \nonumber\\
	\GF(p)^m\setminus\{\bzero\}}}\frac1{q^K}\sum_{\bk\in\GF(p)^{mK}}  \nonumber\\
&\hphantom{=}	\prod_{i=1}^\rho \left(\sum_{\bh\in\GF(p)^{mK}}
	e^{-j\frac{2\pi}{p}\sum_{\ell=0}^{mK}k_\ell 
	h_\ell}x\left[\frac\bh{\be_i}\right]\right) \nonumber \\
&  \stackrel{(a)}{=}  \frac{(q-1)^\rho}{q^K} + \frac1{q^K}\sum_{\bk\in\GF(p)^{mK}\setminus\{\bzero\}} 
	\left(\sum_{\be\in\GF(p)^m\setminus\{\bzero\}} \right.  \nonumber\\
&\hphantom{==}	\left.\left(\sum_{\bh\in\GF(p)^{mK}}
	e^{-j\frac{2\pi}{p}\sum_{\ell=0}^{mK}k_\ell h_\ell}
	x\left[\frac\bh{\be}\right]\right)\right)^\rho  \nonumber\\
& \stackrel{(b)}{=} \frac{(q-1)^\rho}{q^K} + \frac1{q^K}\sum_{\bk\in\GF(p)^{mK}\setminus\{\bzero\}} 
	\left(\sum_{\bh\in\GF(p)^{mK}}
	\right.  \nonumber\\
&\hphantom{==}	\left.\cdot e^{-j\frac{2\pi}{p}\sum_{\ell=0}^{mK}k_\ell h_\ell}\left(\sum_{\be\in\GF(p)^m\setminus\{\bzero\}}
	x\left[\frac\bh{\be}\right]\right)\right)^\rho  \nonumber \\
& \stackrel{(c)}{=}  \frac{(q-1)^\rho}{q^K}+ \frac1{q^K}\sum_{\bk\in\GF(p)^{mK}\setminus\{\bzero\}}\left(x[\bzero](q-1)
	\vphantom{\sum_{\bh\in\GF(p)^{mK}}}\right.  \nonumber\\
&\hphantom{==}	\left.+(1-x[\bzero])\sum_{\substack{\bh\in\\\GF(p)^{mK}\setminus\{\bzero\}}}
	e^{-j\frac{2\pi}{p}\sum_{\ell=0}^{mK}k_\ell h_\ell}\right)^\rho   \nonumber \\
& \stackrel{(d)}{=}  \frac{(q-1)^\rho}{q^K}+\frac{(q-1)^\rho}{q^K}\left(\sum_{\bk\neq\bzero}\left(x[\bzero]
	\vphantom{\frac1{(q-1)^\rho}}	+(1-x[\bzero])\right.\right.   \nonumber\\
&	\left.\left.\hphantom{\frac{(q-1)^\rho}{q^K}}
	\cdot \frac{\sum_{\bh\neq\bzero}e^{-j\frac{2\pi}{p}\sum_{\ell=0}^{mK}k_\ell h_\ell}}{q-1}
	\right)^\rho\right), \label{eqn:Ax_BR_proof}
\end{align}
where $(a)$ follows by first separating the term $\bk=\bzero$ and noting $\sum_{g\in\cQ} \bx[g]=1$, then interchanging the order of summation, and noting that the product term is identical for all $\be_i, i \in[\rho]$; $(b)$ holds since the exponential term is independent of $\be$; $(c)$ follows by separating the summation over $\bh\in \GF(q)^{mK}$ into the case where $\bh = \bzero$ and the case where $\bh\neq \bzero$; $(d)$ follows from taking a factor of $(q-1)^\rho$ out of the summation over $\bk$.

To bound the final term, notice that for each $\bk \in \GF(p)^{mK}$
\begin{eqnarray*}
\lefteqn{\left|\frac1{q-1}\sum_{\bh\neq\bzero}
e^{-j\frac{2\pi}{p}\sum_{\ell=0}^{mK}k_\ell h_\ell}\right|^2} \\
& = & \left|\frac1{q-1}\sum_{\hp=0}^{p-1}
	\sum_{\bh\neq\bzero:\sum_\ell k_\ell h_\ell=\hp}e^{-j\frac{2\pi}{p}\hp}\right|^2 \\
& \stackrel{(e)}{\leq} & \max_{\hp\in[p-1]}\left|
	\frac{|\{\bh\neq\bzero:\sum_\ell k_\ell h_\ell=0\}|}{q-1}\right. \\
&&	\left.+\frac{|\{\bh\neq\bzero:\sum_\ell k_\ell h_\ell\neq0\}|}{q-1}
	e^{-j\frac{2\pi}{p}\hp}\right|^2 \\
& = & \max_{\hp\in[p-1]}\left|(1-\lambda_\bk)
	+\lambda_\bk e^{-j\frac{2\pi}{p}\hp}\right|^2, 
\end{eqnarray*}
where $(e)$ holds by separating the summation over $\bh\neq 0$ for which $\sum_l k_l h_l=\hp$ into cases where $\hp=0$ and the cases where $\hp\neq0$ and using the maximum term for the $\hp \neq 0$ group; finally, in the last term, we let
\[
\lambda_\bk=\frac{|\{\bh\neq\bzero:\sum_\ell k_\ell h_\ell\neq0\}|}{q-1};
\]
notice that $\lambda_\bk$ is a function of $\bk$ and $q$.

Therefore, setting 
\[
\tau\defeq\max_{\hp\in[q-1]}
	{\rm Re}(e^{-j\frac{2\pi\hp}p})
\]
gives the following upper bound
\begin{eqnarray*}
\lefteqn{\left|\frac1{q-1}\sum_{\bh\neq\bzero}
e^{-j\frac{2\pi}{p}\sum_{\ell=0}^{mK}k_\ell h_\ell}\right|^2} \\
& \leq & \max_{\hp\in[p-1]}
	\left[(1-\lambda_\bk)^2+\lambda_\bk^2
	+ 2\lambda_\bk(1-\lambda_\bk){\rm Re}(e^{-j\frac{2\pi\hp}p})\right] \\
&=&	(1-\lambda_\bk)^2+\lambda_\bk^2+2\tau\lambda_\bk(1-\lambda_\bk) \\
&=& 1-(1-\tau)2\lambda_\bk(1-\lambda_\bk) \\
&\leq& \psi^2,
\end{eqnarray*}
% which is strictly less than one since $\lambda_\bk\in (0,1)$ for all $\bk\neq\bzero$. 
% Therefore, under the theorem assumption 
% that $x[\bzero]\in[\delta,1-\delta]$, for all $\bk\neq\bzero$ 
where
\[
% \left|\frac1{q-1}\sum_{\bh\neq\bzero}
% e^{-j\frac{2\pi}{p}\sum_{\ell=0}^{mK}k_\ell h_\ell}\right|
% \leq 1-(1-\rho)2\lambda_\bk(1-\lambda_{\bk})
\psi^2\defeq 
\max_{\bk\neq\bzero}[1-(1-\tau)2\lambda_\bk(1-\lambda_\bk)].
\]
% where $\phi$ is a function of $\delta$ and $p$ but not of $\bx$.  

% Therefore, from (\ref{eqn:Athe}), we get
Notice that $\psi^2$ depends on $q$ but does not vary with $\bx$.  
Notice further that  $\psi^2$ lies in $(0,1)$ since 
$\tau\in(0,1)$ for all $q$ and 
$\lambda_\bk\in (0,1)$ for all $\bk\neq\bzero$; therefore, $2\lambda_k(1-\lambda_k)\in(0,\frac12]$.
Noting that $x[\bzero]\in[0,1-\sigma]$ by assumption ($\bx \in J_\sigma$), we have
\begin{multline*}
\left|x[\bzero]+(1-x[\bzero])
      \frac{\sum_{\bh\neq\bzero}e^{-j\frac{2\pi}{p}\sum_{\ell=0}^{mK}k_\ell h_\ell}}{q-1}
      \right|^2 \\
\leq  (x[\bzero]+\psi(1-x[\bzero]))^2  <1 ,
\end{multline*}
% \begin{eqnarray*}
% A(\bx)
% & = & \sum_{\be_1,\ldots,\be_\rho\in\GF(p)^m\setminus\{\bzero\}}\frac1{q^K}
% 	\sum_{\bk\in\GF(p)^{mK}} \prod_{i=1}^\rho f(\bk,\be_i) \\
% & = & \sum_{\substack{\be_1,\ldots,\be_\rho\\\in\GF(p)^m\setminus\{\bzero\}}}
% 	\frac1{q^K}
% 	\left[\prod_{i=1}^\rhof(\bzero,\be)\vphantom{\sum_{\bk\in\GF(p)^{mK}}}\right. \\
% &&	\left.\hphantom{\sum_{\be_1,\ldots,\be_\rho\in\GF(p)^m}}
% 	+\sum_{\bk\in\GF(p)^{mK}\setminus\{\bzero\}}
% 		\prod_{i=1}^\rho f(\bk,\be_i)\right]  \\
% & = & \sum_{\substack{\be_1,\ldots,\be_\rho\in\\\GF(p)^m\setminus\{\bzero\}}}\frac1{q^K}
% 	\left[1+\sum_{\bk\in\GF(p)^{mK}\setminus\{\bzero\}}
% 		\prod_{i=1}^\rho f(\bk,\be_i)\right] 
% % \lefteqn{\left|x[\bzero]+(1-x[\bzero])
% %       \frac{\sum_{\bh\neq\bzero}e^{-j\frac{2\pi}{p}\sum_{\ell=0}^{mK}k_\ell h_\ell}}{q-1}
% %       \right|^2} \\
% % & \leq & (x[\bzero]+\psi(1-x[\bzero]))^2  <1 
% \end{eqnarray*}
Taking the square root of both sides gives
\begin{multline*}
\left|x[\bzero]+(1-x[\bzero])
      \frac{\sum_{\bh\neq\bzero}e^{-j\frac{2\pi}{p}\sum_{\ell=0}^{mK}k_\ell h_\ell}}{q-1}
      \right| \\
\leq  (x[\bzero]+\psi(1-x[\bzero])).
\end{multline*}

Returning to our equation for $A(\bx)$ in \eqref{eqn:Ax_BR_proof}, we have
% Old
% \begin{eqnarray*}
% \lefteqn{\left|A(\bx)-\frac{(q-1)^{d}}{q^K}\right|} \\
% & \leq & \left|\sum_{\be_1,\ldots,\be_\rho\in\GF(p)^m\setminus\{\bzero\}}\frac1{q^K}
% 	\sum_{\bk\in\GF(p)^{mK}\setminus\{\bzero\}}
% 		\prod_{i=1}^\rho f(\bk,\be_i)\right| \\
% & \leq & \frac{(q-1)^\rho(q^K-1)}{q^K}(1+\psi^\rho),
% \end{eqnarray*}
% New
% \begin{eqnarray*}
% \lefteqn{\left|A(\bx)-\frac{(q-1)^{d}}{q^K}\right|} \\
% & \leq & \left|\frac{(q-1)^\rho}{q^K}\sum_{\bk \in \GF(p)^{mK}\setminus\{\bzero\}} (\bx[\bzero]+(1-\bx[\bzero])\psi)^\rho\right| \\
% & \leq & \frac{(q-1)^\rho(q^K-1)}{q^K}(1+\psi^\rho),
% \end{eqnarray*}
% where $\phi^\rho$ approaches zero as $\rho$ grows without bound.
% noting that $x[\bzero]\in[0,1-\sigma]$ by assumption ($\bx \in J_\sigma$),
\begin{eqnarray*}
\frac{\left|A(\bx)-\frac{(q-1)^\rho}{q^K}\right|}{\frac{(q-1)^\rho}{q^K}}& \leq & \left|\sum_{\bk\neq\bzero}
      (x[\bzero]+\psi(1-x[\bzero]))^\rho\right|,
\end{eqnarray*}
which gives an upper bound on $A(\bx)$
\begin{align}
    A(\bx)  &\leq \frac{(q-1)^\rho}{q^K}\left(1+\sum_{\bk\neq\bzero}
      (x[\bzero]+\psi(1-x[\bzero]))^\rho\right)  .
\end{align}
%&\leq \frac{(q-1)^\rho}{q^K} + \frac{(q-1)^\rho}{q^K}\sum_{\bk\neq\bzero}
    %  (x[\bzero]+\psi(1-x[\bzero]))^\rho \\
Therefore, we obtain
\begin{align}
    \log A(\bthe) &\leq \log\left(\frac{(q-1)^\rho}{q^K}\right) \nonumber \\
    &\hphantom{=}+\log\left( 1+\sum_{\bk\neq\bzero}
      (\bthe[\bzero]+\psi(1-\bthe[\bzero]))^\rho\right), \label{eqn:Abthe}
\end{align}
where the second term approaches $0$ as $\rho$ increases uniformly for all $\bthe \in J_\delta$.
% {\color{red}
Returning to \eqref{eqn:bthe_3}, fixing $\lambda = \rho(1-R)$ and letting $\rho = \kappa n$ for some constant $\kappa$ gives
\begin{align*}
\lefteqn{\lim_{n\rightarrow\infty}\oS_L(\bthe)} \\
& \leq  \lim_{n\rightarrow\infty} \left[ H(\bthe)-\lambda\log(q-1) +\frac{\lambda}{\rho}\log\left(\frac{(q-1)^\rho}{q^K}\right)\right.\\ 
&\hphantom{=====} \left.+\frac{\lambda}{\rho} \log\left( 1+\sum_{\bk\neq\bzero}
      (\bthe[\bzero]+\psi(1-\bthe[\bzero]))^\rho\right)  \right] \\
&= H(\bthe) - \frac{\lambda}{\rho}K \\
&\hphantom{=} +\lim_{n\rightarrow\infty}\left[\frac{\lambda}{\rho}\log\left( 1+\sum_{\bk\neq\bzero}
      (\bthe[\bzero]+\psi(1-\bthe[\bzero]))^{\kappa n}\right) \right]  \\
& = H(\bthe)-K(1-R) + 0\\
&=  \oS_U(\bthe). 
\end{align*}
Note that the choice of $\kappa$ in $\rho = \kappa n$ should be much smaller than $\frac{q-1}{q}$ to maintain some of the sparsity of LDPC codes. The upper bound $\frac{q-1}{q}$ is chosen to ensure that the edges values in the Tanner graph can be chosen from $\GF(q)\setminus\{0\}$ instead of $\GF(q)$.

If, instead of setting $\rho = \kappa n$ for some constant $\kappa$, we set $\rho = \kappa(n)n$ for some function $\kappa(n)$ that satisfies $\kappa(n) n \rightarrow \infty$ as $n\rightarrow \infty$, we again find that $\lim_{n\rightarrow\infty}\oS_L(\bthe) = \oS_U(\bthe)$. In Appendix~\ref{app:ePe}, we show that in order for $\frac{1}{n}\log \alpha_{\rm \scalebox{0.4}{MAC}}$ to behave as $O(\frac{\log n}{n})$, $\kappa(n)$ should decay no more quickly than $\bThe(\frac{\log n}{n})$.
% }

% In fact, a constant $\kappa$ is a special case, any $\kappa$ that satisfies $\kappa n \rightarrow \infty$ as $n\rightarrow \infty$ suffices to prove $\lim_{n\rightarrow\infty}\oS_L(\bthe) = \oS_U(\bthe)$. However, it is shown in Appendix~\ref{app:ePe} that $\kappa$ should decay no more quickly than $O(\frac{\log n}{n})$ so that $\frac{1}{n}\log \alpha_{\rm \scalebox{0.4}{MAC}}$ behaves as $O(\frac{\log n}{n})$.
% % }

\hfill$\blacksquare$

% \section{Proof of Theorem~\ref{thm:BRthe}}\label{app:BRthe}

\section{Probability of Small Minimum Distance Codes in the ${\rm LDPC}_K(\lambda,\rho;n)$ ensemble} \label{app:dMin}
Since Theorem~\ref{thm:BRthe} bounds the difference between $\oS_L(\bthe)$ and $\oS_U(\bthe)$ only when $\bthe \in J_\sigma$, it does not eliminate the possibility that $\alpha_{\rm \scalebox{0.4}{MAC}}$ (defined in \eqref{eqn:alpha}) may be large for all values of $\lambda$ and $\rho$ if we consider all possible values of $\bthe$.

To resolve this problem, Theorem~\ref{thm:ePe} removes from the ${\rm LDPC}_K({\rm Full},\lambda,\rho;n)$ ensemble all codes for which the minimum distance between codematrices is less than or equal to $\gamma n$. Recall that the distance between two codematrices $\bd_1$ and $\bd_2$ with dimension $n\times K$ is the number of rows they differ,
\[d(\bd_1,\bd_2)=\sum_{i=1}^n \ind(\bd_1[i,*] \neq \bd_2[i,*]);\]
that is, $d(\bd_1,\bd_2)$ is the number of time slots in which the transmissions for codematrices $\bd_1$ and $\bd_2$ differ. The minimum distance of codebook $\bd$ is 
\[d_{\min}(\bd) = \min_{\bm \neq \bm'} d(\bd_\bm,\bd_{\bm'}).\]

 In \cite[Th. 6]{BennatanB:04}, Bennatan et al. prove that if $\bC$ is a randomly chosen code from the ${\rm LDPC}({\rm Full}, \lambda,\rho;n)$ ensemble (using the full collection of legitimate codewords, rather than our possibly reduced collection of codewords), then there exists some $\gamma \in (0,1/2]$ that depends only on $R$ and $q$ such that 
 \begin{align}
    \Pr[d_{\min}(\bC) \leq \gamma n] = O(n^{-\left(\frac{\lambda}{2}-1\right)}), \label{eqn:dmin_single_TX}
\end{align} 
where the distance between two codewords is the number of positions at which the corresponding symbols are different.

Lemma~\ref{lem:dMin} builds on this result in order to bound the probability of codes with small minimum distance under the random LDPC code design. This bound is later employed in the proof of Theorem~\ref{thm:ePe} (see Appendix~\ref{app:ePe}) to bound the change in ensemble-average number of codematrices due to expurgation. 

\begin{lem} \label{lem:dMin}
    Fix the rate $R = 1 - \frac{\lambda}{\rho}$. Let $\lambda \geq 3$ and  fix some prime power $q$. Let $\bD$ be a randomly chosen code from the ${\rm LDPC}_K(\lambda,\rho;n)$ ensemble. Then there exists some $\gamma \in (0,1/2]$ that depends only on $R$ and $q$ such that
    \begin{equation*}
        \Pr[d_{\min}(\bD) \leq \gamma n] = O(n^{-\left(\frac{\lambda}{2}-1\right)}).
    \end{equation*}
    
    % ${\rm LDPC}(\lambda,\rho,\delta;n)$ ensemble of $(\lambda,\rho)$-regular LDPC MAC codes.
\end{lem}

{\em Proof:} 
We begin by noting that Bennatan et al.'s  single-transmitter bound \eqref{eqn:dmin_single_TX} on the minimum distance continues to hold if one restricts code $\bC$ to exactly $q^{nR}$ codewords through random codeword selection; this follows because removing codewords from the codebook cannot decrease the pairwise minimum distance between the codewords that remain. 

% For every code $\bC$, there exists a pair of distinct indices $i,j\in [M]$ for which $d_{\min}(\bC) = d(\bC_i,\bC_j)$, where $\bC_i$ and $\bC_j$ are the $i$-th and $j$-th codeword of $\bC$.

We next show that $d_{\min}(\bD) = d_{\min}(\bC)$, where $\bC$ is the underlying single-transmitter code for $\bD$.

First note that $d_{\min}(\bD) \leq d(\bD_\bm,\bD_{\bm'})$, where $\bm$ and $\bm'$ are any pair of index vectors from $[M]^K$ that differ in exactly one component. Choosing the element in that differing component to be any pair $(i,j)$ for which $d(\bC_i,\bC_j) = d_{\min}(\bC)$ shows that $d_{\min}(\bD) \leq d_{\min}(\bC)$; that is, since $\bm$ and $\bm'$ differ in exactly one component, say $\bm = (i,1,\ldots,1)$ and $\bm' = (j,1,\ldots,1)$, the time slots in which $\bD_\bm$ and $\bD_{\bm'}$ differ are exactly the time slots in which $\bC_i$ and $\bC_j$ differ, giving
\begin{align}
    d_{\min}(\bD) \leq d(\bD_\bm,\bD_{\bm'}) = d(\bC_i,\bC_j) = d_{\min}(\bC).
\end{align}
% To prove that this bound is tight, note that the distance $d(\bD_\bm,\bD_{\bm'})$ between any pair of distinct codematrices $\bD_\bm,\bD_{\bm'} \in \bD$ is the size of the set found by taking the union of all positions in which codewords $\bC_{m(i)}$ and $\bC_{m'(j)}$ differ for some $i,j \in [K], m(i)\neq m'(j)$. Since $\bm \neq \bm'$ implies there exists at least one such pair $i,j$,
To prove that this bound is tight, note that the distance $d(\bD_\bm,\bD_{\bm'})$ between any pair of distinct codematrices $\bD_\bm,\bD_{\bm'} \in \bD$ is
% the size of the set found by taking the union of all positions in which codewords $\bC_{m(i)}$ and $\bC_{m'(i)}$ differ for all $i \in [K]$ such that $m(i) \neq m'(i)$.
\begin{align}
    d(\bD_\bm,\bD_{\bm'}) = \left|\bigcup_{k\in [K]:m(k)\neq m'(k)}\mkern-40mu \{i \in[n]: \bC_{m(k),i}\neq \bC_{m'(k),i}\} \right|
\end{align}
Since $\bm \neq \bm'$ implies there exists at least one such $k$,
\begin{align}
    d_{\min}(\bD) &\geq \min_{\bm,\bm':\bm\neq\bm'} \min_{k:m(k)\neq m'(k)} d(\bC_{m(k)},\bC_{m'(k)}) \\
    &= d_{\min}(\bC).
\end{align}
Combining the two sides of the argument gives $$d_{\min}(\bD) = d_{\min}(\bC).$$
Thus $ \Pr[d_{\min}(\bD) \leq \gamma n] = \Pr[d_{\min}(\bC) \leq \gamma n] = O(n^{-\left(\frac{\lambda}{2}-1\right)})$ gives the desired result.

\hfill$\blacksquare$

% Given the ensemble ${\rm LDPC}_K(\lambda,\rho;n)$ of $(\lambda,\rho)$-regular LDPC MAC codes of dimension $n\times K$, and given some $\sigma >0$, removing all codes of $d_{\min}(\bD) \leq \sigma n$ gives the expurgated MAC LDPC ensemble ${\rm LDPC}_K$-${\rm Ex}_\sigma(\lambda,\rho;n)$. If the number $\sigma$ is chosen to satisfy $\sigma \leq \gamma$ for the value of $\gamma$ shown to exist by Theorem \ref{lem:dMin}, and $n$ is large enough, then this expurgation operation does not reduce the size of the ensemble by a factor greater than 2 since Lemma~\ref{lem:dMin} shows probability of obtaining codes with small minimum distance decreases exponentially in blocklength $n$. Therefore, the average spectrum of the expurgated ensemble, denoted by $\obS^{ex,n}$, satisfies
% The analysis of decoding error probability over the expurgated ensemble is shown in the following theorem.

\section{Proof of Theorem~\ref{thm:ePe}}\label{app:ePe}
Lemma~\ref{lem:dMin} of Appendix~\ref{app:dMin} shows that for $R = 1 - \frac{\lambda}{\rho}$ and $\lambda \geq 3$, there exist some $\gamma \in (0,1/2]$ for which $\Pr[d_{\min}(\bD)\leq \gamma n]=O(n^{-\left(\frac{\lambda}{2}-1\right)})$ under our ${\rm LDPC}_K(\lambda,\rho;n)$ ensemble.
Fix any $\sigma$ smaller than $\gamma$. We first particularize Theorem \ref{thm:Pe} to the expurgated ensemble ${\rm LDPC}_K-{\rm Ex}_\sigma(\lambda,\rho;n)$ to bound the ensemble-average error probability $E_{{\rm ex},\sigma}\left[P_e^{(n)}\right]$ as
\begin{align}
E_{{\rm ex},\sigma}\left[P_e^{(n)}\right]\leq \sum_{\bt\in \Tau}\oS_{{\rm ex},\sigma}^{n}(\bt)\bcD^\bt
+q^{-nE_p(KR+(\log\alpha_{\rm \scalebox{0.4}{MAC}})/n)}, \label{eqn:error_exp}
\end{align}
where $E_{{\rm ex},\sigma}[\cdot]$ denotes expectation under the expurgated ensemble, $\obS_{{\rm ex},\sigma}^{n}$ is the ensemble-average spectrum under the expurgated ${\rm LDPC}_K-{\rm Ex}_\sigma(\lambda,\rho;n)$ ensemble, and
\begin{equation*}
\Tau \defeq \{\bt \in  \cT_\cQ^n: 0 < \mbox{wt}(\bt) \leq \sigma n\},
\end{equation*}
where for all $\bt \in \cT_\cQ^n$, $\mbox{wt}(\bt)$ is the number of nonzero rows in a matrix of type $\bt$. 
% We next bound each of the elements in \eqref{eqn:error_exp} in turn.

% {\color{red}
Before bounding each of the elements in \eqref{eqn:error_exp}, we first prove that there exist some finite integer $n_0$ such that 
% two properties of the ${\rm LDPC}_K-{\rm Ex}_\sigma(\lambda,\rho;n)$ ensemble-average spectrum $\obS_{{\rm ex}, \sigma}^{n}$, shown in \eqref{eqn:obSe}
\begin{eqnarray} \label{eqn:obSe}
\begin{aligned}[c]
\oS_{{\rm ex},\sigma}^{n}(\bt) &= 0\\
\oS_{{\rm ex},\sigma}^{n}(\bt) &\leq 2\oS_L^n(\bt)
\end{aligned}
\qquad
\begin{aligned}[c]
&\mbox{if } 0<\mbox{wt}(\bt)\leq \sigma n \\
&\mbox{if }\mbox{wt}(\bt)>\sigma n \mbox{ and } n>n_0.
\end{aligned}
\end{eqnarray}
The first property follows immediately from the definition of the expurgated code. To prove the second property, recall that $\sigma <\gamma$ by assumption. Therefore, the probability that the minimum distance of a randomly chosen code from the ${\rm LDPC}_K-{\rm Ex}_\sigma(\lambda,\rho;n)$ ensemble is less than $\sigma n$ decays as $O(n^{-(\frac{\lambda}{2}-1)})$. In other words, there exist constants $a\in \mathbb{R}$ and $n_0'\in \mathbb{Z}$ such that for all $n>n_0'$,
\begin{align}
    \Pr[d_{\min}(\bD)\leq \gamma n] \leq an^{-(\frac{\lambda}{2}-1)}.
\end{align}To guarantee that the support set of the expurgated ensemble is at least half the size of the support set of the original ensemble, we choose $n$ sufficiently large so that $an^{-(\frac{\lambda}{2}-1)} \leq \frac{1}{2}$. Under this assumption, no more than half of the support are expurgated. This gives
\begin{align}
    an^{-(\frac{\lambda}{2}-1)} &\leq \frac{1}{2} \\
    -\left(\frac{\lambda}{2}-1\right) \log (an) &\leq \log \frac{1}{2} \\
    \log (an) &\geq \left(\frac{1}{\lambda/2-1}\right)\log 2 \\
    n&\geq \frac{q^{\frac{\log 2}{\lambda/2-1}}}{a}.
\end{align}
Therefore choosing $n_0> \max\{n_0',q^{\frac{\log 2}{\lambda/2-1}}/a \}$ ensures that $\oS_{{\rm ex},\sigma}^{n}(\bt) \leq 2\oS_L^n(\bt)$ for any $n>n_0$ and any $\bt$ with $\mbox{wt}(\bt)>\sigma n$. While both $n_0'$ and $a$ are unknown, the existence of such values proves that the desired property holds for all $n$ sufficiently large.

% In addition, note that the minimum weight equals the minimum distance in any linear code. Therefore, removing all codes with minimum distance less than $\delta n$ from the ${\rm LDPC}_K({\rm Full},\lambda,\rho;n)$ ensemble ensures the minimum weight of each remaining code in the ${\rm LDPC}_K-{\rm Ex}_\sigma({\rm Full},\lambda,\rho;n)$ ensemble, and consequently the minimum weight of each code in the ${\rm LDPC}_K-{\rm Ex}_\sigma(\lambda,\rho;n)$ ensemble, to be bigger than $\delta n$.

% Since $\sigma <\gamma$ by assumption, choosing $n$ greater than some function that grows as $O\left(q^{\frac{2}{\lambda/2-1}}\right)$ guarantees that the support set of the expurgated ensemble is at least half the size of the support set of the original ensemble. This follows because  setting this probability to $\frac{1}{2}$ (so that no more than half of the support are expurgated) gives $O\left(q^{\frac{2}{\lambda/2-1}}\right)$. In addition, note that the minimum weight equals the minimum distance in any linear code. Therefore, removing all codes with minimum distance less than $\delta n$ from the ${\rm LDPC}_K({\rm Full},\lambda,\rho;n)$ ensemble ensures the minimum weight of each remaining code in the ${\rm LDPC}_K-{\rm Ex}_\sigma({\rm Full},\lambda,\rho;n)$ ensemble, and consequently the minimum weight of each code in the ${\rm LDPC}_K-{\rm Ex}_\sigma(\lambda,\rho;n)$ ensemble, to be bigger than $\delta n$. 

By \eqref{eqn:obSe}, $\oS_{{\rm ex},\sigma}^{n}(\bt)=0$ for any $\bt$ with $\mbox{wt}(\bt) \leq \sigma n$. Therefore, the term $\sum_{\bt\in \Tau}\oS_{{\rm ex},\sigma}^{n}(\bt)\bcD^\bt$ in \eqref{eqn:error_exp} equals $0$.

% disappears as a result of removing codes with minimum distance less than $\delta n$ to form the expurgated LDPC ensemble. 

For the rate offset $\frac{\log \alpha_{\rm \scalebox{0.4}{MAC}}}{n}$ in the second term of \eqref{eqn:error_exp}, recall that
\begin{align}
\Tau^c &= \{n\bthe:\bthe \in J_{\sigma}\} \\
\alpha_{\rm \scalebox{0.4}{MAC}} &= \max_{\bt \in \Tau^c} \frac{\oS_{{\rm ex},\sigma}^{n}(\bt)}{(M^K-1)B(n,\bt)q^{-nK}}\nonumber \\
&=\max_{\bthe \in J_\sigma} \frac{\oS_{{\rm ex},\sigma}^{n}(\bt)}{(M^K-1)B(n,\bt)q^{-nK}},
\end{align}
$J_\sigma$ is defined in \eqref{eqn:J_sigma}, and  and $B(n,\bt) = n!/(\prod_g(t_g)!)$ is the number of distinct possible codematrices of type $n\bthe$. Therefore
\begin{eqnarray}
\lefteqn{\frac{1}{n} \log \alpha_{\rm \scalebox{0.4}{MAC}}} \nonumber \\
& = & \frac{1}{n} \log \max_{\bthe \in J_\sigma} \frac{\oS_{{\rm ex},\sigma}^{n}(n\bthe)}{(M^K-1)B(n,n\bthe)q^{-nK}} \nonumber \\
& = & \max_{\bthe \in J_\sigma} \left[\frac{1}{n}\log \oS_{{\rm ex},\sigma}^{n}(n\bthe) - \right. \nonumber\\
&& \left. \hphantom{===}\frac{1}{n} \log \left((M^K-1)B(n,n\bthe)q^{-nK} \right) \right] \nonumber \\
& \stackrel{(a)}\leq & \max_{\bthe \in J_\sigma} \left[\frac{1}{n}\log \oS_{{\rm ex},\sigma}^{n}(n\bthe) -\frac{1}{n}\log \oS_L^{n}(n\bthe) \right] \nonumber \\
&& +  \max_{\bthe \in J_\sigma} \left[\frac{1}{n}\log \oS_L^{n}(n\bthe) - \oS_L(\bthe) \right] \nonumber \\
&& + \max_{\bthe \in J_\sigma}  \left[ \oS_L(\bthe)  - \oS_U(\bthe) \right] \nonumber \\
&& \mkern-30mu +\max_{\bthe \in J_\sigma}  \left[  \oS_U(\bthe) - \frac{1}{n} \log \left((M^K-1)B(n,n\bthe)q^{-nK} \right) \right] \nonumber,\\
\hphantom{=} \label{eqn:loga_n_4terms}
\end{eqnarray}
where $(a)$ follows from triangle inequality for the $\max$ function.

% Using \eqref{eqn:obSe}, we know $\obS^{ex,n}(n\bthe)$ and $\obS^{n}(n\bthe)$ differ by a factor bounded by $2$ for $\bthe\in J_\sigma$, therefore the first element of the above sum is $O(1/n)$, goes to $0$ as $n$ grows without bound. The second element also goes to zero from our result in Appendix \ref{app:BRthe} as $\rho$ grows without bound. To make the result more precise, a finite-$n$ upper bound for the second term provided
By \eqref{eqn:obSe}, $\oS_{{\rm ex},\sigma}^{n}(n\bthe)\leq 2\oS_L^{n}(n\bthe)$ for all $\bthe\in J_\sigma$; therefore,
% the first element of the above sum is $O(1/n)$.
\begin{align}
    \max_{\bthe \in J_\sigma} \left[\frac{1}{n}\log \oS_{{\rm ex},\sigma}^{n}(n\bthe) -\frac{1}{n}\log \oS_L^{n}(n\bthe) \right] &\leq \frac{\log 2}{n}\\
    &=O\left(\frac{1}{n}\right). \label{eqn:loga_n_1st}
\end{align}
% The second element also goes to zero from our result in \ref{thm:BRthe} as $\rho$ grows without bound. To make the result more precise, a finite-$n$ upper bound for the second term provided

To bound the second element in \eqref{eqn:loga_n_4terms}, note that
 %Using [Theorem 10 and Lemma 1 in 2004 Paper] and $\Pr(\bd_{n\bthe}) \in C$ from Appendix \ref{app:BRthe}, we bound the second element for a finite $n$ as:
\begin{align}
% \lefteqn{\frac{1}{n}\log \oS_L^{n}(n\bthe)} \\
\frac{1}{n}\log \oS_L^{n}(n\bthe)&=\frac{1}{n}\log\left[B(n,n\bthe) \Pr[\bd_{n\bthe} \in \bD]\right] \nonumber \\
&=\frac{1}{n}\log B(n,n\bthe)+\frac{1}{n}\log \Pr[\bd_{n\bthe} \in \bD] \label{eqn:S_L}
\end{align}
By \cite[Th. 17.4.3]{Cover:2012}, the multinomial coefficient can be bounded as 
\begin{align}
    \frac{1}{(n+1)^{q^K}}q^{nH(\bthe)} \leq B(n,n\bthe) \leq q^{nH(\bthe)}. \label{eqn:multi_coe}
\end{align}
Further, recall from \eqref{eqn:d_nbthe} that $\Pr[\bd_{n\bthe}\in\bD]$ is the probability that a type-$n\bthe$ matrix $\bd_{n\bthe}$ is in the codebook $\bD$ of a randomly drawn code from the ${\rm LDPC}_K(\lambda,\rho;n)$ ensemble and that
\begin{eqnarray}
\Pr[\bd_{n\bthe}\in\bD]= \frac{\lfloor((A(\bx))^{n\lambda/\rho})\rfloor_{n\lambda\bthe}}{B(n\lambda,n\lambda\bthe)(q-1)^{n\lambda}}. \label{eqn:dm_1}
\end{eqnarray}
% Applying \eqref{eqn:multi_coe} to bound the denominator in \eqref{eqn:dm_1} gives
% \begin{align}
%     \frac{1}{(n\lambda+1)^{q^K}}q^{n\lambda H(\bthe)}\leq B(n\lambda,n\lambda\bthe) \leq q^{n\lambda H(\bthe)}. \label{eqn:dm_3}
% \end{align}
Applying Lemma~\ref{lem:MN} to bound the numerator in \eqref{eqn:dm_1} gives
\begin{align}
    \frac{1}{n}\log \lfloor((A(\bx))^{n\lambda/\rho})\rfloor_{n\lambda\bthe} \leq \frac{\lambda}{\rho} \log \inf_{\substack{\bx:\sgn(\bx)\\=\sgn(\bthe)}} \frac{A(\bx)}{\bx^{\rho\bthe}}. \label{eqn:dm_2}
\end{align}
Applying the lower bound of \eqref{eqn:multi_coe} to bound $B(n\lambda,n\lambda\bthe) $, combining it with \eqref{eqn:dm_1} and \eqref{eqn:dm_2} yields
\begin{align}
\lefteqn{\Pr[\bd_{n\bthe} \in \bD]} \nonumber \\
&\leq (n\lambda+1)^{q^K} q^{n\left(-\lambda\log(q-1)-\lambda H(\bthe) +(1-R)\log\inf A(\bx)/\bx^{\rho \bthe}\right)}, \label{eqn:dm_4}
\end{align}
where we take the infimum in \eqref{eqn:dm_4} over all $\bx$ for which $\sgn(\bx)=\sgn(\bthe)$.

Therefore, we obtain the bound 
\begin{align}
\frac{1}{n}\obS_L(n\bthe)&\leq H(\bthe) + \frac{1}{n}\log(n\lambda+1)^{q^K} -\lambda H(\bthe)  \\
& -\lambda \log(q-1)+\frac{\lambda}{\rho}\log\inf_{\bx:\sgn(\bx)=\sgn(\bthe)} \frac{A(\bx)}{\bx^{\rho\bthe}},
% \\
% &= \oS_L(\bthe) + \frac{1}{n}\log(n\lambda+1)^{q^K},
\end{align}
% on $\frac{1}{n}\obS_L(n\bthe)$,
giving
\begin{align}
\max_{\bthe \in J_\sigma} \left[\frac{1}{n}\log \oS_L^{n}(n\bthe) - \oS_L(\bthe) \right] &\leq \frac{1}{n}\log(n\lambda+1)^{q^K}\\
&=O\left(\frac{\log n}{n}\right). \label{eqn:loga_n_2nd}
 \end{align}
% where $(b)$ follows from combinatorial upper bound on $B(n,n\bthe)$ and an upper bound result of $\Pr[\bd_{n\bthe} \in \bD]$ shown in \eqref{eqn:dm_4}. Therefore
% \begin{align}
%      \max_{\bthe \in J_\sigma} \left[\frac{1}{n}\log \oS_L^{n}(n\bthe) - \oS_L(\bthe) \right]=O\left(\frac{\log n}{n}\right).
% \end{align}

% Follow the approach in Lemma 1 from~\cite{BennatanB:04} to bound $\Pr[\bd_\bm \in \bD]$, 

% Using \eqref{eqn:MN_2}, the numerator can be upper bounded as

% Setting $\bx = \bthe$ (similar to \eqref{eqn:bthe_3}) gives
% \begin{eqnarray*}
% \lefteqn{\Pr[\bd_{n\bthe} \in \bD]} \\
% &\leq& (n\lambda+1)^{q^K} q^{n(-\lambda\log(q-1) +(1-R)\log A(\bthe))}.
% \end{eqnarray*}

By Theorem \ref{thm:BRthe}, there exist $n$ and $\rho$ such that $ \oS_L(\bthe)  - \oS_U(\bthe) <\epsilon$ for any $\epsilon>0$. To make the statement more precise, recall from \eqref{eqn:Abthe} that
\begin{align}
    \oS_L(\bthe)  - \oS_U(\bthe)< \log\left( 1+\sum_{\bk\neq\bzero}
      (\bthe[\bzero]+\psi(1-\bthe[\bzero]))^\rho\right),
\end{align}
where $\bthe[\bzero]+\psi(1-\bthe[\bzero])<1$ is some constant that depends on $q$ and $\bthe$.

Using a power series expansion on the function $\log(1+x)$ reveals that $ \oS_L(\bthe)  - \oS_U(\bthe)$ decreases exponentially in $\rho$. Specifically, assume that $\rho = \kappa n$, where $\kappa$ is some constant that is much smaller than $\frac{q-1}{q}$; here $\kappa$ captures the density of the LDPC code which we treat as a fixed proportion of the blocklength $n$, with low $\kappa$ yielding low density and therefore low LDPC decoding complexity and high $\kappa$ yielding improvements in LDPC code performance at the cost of higher complexity. Using this choice of $\rho$ gives
\begin{align}
    \max_{\bthe \in J_\sigma}  \left[ \oS_L(\bthe)  - \oS_U(\bthe) \right] = O\left(c_0^{\kappa n}\right), \label{eqn:loga_n_3rd}
\end{align}
where $c_0<1$ is some constant that depends on $q$ and $\delta$.
%makes the third element decrease exponentially in $n$, which is some higher order term than $O(\frac{1}{n})$, therefore

To bound the final term in \eqref{eqn:loga_n_4terms}, recall that $\obS_U(\bthe) = H(\bthe) - K(1-R)$ and $M = q^{nR}$. Therefore,
\begin{align}
  &\hphantom{=}\oS_U(\bthe) - \frac{1}{n} \log \left((M^K-1)B(n,n\bthe)q^{-nK} \right) \nonumber \\
    &= H(\bthe) - K(1-R) \nonumber \\
    &\hphantom{===} -\left[\frac{1}{n}\log(q^{nRK}-1) + \frac{1}{n}\log B(n,n\bthe) -K\right] \\
    &= H(\bthe)-\frac{1}{n}\log B(n,n\bthe) +\left[KR -\frac{1}{n}\log(q^{nRK}-1)\right]\\
    &=  H(\bthe)-\frac{1}{n}\log B(n,n\bthe)+ \frac{\log(1- 1/q^{nRK})}{n} \\
    &= H(\bthe)-\frac{1}{n}\log B(n,n\bthe)+ O\left( \frac{1}{nq^{nRK}}\right) \label{eqn:q_NRK_1},
\end{align}
where \eqref{eqn:q_NRK_1} follows from the power series expansion on the function $\log(1-x)$.

Applying the following Stirling's bound on $n!$ \cite{Robbin:55}, which is valid for any positive integers $n$,
\begin{align}
    \sqrt{2\pi}n^{n+\frac{1}{2}}e^{-n}e^{\frac{1}{12n+1}}<n!<\sqrt{2\pi}n^{n+\frac{1}{2}}e^{-n}e^{\frac{1}{12n}}
\end{align}
gives the expression $\log n! =n\log n-n\log e +\frac{1}{2}\log(2\pi n)+ O(1/n)$. Therefore, 
% where the constant in the $O(\log n)$ term is $\frac{1}{2}\log(2\pi n)$,
\begin{align}
    &\hphantom{=}H(\bthe) - \frac{1}{n}\log B(n,n\bthe) \nonumber\\
    &= H(\bthe) - \frac{1}{n}\left[\vphantom{\sum_{g\in\cQ}}n\log n - n\log e + \frac{1}{2}\log (2\pi n)  \nonumber \right.\\
    &\hphantom{===}\left.- \sum_{g \in \cQ:\theta_g\neq0} \mkern-10mu (n\theta_g\log (n\theta_g) -n\theta_g\log e +\frac{1}{2}\log (2\pi n\theta_g)) \right. \nonumber \\
    &\mkern260mu \left. +O\left(\frac{1}{n}\right) \right]\\
    &= H(\bthe) - \frac{1}{n}\left[n\log n - 
\sum_{g \in \cQ:\theta_g\neq0}\mkern-10mu n\theta_g\log (n\theta_g) \right. \\
&\hphantom{==} \left. + \frac{1}{2}\log (2\pi n) - \frac{1}{2} \log\left(\prod_{g \in \cQ:\theta_g\neq0}\mkern-10mu 2\pi n \theta_g \right)+O\left(\frac{1}{n}\right) \right]\\
&= H(\bthe) - \left[\log n - 
\sum_{g \in \cQ:\theta_g\neq0}\mkern-10mu \theta_g\log (n\theta_g) \right] \nonumber \\
&\mkern 180mu +O\left(\frac{\log n}{n}\right)+ O\left(\frac{1}{n^2}\right)\\
&= H(\bthe) - \left[\log n - 
\sum_{g \in \cQ:\theta_g\neq0}\mkern-10mu \theta_g\log (n\theta_g) \right]+O\left(\frac{\log n}{n}\right)\\
&= H(\bthe) -
\sum_{g \in \cQ:\theta_g\neq0} (\theta_g\log \theta_g) +O\left(\frac{\log n}{n}\right)\\
  &= O\left(\frac{\log n}{n}\right).
\end{align}
Returning to \eqref{eqn:q_NRK_1},
\begin{align}
  &\hphantom{=}\max_{\bthe \in J_\sigma}  \left[  \oS_U(\bthe) - \frac{1}{n} \log \left((M^K-1)B(n,n\bthe)q^{-nK} \right) \right] \nonumber \\
  &= O\left(\frac{\log n}{n}\right). \label{eqn:loga_n_4th}
\end{align}
Combining \eqref{eqn:loga_n_4terms}, \eqref{eqn:loga_n_1st}, \eqref{eqn:loga_n_2nd}, \eqref{eqn:loga_n_3rd} and \eqref{eqn:loga_n_4th} gives
\begin{align}
    \frac{\log\alpha_{\rm \scalebox{0.4}{MAC}}}{n} = O\left(\frac{1}{n}\right)+O\left(\frac{\log n}{n}\right)+O\left(c_0^{\kappa n}\right) + O\left(\frac{\log n}{n}\right), \label{eqn:loga_n_combine}
\end{align}
where $c_0<1$  is some constant that depends on $q$ and $\delta$.

To conclude, $\frac{1}{n}\log \alpha_{\rm \scalebox{0.4}{MAC}}$ decays to zero as $O(\frac{\log n}{n})$ for large enough $\rho$ (or, as a special case, for a constant $\kappa$ such that $\frac{q-1}{q}>\kappa >0$ and $\rho =\kappa n$).
% }
% In addition, there exists positive integers $n_0, \rho_0$, such that for all $\lambda,\rho,n$ with $\rho>\rho_0, n>n_0$, and$ R = 1- \frac{\lambda}{\rho}$ such that
% \begin{equation} \label{eqn:loga}
%     \frac{1}{n}\log \alpha < \epsilon^*. 
% \end{equation}

% Substituting \eqref{eqn:loga} into the expression for $E\left[P_e^{e,(n)}\right]$ gives the desired result.
\begin{remark}
To achieve even lower density, we can set $\rho = \kappa(n)n$ for some function $\kappa(n)$ that decays with $n$. When $\kappa(n)\rightarrow 0$ no more quickly than $\bThe(\frac{\log n}{n})$, we again find that $\frac{1}{n}\log \alpha_{\rm \scalebox{0.4}{MAC}}$ decays to zero as $O(\frac{\log n}{n})$. To see this, note from \eqref{eqn:loga_n_combine} that 
\begin{align}
    \frac{1}{n}\log \alpha_{\rm \scalebox{0.4}{MAC}} = O\left(c_0^{\kappa(n) n}\right) + O\left(\frac{\log n}{n}\right).
\end{align}
To find the fastest decay rate of $\kappa(n)$ for which $O(c_0^{\kappa(n) n})$ behaves as $O\left(\frac{\log n}{n}\right)$, we set
\begin{align}
    c_0^{\kappa(n)  n} &= \frac{\log n}{n} \\
    \kappa(n)  n \log c_0 &= \log \log n - \log n \\
    \kappa(n)  &= \frac{\log\log n-\log n}{n\log c_0} \label{eqn:kappa_1}.
\end{align}
Therefore, from \eqref{eqn:kappa_1} we conclude that when $\kappa(n) $ decays no more quickly than $\bThe\left(\frac{\log n}{n}\right)$, then $O(c_0^{\kappa(n) n})$ does not dominate $ O\left(\frac{\log n}{n}\right)$, which in turn makes $\frac{\log \alpha_{\rm \scalebox{0.4}{MAC}}}{n}$ behave as $O\left(\frac{\log n}{n}\right)$. 
\end{remark}
\hfill$\blacksquare$

\section{Proof of Theorem~\ref{thm:full_R}}\label{app:full_R}
The proof of Theorem~\ref{thm:full_R} is similar to the proof of \cite[Lemma 7]{MeassonM:04}. For any code $\bC$ from the ${\rm LDPC}(\lambda,\rho;n)$ ensemble before random codeword removal, we have $R_\bC \geq R$. If the expected value of the actual rate is close to the design rate, then one can apply Markov's inequality to demonstrate that most codes have rates close to the design rate. 

From Theorem~\ref{thm:BML}, we have for any $\bthe=(\theta(g):g\in\cQ)$
\begin{align*}
\oS_L(\bthe)&\defeq \lim_{n\rightarrow\infty}\frac1n\log\oS_L^n(n\bthe) \\
&= (1-\lambda)H(\bthe)-\lambda\log(q-1) 
	+\frac{\lambda}{\rho}\log\inf_{\substack{\bx:\sgn(\bx)\\=\sgn(\bthe)}}
	\frac{A(\bx)}{\bx^{\rho\bthe}}.
\end{align*}

We want to determine the expected rate $\lim_{n \rightarrow \infty} \frac{1}{n} \log \oS^n_{\rm all}$, where $\oS^n_{\rm all} =\sum_{n\bthe \in \cT_\cQ^n} \oS_L^n(n\bthe)$ is the ensemble-average number of codematrices and $\cT_\cQ^n$ is the set of possible types at length $n$. Since there is only a polynomial number of types $|\cT_\cQ^n|$ and the number of codematrices increases exponentially in $n$, the expected rate is equal to the supremum of $\oS_L(\bthe)$ over all $\bthe$. Setting up the Lagrangian of $\oS_L(\bthe)$ with the constraint $\sum_{g\in \cQ} \theta_g =1$ gives
\[
\theta_g =\Lambda x_g^{\frac{\lambda}{\lambda-1}}, \forall g\in \cQ,
\]
where $\Lambda$ is a constant chosen to satisfy the constraint $\sum_{g\in \cQ} \theta_g =1$. Substituting the value of $\theta_g$ back into $\oS_L(\bthe)$, taking its partial derivative with respect to each $x_g$, and using the symmetry of $\oS_L(\bthe)$ with respect to each $x_g$, we find that the stationary point happens when 
\begin{align}
 x_g=x_\bzero , \forall g\in \cQ. \label{eqn:x_0}    
\end{align}

Therefore
\[
\theta_g =\theta_\bzero = \frac{1}{|\cQ|} = \frac{1}{q^K}, \forall g \in \cQ,
\]
and the supremum of $\oS_L(\bthe)$ is
\begin{eqnarray*}
\oS_L(\bthe)
& = & (1-\lambda) \log q^K -\lambda\log(q-1) +   \\
&&  \frac{\lambda}{\rho}\inf_{\substack{\bx:\sgn(\bx)\\=\sgn(\bthe)}}  \left[\log A(\bx) - \log \prod_g x_g^{\rho/q^K} \right] \\
& \stackrel{(a)}{=} & K(1-\lambda) - \lambda\log (q-1) + \frac{\lambda}{\rho} \inf_{\substack{\bx:\sgn(\bx)\\=\sgn(\bthe)}} \\
&& \left[\log \frac{(q-1)^\rho}{q^K}(q^Kx_\bzero)^\rho - \log x_\bzero^\rho\right] \\
&=& K(1-\lambda) - \lambda\log (q-1) + \frac{\lambda}{\rho} \inf_{\substack{\bx:\sgn(\bx)\\=\sgn(\bthe)}} \\
&& [\rho\log (q-1) -K + \rho K + \rho \log x_\bzero - \rho\log x_\bzero] \\
&=& K(1-\lambda) - \frac{\lambda}{\rho}K +\lambda K \\
&=&KR,
\end{eqnarray*}
where  $(a)$ follows from \eqref{eqn:x_0} and the fact that the DFT of a constant sequence is only non-zero at zero. That is, the components of $A(\bx)$ (defined in \eqref{eqn:A_x}) is non-zero only when $\bk = \bzero$, and there are $(p^m-1)^\rho = (q-1)^\rho$ of such terms.

In summary, the expected rate satisfies 
\begin{align*}
    \frac{1}{n} \log \oS^n_{\rm all} =KR + \omega_n,
\end{align*}
where $\omega_n =o(1)$. Let $S_\bD^n$ denote the number of codematrices in the randomly drawn MAC codebook corresponding to the underlying LDPC code $\bC$. Applying Markov's inequality gives
\begin{align*}
    \Pr[R_\bC\geq R + \epsilon] &= \Pr\left[q^{nKR_\bC}\geq q^{nKR}\cdot q^{nK\epsilon}\right] \\
    &= \Pr\left[S_\bd^n\geq \oS^n_{\rm all} q^{n(\epsilon-\omega_n)}\right]\\
    &\leq \frac{\bbE[S_\bd^n]}{\oS^n_{\rm all} q^{n(\epsilon-\omega_n)}} \\
    &\leq q^{-n\epsilon/2},
\end{align*}
for any $\epsilon>0$ and $n\geq n(\epsilon)$, where $n(\epsilon)$ is chosen so that $\omega_n \leq \epsilon/2$ for all $n\geq n(\epsilon)$. This completes the proof of the first claim \eqref{eq:full_R}.

To prove the second claim \eqref{eq:full_R_2}, notice that $R_\bC \leq 1$, hence
\begin{eqnarray*}
\lefteqn{\bbE[R_\bC - R]}\\
     &=& \bbE[R_\bC - R | R_\bC-R \leq \epsilon]\cdot \Pr[R_\bC-R \leq \epsilon] \\
    &&+ \bbE[R_\bC - R | R_\bC-R > \epsilon]\cdot \Pr[R_\bC-R > \epsilon]  \\
    &\leq & \epsilon\cdot 1 + 1 \cdot q^{-n\epsilon /2},
\end{eqnarray*}
and the second claim follows by choosing $\epsilon = \frac{2\log n}{n}$.
\hfill$\blacksquare$

\section{Proof of Theorem~\ref{thm:Pe_2mac}} \label{app:Pe_2mac}
The proof of this theorem is very similar to the proof of Theorem~\ref{thm:Pe}. The key difference is that when $(X_1,X_2) = (\bc_{1,1},\bc_{2,1})$ is transmitted, the set of codeword pairs for which the ML decoder fails to decode is separated into three groups:
\begin{align*}
    (\bc_{1,i},\bc_{2,1})&, \mbox{ for some } i\neq 1, \\
    (\bc_{1,1},\bc_{2,j})&, \mbox{ for some } j\neq 1, \\
    (\bc_{1,i},\bc_{2,j})&, \mbox{ for some } (i,j)\neq (1,1).
\end{align*}

By the given code construction, 
transmitter $i$ employs a random code from the ${\rm LDPC}(\lambda_i,\rho_i,\delta_i;n)$ ensemble. Notice that the codebook is restricted by our code design to include precisely $M_i = q^{nR_i}$ codewords for each transmitter, where the design rate $R_i = 1 - \frac{\lambda_i}{\rho_i}$ for $i\in \{1,2\}$.

Denote $$\bc_{(i)}=\{\bc_{i,1},\ldots,\bc_{i,M_i}\}$$ as the codebook for transmitter $i$ before applying coset vector and quantization.
% , and $\bd=\{\bd_\bm:\bm\in [M]^K\}$ describes the corresponding MAC codebook; 
% here for any $\bm=(m(1),\ldots,m(K))$, $\bd_\bm=(\bc_{m(1)},\ldots,\bc_{m(K)})$.
Given the coset vectors $\bv_1$ and $\bv_2$ and quantizers $\delta_1$ and $\delta_2$, 
the resulting set of channel inputs is $$\{(\delta_1(\bc_{1,k_1}+\bv_1),\delta_2(\bc_{2,k_2}+\bv_2)):(k_1,k_2)\in [M_1]\times[M_2]\}.$$

For notational simplicity, let $\bd=\{\bd_\bm:\bm\in [M_1] \times [M_2]\}$ describe the corresponding MAC codebook; 
here for any $\bm=(m(1),m(2))$, $\bd_\bm=(\bc_{1,m(1)},\bc_{2,m(2)})$. The corresponding channel input is
\[
\bdel(\bd_\bm+\bv) \defeq (\delta_1(\bc_{1,m(1)}+\bv_1),\delta_2(\bc_{2,m(2)}+\bv_2)).
\]
The expected value under our random code construction 
of the average error probability is
\begin{eqnarray*}
E[P_e^{(n)}] 
	& = & \sum_\bm\sum_\bd\sum_\bv P_\bM(\bm)P_\bD(\bd)P_\bV(\bv)P_{e|\bm,\bd,\bv}^{(n)} \\
	& = & E_{\bM\bD\bV}\left[P_{e|\bM,\bD,\bV}^{(n)}\right],
\end{eqnarray*}
where $P_{e|\bm,\bd,\bv}^{(n)}$ is the conditional error probability 
under fixed values of the message vector $\bm=(m(1),m(2))$, codebook $\bd = \bc_{(1)}\times \bc_{(2)}$, and coset matrix $\bv=(\bv_1,\bv_2)$, 
$P_\bM(\bm)$, $P_\bD(\bd)$, and $P_\bV(\bv)$ capture the (independent, uniform) 
distributions on the vectors of possible messages over $[M_1]\times [M_2]$, set of possible codebooks, and cosets over $\GF(q)^n \times \GF(q)^n$, respectively, and 
$E_{\bM\bD\bV}[\cdot]$ is the resulting expectation. 

We begin by bounding the conditional error probability 
$P_{e|\bm,\bd,\bv}^{(N)}$.  
Let 
\begin{eqnarray*}
\bcY_{\bm,\bd,\bv}  & =  & \left\{\by:\exists\ \bm'\in [M_1]\times[M_2]\setminus\{\bm\} \mbox{ s.t. }\right. \\
&& \left.\Pr\left[\by|\bdel(\bd_{\bm'}+\bv)\right]\geq\Pr\left[\by|\bdel(\bd_\bm+\bv)\right]\right\}.
\end{eqnarray*}
Then 
\[
P_{e|\bm,\bd,\bv}^{(n)} \leq \Pr\left[\left.\bcY_{\bm,\bd,\bv}\right|\bm,\bd,\bv\right], 
\]
which is an inequality rather than an equality since an error is not guaranteed for the case of a tie.  

The set $\bcY_{\bm,\bd,\bv}$ can be equivalently written as the union of the following sets
\begin{align*}
\bcY^1_{\bm,\bd,\bv}  & = \left\{\by:\exists\ i\in [M_1]\setminus\{m(1)\} \mbox{ s.t. }\right. \\
& \left.\Pr\left[\by|\bdel(\bd_{(i,m(2))}+\bv)\right]\geq\Pr\left[\by|\bdel(\bd_\bm+\bv)\right]\right\},\\
\bcY^2_{\bm,\bd,\bv}  & = \left\{\by:\exists\ j\in [M_2]\setminus\{m(2)\} \mbox{ s.t. }\right. \\
& \left.\Pr\left[\by|\bdel(\bd_{(m(1),j)}+\bv)\right]\geq\Pr\left[\by|\bdel(\bd_\bm+\bv)\right]\right\},\\
\bcY^{12}_{\bm,\bd,\bv}  & = \left\{\by:\exists\ i\in [M_1]\setminus\{m(1)\}, j\in [M_2]\setminus\{m(2)\} \right. \\
& \left.\mbox{ s.t. }\Pr\left[\by|\bdel(\bd_{(i,j)}+\bv)\right]\geq\Pr\left[\by|\bdel(\bd_\bm+\bv)\right]\right\}.
\end{align*}
Therefore, by the union bound
\begin{align}
    \Pr[\bcY_{\bm,\bd,\bv}|\bm,\bd,\bv] &\leq \sum_{i \in \{1,2,12\}} \Pr[\bcY^i_{\bm,\bd,\bv}|\bm,\bd,\bv]. \label{eqn:Pe_2mac_union}
\end{align}
For the first term in the summation, abbreviating $i \in [M_1]\setminus\{m(1)\}$ to $i \neq m(1)$ and taking the expectation over $\bm, \bd, \bv$ gives
\begin{align*}
\lefteqn{E[\Pr[\bcY^{1}_{\bM,\bD,\bV}|\bM,\bD,\bV]]} \\
& =  \sum_{\bm,\ba,\by}P_\bM(\bm)P_{\bD_\bm+\bV}(\ba)\Pr[\by|\delta(\ba)] \\
& \hphantom{ \sum} \cdot\Pr[\exists i\in [M_1] \setminus\{m(1)\}: \Pr[\by|\delta(\bD_{(i,m(2))}+\bV)] \\
& \hphantom{\cdot\Pr[]=}\geq\Pr[\by|\delta(\bD_\bm+\bV)] |\bD_\bm+\bV=\ba]\\
& = \sum_{\bm,\ba,\by}P_\bM(\bm)P_{\bD_\bm+\bV}(\ba)\Pr[\by|\delta(\ba)] \\
&\hphantom{ \sum} \cdot\Pr[\exists i\in [M_1] \setminus\{m(1)\}: \bD_{(i,m(2))}+\bV=\ba', \\
& \hphantom{\cdot\Pr[]}\Pr[\by|\delta(\ba')]\geq \Pr[\by|\delta(\ba)]|\bD_\bm+\bV=\ba] \\
& \stackrel{(a)}{\leq}  \sum_{\bm,\ba,\by}P_\bM(\bm)P_{\bD_\bm+\bV}(\ba)\Pr[\by|\delta(\ba)]\min\left\{1,\sum_{i\neq m(1)}\right.\\
& \hphantom{ \sum} \left. \sum_{\Pr[\by|\delta(\ba')]\geq \Pr[\by|\delta(\ba)]}
	\mkern-45mu \Pr[\bD_{(i,m(2))}+\bV=\ba'|\bD_\bm+\bV=\ba]\right\} \\
& \stackrel{(b)}{\leq}  \sum_{\bm,\ba,\by}P_\bM(\bm)P_{\bD_\bm+\bV}(\ba)\Pr[\by|\delta(\ba)]\left(\sum_{i\neq m(1)}\right. \\
&\hphantom{ \sum} \sum_{\Pr[\by|\delta(\ba')]\geq \Pr[\by|\delta(\ba)]}
	\mkern-45mu \Pr[\bD_{(i,m(2))}+\bV=\ba'|\bD_\bm+\bV=\ba]\left.\vphantom{\sum_{\bm'\neq\bm}}\right)^\rho \\
& \stackrel{(c)}{=}  \sum_{\by,\ba} P_{\bD_\bone+\bV}(\ba) \Pr[\by|\delta(\ba)]\left(\sum_{\substack{i\neq 1}}\right.\\
&\hphantom{ \sum}\!\!\! \sum_{\Pr\by|\delta(\ba'])\geq \Pr[\by|\delta(\ba)]}
	\mkern-45mu\Pr[\bD_{(i,1)}+\bV=\ba'|\bD_\bone+\bV=\ba]\left.\vphantom{\sum_{\bm'\neq\bm}}\right)^\rho,
\end{align*}
where $(a)$ follows from the union bound and the bounded nature of probabilities, 
$(b)$ follows by a case analysis for any $\rho\in[0,1]$: $\min\{1,a\} = 1 \leq a^\rho$ when $a\geq 1$, and $\min\{1,a\} = a \leq a^\rho$ when $0\leq a<1$; and 
$(c)$ follows for $\bone=(1,1)$ 
by the symmetry of our random code design. 
Under our random code design and coset choice, 
for any $i \neq 1$ 
\begin{eqnarray*}
\lefteqn{\Pr[\bD_\bone+\bV=\ba,\bD_{(i,1)}+\bV=\ba']} \\
& = & \sum_{\bv}\Pr[\bV=\bv,\bD_\bone=\ba-\bv,\bD_{(i,1)}-\bD_\bone=\ba'-\ba] \\
& = & q^{-2n}\sum_{\bv}\Pr[\bD_\bone=\ba-\bv,\bD_{(i,1)}-\bD_\bone=\ba'-\ba] \\
& = & q^{-2n}\Pr[\bD_{(i,1)}-\bD_\bone=\ba'-\ba]\\
&\stackrel{(d)}{=}& q^{-2n}\Pr[\bC_{1,i}-\bC_{1,1}=\ba'[*,1]-\ba[*,1]]\\
& \stackrel{(e)}{\leq} & q^{-2n}\Pr[\ba'[*,1]-\ba[*,1]\in \bC_{(1)}] \\
&& \cdot \Pr[\bC_{1,i}-\bC_{1,1}=\ba'[*,1]-\ba[*,1] \\
&&	\mkern100mu |\ba'[*,1]-\ba[*,1]\in\bC_{(1)}] \\
& \stackrel{(f)}{\leq} & q^{-2n}\frac{\oS^n_1(\cT^n_q(\ba'[*,1]-\ba[*,1]))}{B\left(n,\cT^n_q(\ba'[*,1]-\ba[*,1])\right)} \frac1{M_1-1} \\
& \stackrel{(g)}{\leq} & q^{-2n}\left(\alpha_1 q^{-n}\right),
\end{eqnarray*}
where $\oS^n_1(\bt)$ refers to the ensemble-average number of type-$\bt$ codewords for transmitter $1$, $(d)$ follows since $\bD_{(i,1)}=(\bC_{1,i},\bC_{2,1})$ and $\bD_\bone=(\bC_{1,1},\bC_{2,1})$, and $\ba'[*,1]/\ba[*,1]$ refers to the first column of the $n\times 2$ codematrix $\ba'/\ba$, $(e)$ follows since the difference between two codewords 
is also a codeword in any linear code, and the upper bound still holds even we select $M_1=q^{nR_1}$ codewords for transmitter 1, $(f)$ follows since the number of codewords $M(\bC_{(1)})$ 
in the random codebook for transmitter 1 $M(\bC_{(1)})\geq M_1 = q^{nR_1}$ prior to our random restriction to precisely $M_1$ codewords, 
and $(g)$ follows from the definition of $\alpha_1$ in (\ref{eqn:alp1_2mac}).

Since $P_{\bD_\bone+\bV}(\ba)=q^{-2n}$ 
by the uniformity of random coset matrix $\bV$, 
\[
\Pr[\bD_{(i,1)}+\bV=\ba'|\bD_\bone+\bV=\ba]\leq\alpha_1 q^{-n}.
\]

Therefore
\begin{align*}
\lefteqn{E[\Pr[\bcY^{1}_{\bM,\bD,\bV}|\bM,\bD,\bV]]} \\
& \leq   \sum_{\by,\ba} q^{-2n}\Pr[\by|\delta(\ba)] \\
& \hphantom{====} \cdot\left(\sum_{i=2}^{q^{nR_1}}\right.
	\sum_{	\Pr[\by|\delta(\ba')]\geq \Pr[\by|\delta(\ba)]}
	\!\!\!\!\!\alpha_1 q^{-n}\left.\vphantom{\sum_{\bm'\neq\bone}}\right)^\rho \\
%&& \cdot 1 +  q^{-n\epsilon /2}  \cdot 1\\
& \leq   \alpha_1^\rho \sum_{\by,\ba}q^{-2n}  \Pr[\by|\delta(\ba)] \\
&\hphantom{====} 	\cdot \left((q^{nR_1}-1)\sum_{\ba':\Pr[\by|\delta(\ba')]\geq  \Pr[\by|\delta(\ba)]}
	\!\!\!\!\!\!\!q^{-n}\right)^\rho 
\\ 
& \leq   \alpha_1^\rho q^{nR_1\rho}
	\sum_{\bx_1,\bx_2,\by} P_{\bY|\bX_1,\bX_2}(\by|\bx_1,\bx_2)\sum_{\ba:\bdel(\ba)=(\bx_1,\bx_2)}q^{-2n} \\
&\hphantom{====}	\cdot \left(\sum_{\bx_1': \frac{P_{\bY|\bX_1,\bX_2}(\by|\bx_1',\bx_2)}{P_{\bY|\bX_1,\bX_2}(\by|\bx_1,\bx_2)}\geq1}\sum_{\ba':\bdel(\ba')=(\bx_1',\bx_2)}q^{-n}\right)^\rho \\
& =  \alpha_1^\rho q^{nR_1 \rho}\sum_{\bx_1,\bx_2,\by} P_{\bY|\bX_1,\bX_2}(\by|\bx_1,\bx_2) P_{\bX_1}(\bx_1)P_{\bX_2}(\bx_2)  \\
&\hphantom{====}
	\cdot \left(\sum_{\bx_1': \frac{P_{\bY|\bX_1,\bX_2}(\by|\bx_1',\bx_2)}{P_{\bY|\bX_1,\bX_2}(\by|\bx_1,\bx_2)}\geq1}P_{\bX_1}(\bx_1')\right)^\rho \\
& \stackrel{(h)}{\leq}   \alpha_1^\rho q^{nR_1 \rho}\sum_{\bx_1,\bx_2,\by}P_{\bY|\bX_1,\bX_2}(\by|\bx_1,\bx_2) P_{\bX_1}(\bx_1)P_{\bX_2}(\bx_2) \\
&\hphantom{===}
	\cdot \left(\sum_{\bx_1'}P_{\bX_1}(\bx_1')\left(\frac{P_{\bY|\bX_1,\bX_2}(\by|\bx_1',\bx_2)}{P_{\bY|\bX_1,\bX_2}(\by|\bx_1,\bx_2)}\right)^s\right)^\rho \\
& =   \alpha_1^\rho q^{nR_1\rho}\sum_\by \sum_{\bx_2} P_{\bX_2}(\bx_2) \\
& \hphantom{===} \cdot \left(\sum_{\bx_1} P_{\bX_1}(\bx_1)P_{\bY|\bX_1,\bX_2}(\by|\bx_1,\bx_2)^{1-s\rho}\right) \\
&\hphantom{===}	\cdot \left(\sum_{\bx_1'}P_{\bX_1}(\bx_1')P_{\bY|\bX_1,\bX_2}(\by|\bx_1',\bx_2)^s\right)^\rho, \label{eqn:Pe_2mac_alpha1}
\end{align*}
where  $(h)$ holds for any $s>0$.

When $s=1/(1+\rho)$, rewriting the result \eqref{eqn:Pe_2mac_alpha1} in an exponential form using error exponent from \eqref{eqn:Error_exp_2mac_a1} and \eqref{eqn:Error_exp_2mac_a1_0}, and optimizing over $0\leq \rho \leq 1$ gives
\begin{align}
E[\Pr[\bcY^{1}_{\bM,\bD,\bV}|\bM,\bD,\bV]] \leq q^{-nE_{p_1}(R_1+\frac{\log\alpha_1}{n})}.
\end{align}

Switching the role of transmitter $1$ and transmitter $2$ in the above proof, we obtain
\begin{align}
E[\Pr[\bcY^{2}_{\bM,\bD,\bV}|\bM,\bD,\bV]] \leq q^{-nE_{p_2}(R_2+\frac{\log\alpha_2}{n})}.
\end{align}

Finally, $E[\Pr[\bcY^{12}_{\bM,\bD,\bV}|\bM,\bD,\bV]]$ can be bounded using the same technique as the proof of Theorem~\ref{thm:Pe}, giving
\begin{align}
    E[\Pr[\bcY^{12}_{\bM,\bD,\bV}|\bM,\bD,\bV]] \leq q^{-nE_{p_{12}}(R_1+R_2+\frac{\log\alpha_1\alpha_2}{n})}.
\end{align}

Plugging the three expressions above into \eqref{eqn:Pe_2mac_union} completes the proof.
\hfill$\blacksquare$
\section{Proof of Theorem~\ref{thm:ebd}}\label{app:ebd}
Given a DM-PPC $(\cX,P_{Y|X}(y|x),\cY)$ with capacity achieving distribution $P_X$ and capacity $C$, Gallager's error exponent is defined as
\begin{align}
    E_p(R) & \defeq \max_{0\leq \rho\leq 1} [E_0(\rho,P_X)-\rho R],
\end{align}
where
\begin{align}
    E_0(\rho,P_X) & \defeq  -\log_e\sum_{y\in\cY}\left[\sum_{x\in\cX}P_X(x)P_{Y|X}(y|x)^{1/(1+\rho)}\right]^{1+\rho}, 
\end{align}
Applying a second-order Taylor expansion to $E_0(\rho,P_X)$ at $\rho = 0$ gives
\begin{align}
    E_0(\rho,P_X) &= E_0(0,P_X) +\rho E'_0(0,P_X) + \frac{\rho^2}{2}E''_0(\rho*,P_X) \label{eqn:ebd_pr_1}
\end{align}
for some $\rho* \in [0,\rho]$.

Direct calculation gives $E_0(0,P_X) = -\log 1 =0$, and \cite[Eq. 5.5.30]{Gallager:68} shows $E'_0(0,P_X) = C$. 

Let $\beta$ be an upper bound for $-E''_0(\rho*,P_X)$. Then \eqref{eqn:ebd_pr_1} becomes
\begin{align}
    E_0(\rho,P_X) &= 0 + \rho C + \frac{\rho^2}{2}E''_0(\rho*,P_X) \\
    &\geq \rho C - \frac{\rho^2}{2}\beta.
\end{align}
Therefore,
\begin{align}
    E_p(R) \geq \rho C - \frac{\rho^2}{2}\beta - \rho R \label{eqn:ebd_pr_2}.
\end{align}

The right-hand side of \eqref{eqn:ebd_pr_2} is a concave quadratic function in $\rho$. Taking its derivative and equating the derivative to $0$ yields the following stationary point 
\begin{align}
    \rho = \frac{C-R}{\beta},
\end{align}
giving
\begin{align}
    E_p(R) &\geq \frac{C-R}{\beta}C - \left(\frac{C-R}{\beta}\right)^2 \frac{\beta}{2} - \frac{C-R}{\beta}R \\
    &= \frac{(C-R)^2}{2\beta}, \text{ for } C-R\leq \beta. \label{eqn:ebd_pr_3}
\end{align}

Following the outline in \cite[Exercise~5.23]{Gallager:68}, one can show that (proof omitted)
\begin{align}
    -E''_0(\rho,P_X) \leq \frac{4}{e^2} +\log_e^2|\cY| - [E'_0(\rho,P_X)]^2. \label{eqn:ebd_pr_4}
\end{align}

% Before the comparison of Polyanskiy's and Gallager's approaches, we note that $E_0(\rho,Q)$ has the following properties (\cite[Th.5.6.3]{Gallager:68})
Note that $E_0(\rho,P_X)$ has the following properties (\cite[Th.5.6.3]{Gallager:68})
\begin{align}
    E_0(\rho,P_X) &\geq 0 , ~\rho \geq 0, \label{eqn:E0_1}\\
   E_0'(\rho,P_X) &> 0 , ~\rho \geq 0,\label{eqn:E0_2}\\
    E_0''(\rho,P_X) &\leq 0 , ~\rho \geq 0. \label{eqn:E0_3}
\end{align}

Therefore, 
\begin{align}
    \min_{\rho \in [0,1]} E'_0(\rho,P_X) = E'_0(1,P_X),
\end{align}
and $R_{cr} \defeq E'_0(1,P_X)$ is known as the critical rate \cite[Eq. (5.6.30)]{Gallager:68}.

Plugging $R_{cr}$ into \eqref{eqn:ebd_pr_4} gives
\begin{align}
    -E''_0(\rho,P_X) \leq \frac{4}{e^2} +\log_e^2|\cY| - R_{cr}^2.
\end{align}

The bound in \eqref{eqn:ebd_pr_3} requires $C-R \leq \beta$ so that $\rho$ is within $[0,1]$. This means that $\beta$ can be taken as
\begin{itemize}
    \item $\beta = \frac{4}{e^2} +\log_e^2|\cY|$, which is valid for all $0\leq R\leq C$, or
    \item $\beta = \frac{4}{e^2} +\log_e^2|\cY| -R_{cr}^2$, which is valid for $\max\{0,C- (\frac{4}{e^2} +\log_e^2|\cY| -R_{cr}^2)\} \leq R \leq C$.
\end{itemize}

Finally, substituting $\beta = \frac{4}{e^2} +2\log_e^2|\cY|$ (which is looser than the conservative value $\frac{4}{e^2} +\log_e^2|\cY|$) into \eqref{eqn:ebd_pr_3} gives a lower bound on $E_p(R)$. Invoking Theorem~\ref{thm:Pe_Gal} with this lower bound completes the proof. 
\hfill$\blacksquare$

% \section*{Acknowledgment}


\ifCLASSOPTIONcaptionsoff
  \newpage
\fi



%\bibliographystyle{IEEEtran}

\bibliographystyle{IEEEtran}
\newcommand{\noopsort}[1]{}



% \end{thebibliography}

% \begin{IEEEbiographynophoto}{Author}

%\newpage


% While the rows of $\bc(\bm)+\bv$ can take any value in $\cQ$, 
% the channel's symmetry across transmitters 
% implies that it is the type of each row of $\bc(\bm)+\bv$, rather than the value of that row, 
% that matters in influencing the channel output.  
% Let $\cT_K=\{\cT(g^K):\bg\in\cQ^k\}$ 
% be the set of possible types for $K$-vectors from $\cQ$ 
% and $\cT_N$ 

\end{document}